# The high-order toroidal moments and anapole states in all-dielectric photonics

Egor A. Gurvitz[1,*], Konstantin S. Ladutenko[1], Pavel A. Dergachev[2], Andrey B. Evlyukhin[1,3], Andrey. E. Miroshnichenko[1,4], and Alexander S. Shalin[1]

[1]ITMO University, Kronverkskiy pr., 49, St. Petersburg 197101, Russia

[2]National Research University Moscow Power Engineering Institute, Krasnokazarmennaya, st., 14, 111250 Moscow, Russia

[3]Laser Zentrum Hannover e.V., Hollerithallee 8, D-30419 Hannover, Germany

[4]School of Engineering and Information Technology, University of New South Wales Canberra, 2600, Australia

*Corresponding Author: egorgurvitz@gmail.com

**Abstract**

All-dielectric nanophotonics attracts ever increasing attention nowadays due to the possibility to control and configure light scattering on high-index semiconductor nanoparticles. It opens a room of opportunities for the designing novel types of nanoscale elements and devices, and paves a way to advanced technologies of light energy manipulation. One of the exciting and promising prospects is associated with the utilizing so called toroidal moment being the result of poloidal currents excitation, and anapole states corresponding to the interference of dipole and toroidal electric moments. Here, we present and investigate in details via the direct Cartesian multipole decomposition higher order toroidal moments of both types (up to the electric octupole toroidal moment) allowing to obtain new near- and far-field configurations. Poloidal currents can be associated with vortex-like distributions of the displacement currents inside nanoparticles revealing the physical meaning of the high-order toroidal moments and the convenience of the Cartesian multipoles as an auxiliary tool for analysis. We demonstrate high-order nonradiating anapole states (vanishing contribution to the far-field zone) accompanied by the excitation of intense near-fields. We believe our results to be of high importance for both fundamental understanding of light scattering by high-index particles, and a variety of nanophotonics applications and light governing on nanoscale.

# 1. Introduction

All-dielectric nanophotonics is one of the fastest-growing hot topics in the field of modern researches [1], aimed to the investigation of nearly lossless high refractive index semiconductor and dielectric nanoparticles (for example, Si, Ge, TiO2) for applications in the optical frequency range. Such structures can possess both electric and magnetic multipole resonances opening new and impressive possibilities to control all the components of light on nanoscales [2].

The applications of the all-dielectric nanophotonics are widespread and manifold, for example, waveguides (discrete, unidirectional, lossless, etc.) [3,4], modulators [5], directional sources (e.g. Huygens sources) of light and nanoantennas [6], detectors [7], devices for cloaking and invisibility [8], phase metasurfaces [9], etc. The development of the dielectric nanophotonics in the near future will allow to invent different materials and devices for handling an optical signal of any complexity almost without losses.

Modern electrodynamic toolkit describing the light interaction with high-index dielectric particles is based on the multipole decomposition, which was recently substantially extended by introducing the so-called toroidal moments [10–12]. When the optical size of a particle is large enough, the system can support displacement current oscillations along toroidal surfaces (besides the linear and closed loop charge currents oscillations) leading to the toroidal moments excitation [13].

Toroidal moment may be equivalently represented as a circulation of an effective magnetic current (see **Figure 1**) [14,15]. During last decade, there were some works reporting the systems possessing a dominating toroidal response experimentally. The pioneer work was done for artificial toroidal metamolecules in Zheludev's group for microwave frequency range [16]. Later on, a toroidal electric dipole response was also demonstrated in the terahertz frequency range [17] and for the optical frequencies [18].

Toroidal moments are not limited to the electric type; there are few works showing the existence of the corrections to the magnetic dipole moments, which can be associated with the magnetic toroidal dipoles [19,20]. They are shown to take place in complex materials (e.g., $LiCoPO_4$, {Dy4} - tetranuclear dysprosium, {Dy6} – triethanolamine, etc) [20–24] or in high index nanoparticles [25]. Thereby the electric and magnetic toroidal moments complement the conventional electric and magnetic ones and form a full set of all possible sign permutations under the inversion of space $r \to -r$ and time $t \to -t$ [13].

Figure 1 shows that electric and magnetic toroidal dipoles inherit the same "space" coordinates inversion property as the basic electric and magnetic dipoles. It is not the case for the "time" inversion: the toroidal electric dipole changes its sign and the toroidal magnetic dipole doesn't, in contrast to the basic multipoles (see Figure 1).

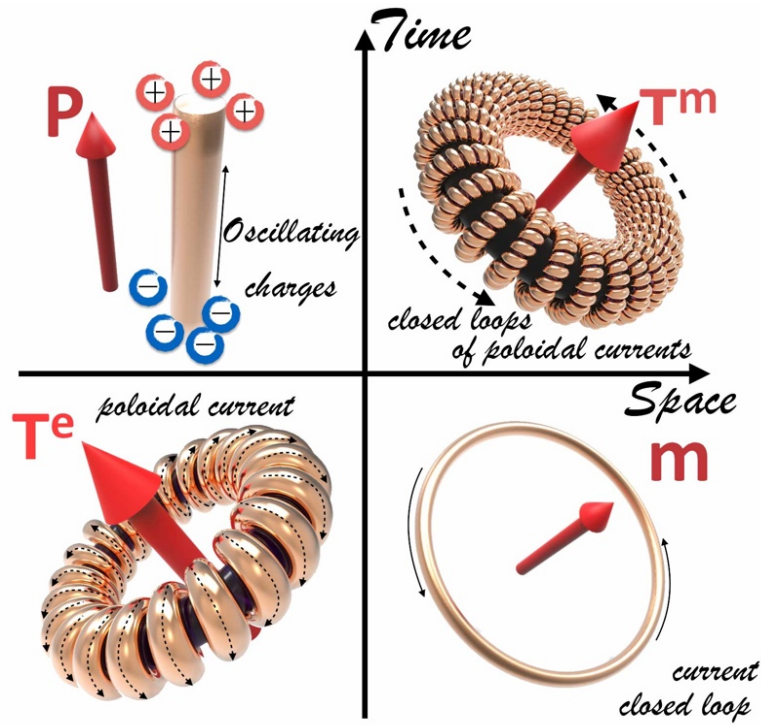

**Figure 1.** Illustration of the space-time coordinate inversion properties for the full set of dipoles: electrical, toroidal electric, magnetic and toroidal magnetic.

The system of currents providing toroidal response can cause oscillations of a vector potential outside in the absence of the electromagnetic field [26], a medium containing molecules with elements of toroidal symmetry can rotate the polarization of light [27] etc. making it to be very interesting for an investigation. On the other hand, the toroidal contributions could be hidden by stronger electric or magnetic dipole responses making their direct observation very difficult. Another problem for the characterization and observation of the toroidal moments is related to the fact that in the far-field zone their contributions are identical to the contributions of basic dipoles, which causes some discussions on the need of their introduction [28,29].

The electromagnetic field generated by toroidal dipole moments can interfere with the fields produced by electric and magnetic dipoles constructively forming super-dipole states [17,30–32], or destructively giving rise to nonradiative current configurations - "anapole" states [16,33,34]. Now, the study of nonradiative current configurations is in its initial stage, and only electric dipole "anapole" states of different orders [30,33–35] and magnetic "anapoles" (destructive interference of a magnetic dipole moment and a corresponding toroidal multipole) [25] are shown. The Mie theory allows to investigate numerous zeros of the scattering coefficients for spherical particles, but do not allow to distinguish between basic and/or toroidal contributions corresponding to different currents configurations [36]. Therefore, neither higher order toroidal multipoles and anapole states were identified nor discussed.

Thus, in this manuscript (section 2) we extend the Cartesian multipole expansion up to the $5^{th}$ order limited by a magnetic 16-pole and electric 32-pole. The primitive multipoles are reduced to symmetric and traceless forms. Among the residual terms the radiative ones form the first 5 toroidal moments (up to the $3^{rd}$ order – toroidal electric octupole and toroidal magnetic quadrupole) and the second order toroidal correction to an electric dipole. We show the formal independency of the toroidal moments using the requirement of irreducibility of the tensors with respect to SO(3) group. In comparison to the $3^{rd}$ order approximation of the Cartesian multipole decomposition, the obtained results allow to analyze in details the light scattering by optically large particles and pave the way to achieving new optical effects due to the additional degrees of freedom provided by new electric, magnetic and toroidal multipoles contributions. In the summary of this section, we present a generalization of the cumbersome algebra for the calculation of a power coefficients of multipole moments and multipole expansion of a scattered power, scattering cross-section and scattering efficiency.

The $3^{rd}$ section demonstrates the verification of the Cartesian multipoles up to the $5^{th}$ order via the comparison with spherical multipoles and numerical calculations of a scattering efficiency. We show that a multipole accompanied by a toroidal moment fully coincides with the corresponding spherical multipole, where the basic Cartesian multipoles differ from the spherical exactly by the toroidal configurations of the displacement currents or fields inside the particles. For the multipoles of $4^{th}$ and $5^{th}$ orders we obtain a mismatch caused by lack of the higher order toroidal moments.

In the $4^{th}$ section we prove Cartesian multipoles to provide additional useful information about the currents configurations in a system. For instance, we for the first time demonstrate the explicit impact of high-order toroidal moments into the first five "anapole" states (actually – zeroes of contributions to the far-field because of the destructive interference between basic and toroidal multipoles, however, with the nontrivial near-fields) and the corresponding near-field maps.

## 2. Irreducible multipole moments

### 2.1 Primitive multipole moments. E-field multipole decompositions

There are several widely-known textbooks devoted to the multipole decomposition of a scattered field [37,38]. They describe the problem using irreducible infinite series of orthogonal sets implicitly comprising toroidal moments [28]. Other approaches [39–42] reveal toroidal moments via delta-function based series or retarded potentials.

Let us follow [42] and consider a Taylor series of the retarded scalar and vector potentials

$$\Phi(\mathbf{R}, t) = \frac{1}{4\pi\varepsilon_0} \int_V \frac{\rho(\mathbf{r}, t' + \Delta t)}{|\mathbf{R} - \mathbf{r}|} dv, \quad (1)$$

$$\mathbf{A}(\mathbf{R}, t) = \frac{\mu_0}{4\pi} \int_V \frac{\mathbf{J}(\mathbf{r}, t' + \Delta t)}{|\mathbf{R} - \mathbf{r}|} dv. \quad (2)$$

Here $\rho$ and $\mathbf{J}$ are charge and current densities in an infinitesimal volume $dv$, $\varepsilon_0$ and $\mu_0$ are vacuum permittivity and permeability. The vector from the origin of coordinates to an observation point is denoted as $\mathbf{R}$, the vector from the origin of coordinates to an arbitrary point of the charge/current distribution area is denoted as $\mathbf{r}$, and retarded time at $dv$ is $(t' + \Delta t) = t - |\mathbf{R} - \mathbf{r}|/c$, where $c$ is speed of light in vacuum and $t' = t - R/c$ is the time retardation at the coordinate origin [41]. The time dependence is assumed to be $e^{-i\omega t}$, where $\omega$ is an angular frequency.

We perform an expansion of the scalar and vector potentials (see, e.g., Supplementary section 1) omitting all near-field components, and apply Lorentz gauge $\mathbf{E} = -\nabla\Phi - \dot{\mathbf{A}}$ to obtain the E-field in the far-field zone:

$$E_i = \frac{k^2}{4\pi\varepsilon_0 R} e^{i\mathbf{k}\mathbf{r}} \left[ (n_i n_j - n^2 \delta_{ij}) \left( -p_j + \frac{ik}{2} \overline{Q}_{jk}^{(e)} n_k + \frac{k^2}{6} \overline{O}_{jkp}^{(e)} n_k n_p + \right. \right.$$
$$\left. + \frac{-ik^3}{24} \overline{S}_{jkpl} n_k n_p n_l + \frac{-k^4}{120} \overline{X}_{jkplt} n_k n_p n_l n_t \right) + \varepsilon_{ikj} n_k \left( \frac{-1}{c} m_j + \frac{ik}{2c} Q_{jp}^{(m)} n_p \right. \quad (3)$$
$$\left. \left. + \frac{k^2}{6c} O_{jpl}^{(m)} n_p n_l + \frac{-ik^3}{24c} Y_{jplt} n_p n_l n_t \right) \right],$$

where **k** – is a wave-vector, $\varepsilon_{ikj}$ is a Levi-Chivita symbol, and repeated indices imply the summation over them in accordance to the Einstein's notation, $n_i = R_i/|\mathbf{R}|$ and the obtained primitive multipole moments are shown in **Table 1**.

**Table 1.** Primitive Cartesian multipoles

| Order | Electric type multipoles | Magnetic type multipoles |
|---|---|---|
| 1 | $p_i = \dfrac{i}{\omega}\int J_i dv$ | - |
| 2 | $\overline{Q}^{(e)}_{ij} = \dfrac{i}{\omega}\int r_j J_i + r_i J_j\, dv$ | $m_j = \dfrac{1}{2}\int (\mathbf{r}\times\mathbf{J})_j\, dv$ |
| 3 | $\overline{O}^{(e)}_{ijk} = \dfrac{i}{\omega}\int r_k r_j J_i + r_k r_i J_j + r_i r_j J_k\, dv$ | $Q^{(m)}_{jp} = \dfrac{2}{3}\int r_p (\mathbf{r}\times\mathbf{J})_j\, dv$ |
| 4 | $\overline{S}_{ijkl} = \dfrac{i}{\omega}\int r_l r_k r_j J_i + r_l r_k r_i J_j +$ $+ r_l r_i r_j J_k + r_i r_k r_j J_k\, dv$ | $O^{(m)}_{jpl} = \dfrac{3}{4}\int r_l r_p (\mathbf{r}\times\mathbf{J})_j\, dv$ |
| 5 | $\overline{X}_{ijklp} = \dfrac{1}{\omega}\int r_p r_l r_k r_j J_i + r_p r_l r_k r_i J_j +$ $+ r_p r_l r_i r_j J_k + r_p r_i r_k r_j J_l + r_i r_l r_k r_j J_p\, dv$ | $Y_{jplt} = \dfrac{4}{5}\int r_t r_l r_p (\mathbf{r}\times\mathbf{J})_j\, dv$ |

The expansion of the electric field and the derivation of multipoles are given in the Supplementary Materials. In contrast to spherical multipoles, the order of Cartesian multipoles coincides with the order of the Taylor series expansion. Therefore, an electric dipole individually forms the 1st order, while an electric quadrupole and a magnetic dipole form the 2nd order, etc.

Starting with an electric quadrupole, all electric type primitive multipoles are overlined (see **Table 2**) in our notation, this denotes a symmetrical tensor. All the integral expressions of overlined multipoles consist of the sum of tensor products of vectors with a cyclic permutation of indices. Later on double overlined multipoles will correspond to symmetrical, traceless tensors, but primitive multipoles are not traceless in general.

Therefore, the multipole moments in Table 1 do not form a fully independent set [41,42]: the detailed explanation is given in the next section.

**2.2 Formulation of irreducible basic multipoles and toroidal terms**

2.2.1. Irreducible multipoles

Starting from Racah's works devoted to the tensor algebra operations and Wigner's developments on the group theory [43,44], which are usually applied to the reduction of tensors in quantum mechanics (spin operators, angular moments, Hamiltonians, etc), the irreducibility of tensors in physics became an inherent part of complex systems analysis [45–48]. The irreducibility appears from the requirement of a tensor defined on a group to be invariant under transformations defined for this group.

The Cartesian multipoles must be invariant with the respect to the mirror reflections and the rotations of the SO(3) group [49], which implies symmetrization and detracing of the Cartesian tensors [47,48,50].

The tensor symmetrization is performed using this a simple relation:

$$\bar{A}_{\alpha_1..\alpha_n} = \frac{1}{n!} \sum_{[\alpha_1..\alpha_n]} A_{\alpha_1..\alpha_n}. \tag{4}$$

In the left part an overlined tensor $\bar{A}$ defines a symmetrized tensor and the expression in the right part denotes the sum with all possible permutations of indices, which is multiplied by an inverse number of combinations.

Next, we should perform a tensor detracing procedure. Basically, the trace is well defined only for the 2-rank tensors. After detracing the sum of the matrix main diagonal elements should zero. For tensors with rank>2 the detracing operation is not so obvious. Following the Applequist's theorem [51] we find traces of symmetrized multipoles using:

$$\bar{\bar{A}}^{(n)}_{\alpha_1...\alpha_n} = \sum_{m=0}^{\left\lfloor \frac{n}{2} \right\rfloor} (-1)^m (2n-2m-1)!! \sum_{\{\alpha\}} \delta_{\alpha_1\alpha_2}...\delta_{\alpha_{(2m-1)}\alpha_{(2m)}} \bar{A}^{n:m}_{\alpha_{(2m-1)}\alpha_{(n)}}, \tag{5}$$

where $\left\lfloor \frac{n}{2} \right\rfloor$ is rounded down to the nearest integer, double exclamation sign (!!) denotes double factorial, the summation $\sum_{\{\alpha\}}$ corresponds to a sum over set of $\alpha_1..\alpha_n$ and gives distinct terms, and $\bar{A}^{n:m}$ denotes tensor of rank (n-m) contracted m/2 times (m is even integer), and $\delta_{\alpha_1 \alpha_2}$ is a delta Kronecker symbol. The resulting symmetrized and traceless tensor is denoted using double overline mark such as $\bar{\bar{A}}$.

Therefore, by symmetrizing and detracing Cartesian primitive multipoles we can derive arbitrary toroidal moments (see Supplementary Section 2 and 3). The far-field multipole decomposition up to the magnetic 16-pole and electric 32-pole in dyadic and tensor form is given for the electric field as:

$$\begin{aligned}
\mathbf{E} = \frac{k^2}{4\pi\varepsilon_0} e^{i\mathbf{kr}} \frac{1}{R} & \Bigg( \left[ \mathbf{n} \times \left[ \left( \mathbf{p} + \frac{ik}{c} \mathbf{T}^{(e)} + \frac{ik^3}{c} \mathbf{T}^{(2e)} \right) \times \mathbf{n} \right] \right] \\
& + \frac{ik}{2} \left[ \mathbf{n} \times \left[ \mathbf{n} \times \left( \left( \hat{\bar{\bar{Q}}}^{(e)} + \frac{ik}{c} \hat{\bar{\bar{T}}}^{(Qe)} \right) \cdot \mathbf{n} \right) \right] \right] + \\
& + \frac{k^2}{6} \left[ \mathbf{n} \times \left[ \mathbf{n} \times \left( \left( \hat{\bar{\bar{O}}}^{(e)} + \frac{ik}{c} \hat{\bar{\bar{T}}}^{(Oe)} \right) \cdot \mathbf{n} \cdot \mathbf{n} \right) \right] \right] + \\
& + \frac{ik^3}{24} \left[ \mathbf{n} \times \left[ \left( \hat{\bar{\bar{S}}} \cdot \mathbf{n} \cdot \mathbf{n} \cdot \mathbf{n} \right) \times \mathbf{n} \right] \right] + \frac{k^4}{120} \left[ \mathbf{n} \times \left[ \left( \hat{\bar{\bar{X}}} \cdot \mathbf{n} \cdot \mathbf{n} \cdot \mathbf{n} \cdot \mathbf{n} \right) \times \mathbf{n} \right] \right] \\
& + \frac{1}{c} \left[ \left( \mathbf{m} + \frac{ik}{c} \mathbf{T}^{(m)} \right) \times \mathbf{n} \right] + \frac{ik}{2c} \left[ \mathbf{n} \times \left( \left( \hat{\bar{\bar{Q}}}^{(m)} + \frac{ik}{c} \hat{\bar{\bar{T}}}^{(Qm)} \right) \cdot \mathbf{n} \right) \right] + \\
& + \frac{k^2}{6c} \left[ \mathbf{n} \times \left( \hat{\bar{\bar{O}}}^{(m)} \cdot \mathbf{n} \cdot \mathbf{n} \right) \right] + \frac{k^3}{24c} \left[ \left( \hat{\bar{\bar{Y}}} \cdot \mathbf{n} \cdot \mathbf{n} \cdot \mathbf{n} \right) \times \mathbf{n} \right] \Bigg)
\end{aligned}$$
(6)

and

$$\begin{aligned}
E_i = \frac{k^2}{4\pi\varepsilon_0 R} e^{i\mathbf{kr}} & \Bigg[ (n_i n_j - n^2 \delta_{ij}) \left( -1 \left( p_j + \frac{ik}{c} T_j^{(e)} + \frac{ik^3}{c} T_j^{(2e)} \right) + \\
& + \frac{ik}{2} \left( \bar{\bar{Q}}_{jk}^{(e)} + \frac{ik}{c} \bar{\bar{T}}_{jk}^{(Qe)} \right) n_k + \frac{k^2}{6} \left( \bar{\bar{O}}_{jkp}^{(e)} + \frac{ik}{c} \bar{\bar{T}}_{jkp}^{(Oe)} \right) n_k n_p + \frac{-ik^3}{24} \bar{\bar{S}}_{jkpl} n_k n_p n_l + \\
& + \frac{-k^4}{120} \bar{\bar{X}}_{jkplt} n_k n_p n_l n_t \right) + \varepsilon_{ikj} n_k \left( -\frac{1}{c} \left( m_j + \frac{ik}{c} T_j^{(m)} \right) + \frac{ik}{2c} \left( \bar{\bar{Q}}_{jp}^{(m)} + \frac{ik}{c} \bar{\bar{T}}_{jp}^{(Qm)} \right) n_p + \\
& + \frac{k^2}{6c} \bar{\bar{O}}_{jpl}^{(m)} n_p n_l + \frac{-ik^3}{24c} \bar{\bar{Y}}_{jplt} n_p n_l n_t \right) \Bigg].
\end{aligned}$$
(6.1)

The Equation (6) shows the contributions of irreducible multipoles to the far-field. High order tensors are denoted in dyadic notation with hat (e.g. $\hat{Q}$) and vectors such as **E**, **p**, **m** are bold. In tensor form we use subscript indices without the hat. The toroidal moments denoted with $T$ with superscripts (e.g., $\hat{\bar{\bar{T}}}^{(Qm)}$ contributing to the magnetic quadrupole) will be discussed in the next section. The irreducible traceless and symmetrical Cartesian multipoles are double overlined and they are considered hereinafter. The relations seem to be quite cumbersome, and we will write them step by step with a necessary discussion. The dipoles have the same form as previously:

$$p_j = \frac{i}{\omega}\int J_j dv, \qquad (7)$$

$$m_j = \frac{1}{2}\int (\mathbf{r}\times\mathbf{J})_j dv. \qquad (8)$$

The basic electric quadrupole is expressed as

$$\bar{\bar{Q}}^{(e)}_{jk} = \frac{i}{\omega}\int r_j J_k + r_k J_j - \frac{2}{3}\delta_{jk}(\mathbf{r}\cdot\mathbf{J})\,dv. \qquad (9)$$

Equation (8) and (9) form the 2$^{nd}$ order of the Cartesian multipoles. The 3$^{rd}$ order is formed by the electrical octupole:

$$\bar{\bar{O}}^{(e)}_{jkl} = \bar{O}^{(e)}_{jkl} - \frac{1}{5}\left(\delta_{jk}\bar{O}_{vvl} + \delta_{jl}\bar{O}_{vvk} + \delta_{lk}\bar{O}_{vvj}\right), \qquad (10)$$

where

$$\bar{O}^{(e)}_{vvj} = \frac{i}{\omega} \qquad (10.1)$$

and the magnetic quadrupole:

$$\bar{\bar{Q}}^{(m)}_{jp} = \frac{1}{3}\int \left((\mathbf{r}\times\mathbf{J})_j r_p + (\mathbf{r}\times\mathbf{J})_p r_j\right) dv. \qquad (11)$$

The 4$^{th}$ order of the basic Cartesian multipoles consist of the electric 16-pole and the magnetic octupole. The basic electric 16-pole is obtained by detracing the 4$^{th}$ rank tensor and then detracting again its residual 2$^{nd}$ rank resulting tensors:

$$\bar{\bar{S}}_{jkpl} = \bar{S}_{jkpl} - \frac{1}{7}(\delta_{pj}\bar{\bar{S}}_{vvkl} + \delta_{pk}\bar{\bar{S}}_{vvjl} + \delta_{jk}\bar{\bar{S}}_{vvpl} + \delta_{jl}\bar{\bar{S}}_{vvpk} +$$
$$+\delta_{kl}\bar{\bar{S}}_{vvpj} + \delta_{pl}\bar{\bar{S}}_{vvjk}) - \frac{1}{15}(\delta_{pj}\delta_{kl}\bar{S}_{uuvv} + \delta_{pk}\delta_{jl}\bar{S}_{uuvv} + \delta_{pl}\delta_{jk}\bar{S}_{uuvv}),$$
(12)

where the definitions of the lower rank constituent tensors are:

$$\bar{\bar{S}}_{vvjp} = \frac{i}{\omega}\int 2(\mathbf{r}\cdot\mathbf{J})r_j r_p + r^2 J_j r_p + r^2 J_p r_j - \frac{4}{3}(\mathbf{r}\cdot\mathbf{J})r^2 dv,$$
$$\bar{S}_{uuvv} = \frac{i}{\omega}\int 4(\mathbf{r}\cdot\mathbf{J})r^2 dv.$$
(12.1)

The irreducible magnetic octupole is obtained after the symmetrization and detracing of the 3$^{rd}$ rank tensor and the symmetrization of the residual 2$^{nd}$ rank tensors:

$$\bar{\bar{O}}_{jpl}^{(m)} = \bar{O}_{jpl}^{(m)} - \frac{1}{5}(\delta_{jp}\bar{O}_{vvl}^{(m)} + \delta_{jl}\bar{O}_{vvp}^{(m)} + \delta_{pl}\bar{O}_{vvj}^{(m)}) +$$
$$-\frac{1}{3}\varepsilon_{\alpha jp}\left(\bar{\bar{R}}_{\alpha l} + \frac{1}{2}\varepsilon_{\beta\alpha l}G_\beta\right) - \frac{1}{3}\varepsilon_{\alpha jl}\left(\bar{\bar{R}}_{\alpha p} + \frac{1}{2}\varepsilon_{\beta\alpha p}G_\beta\right),$$
(13)

with the lower rank tensors

$$\bar{O}_{jpl}^{(m)} = \frac{1}{3}\left(O_{jpl}^{(m)} + O_{pjl}^{(m)} + O_{ljp}^{(m)}\right),$$
$$\bar{O}_{vvj}^{(m)} = \frac{1}{4}\int r^2(\mathbf{r}\times\mathbf{J})_j dv,$$
$$\bar{\bar{R}}_{\alpha l} = \frac{3}{8}\int r^2(J_l r_\alpha + J_\alpha r_l) - 2(\mathbf{J}\cdot\mathbf{r})r_\alpha r_l \, dv,$$
$$G_\beta = -\frac{3}{4}\int r^2 (\mathbf{r}\times\mathbf{J})_j \, dv.$$
(13.1)

The electric 32-pole and magnetic 16-pole form the 5$^{th}$ order of the basic Cartesian multipoles. The irreducible electric 32-pole is obtained after detracing the 5$^{th}$ rank tensor and detracing resulting 3$^{rd}$ rank tensors:

$$\bar{\bar{X}}_{jkplt} = \bar{X}_{jkplt} - \frac{1}{9}\sum_{\{jkplt\}}\delta_{jk}\bar{\bar{X}}_{vvplt} - \frac{1}{35}\sum_{\{jkplt\}}\delta_{jk}\delta_{pl}\bar{X}_{vvuut},$$
(14)

taking into account the combination with all distinct indices repetitions in sums same as in Equation (5) and the definition of the 3$^{rd}$ and 1$^{st}$ rank tensors obtained after the detracing process:

$$\bar{\bar{X}}_{vvjkl} = \frac{i}{\omega}\int 2(\mathbf{r}\cdot \mathbf{J})r_j r_k r_l + r^2(r_j r_k J_l + r_l r_k J_j + r_j r_l J_k) \tag{14.1}$$
$$-\frac{1}{5}\left[(4r^2(\mathbf{r}\cdot \mathbf{J})r_p + r^4 J_p)(\delta_{jk}\delta_{lp} + \delta_{jl}\delta_{kp} + \delta_{pk}\delta_{jp})\right]dv,$$

$$\bar{X}_{vvuuj} = \frac{i}{\omega}\int 4(\mathbf{r}\cdot \mathbf{J})r^2 r_j + r^4 J_j\, dv.$$

The magnetic 16-pole is the most cumbersome formula in the paper. The symmetrization of the 4$^{th}$ rank tensor produces the 3$^{rd}$ rank tensors. For the resulted 3$^{rd}$ rank tensors we apply the same consequences of the detracing and symmetrization procedures as for the magnetic octupole. After the first symmetrization, the 4$^{th}$ rank tensor obtained is twice reduced:

$$\overline{\overline{\overline{Y_{jplt}}}} = Y_{jplt} - \frac{1}{7}\sum_{\{jplt\}} \delta_{jp}\bar{\bar{Y}}_{vvlt} + \frac{1}{35}\sum_{\{jplt\}}\delta_{jp}\delta_{lt}\bar{Y}_{uuvv} - \frac{1}{4}\sum_{\{plt\}}\varepsilon_{\gamma jp}[\,\overline{\overline{C_{\gamma lt}}} +$$
$$+\frac{1}{5}\left(\delta_{\gamma l}\overline{C_{vvt}} + \delta_{\gamma t}\overline{C_{vvl}} + \delta_{lt}\overline{C_{vv\gamma}}\right) + \frac{1}{3}\varepsilon_{\alpha \gamma l}\left(\bar{\bar{F}}_{\alpha t} + \frac{1}{2}\varepsilon_{\beta\alpha t}H_\beta\right) + \tag{15}$$
$$\frac{1}{3}\varepsilon_{\alpha\gamma t}(\bar{\bar{F}}_{\alpha l} + \frac{1}{2}\varepsilon_{\beta\alpha l}H_\beta) + \overline{\overline{D_{\gamma lt}}} + \frac{1}{5}\left(\delta_{\gamma l}\overline{D_{vvt}} + \delta_{tl}\overline{D_{vv\gamma}} + \delta_{\gamma t}\overline{D_{vvl}}\right)].$$

Its constituent parts are

$$\bar{\bar{Y}}_{vvlt} = \frac{1}{5}\int r^2((\mathbf{r}\times\mathbf{J})_l r_t + (\mathbf{r}\times\mathbf{J})_t r_l)\,dv,$$

$$C_{\gamma lt} = \frac{4}{5}\int r^2 J_\gamma r_l r_t\, dv,$$

$$D_{\gamma lt} = -\frac{4}{5}\int r_\gamma (\mathbf{r}\cdot \mathbf{J})r_l r_t\, dv,$$

$$\overline{D_{vvt}} = -\frac{4}{5}\int (\mathbf{r}\cdot \mathbf{J})r^2 r_t\, dv, \tag{16}$$

$$\bar{C}_{vvt} = \frac{4}{15}\int r^4 J_t + 2(\mathbf{r}\cdot \mathbf{J})r_t\, dv,$$

$$\bar{\bar{F}}_{\alpha t} = -\frac{2}{5}\int r^2[(\mathbf{r}\times\mathbf{J})_t r_\alpha + (\mathbf{r}\times\mathbf{J})_\alpha r_t]\, dv,$$

$$H_\beta = \frac{4}{5}\int \left(r^2(\mathbf{J}\cdot \mathbf{r})r_\beta - r^4 J_\beta\right)dv.$$

2.2.2. Toroidal moments

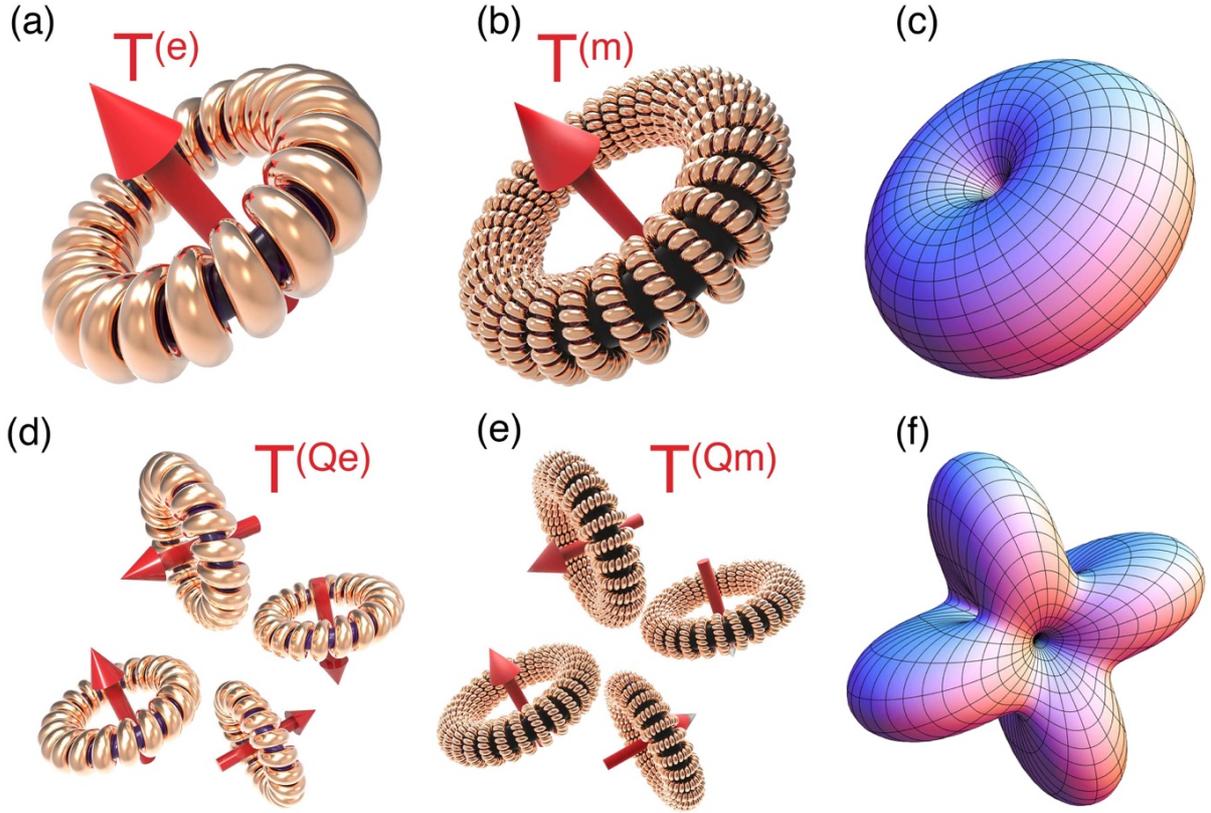

**Figure 2.** Illustration of the toroidal moments family. Electric toroidal moment (a) corresponding to a wire coiled around a torus surface and forming a poloidal current configuration. Toroidal magnetic dipole consisting of circulation of electric toroidal dipoles (b). Electric toroidal quadrupole configuration (d) as a composition of the electric toroidal moment coils. Toroidal magnetic quadrupole (e) formed in a similar manner with the electric quadrupole. The typical scattering patterns of the toroidal dipole moments (c), the toroidal quadrupole moments (f) (the magnetic and electric scattering patterns have the shown usual shapes, but are perpendicular to each other) fully coincide with scattering patterns of the conventional electric and magnetic multipoles.

After the detracing and simmetrization of the primitive Cartesian multipoles, the residual terms form a set of the irreducible toroidal moments limited by the toroidal electric octupole $\widehat{\widehat{\overline{T}}}^{(Oe)}$ and toroidal magnetic quadrupole $\widehat{\widehat{\overline{T}}}^{(Qm)}$ (Table 2). The 1$^{st}$ order toroidal electric dipole $\mathbf{T}^{(e)}$ considered previously in a plenty of papers [25,33,52] is derived from the decomposition of the primitive electric octupole $\widehat{\widehat{O}}^{(e)}$ and the primitive magnetic quadrupole $\widehat{Q}^{(m)}$; the 4$^{th}$ order of the primitive Cartesian multipoles produce two toroidal multipole moments – the toroidal magnetic dipole $\mathbf{T}^{(m)}$ [13,22,24,25] and the toroidal electric quadrupole $\widehat{\widehat{\overline{\overline{T}}}}^{(Qe)}$, which are obtained from the magnetic octupole $\widehat{\widehat{O}}^{(m)}$ and the electric 16-pole $\widehat{\widehat{\overline{S}}}$; the decomposition of the primitive electric 32-pole $\widehat{\widehat{\overline{X}}}$ and magnetic 16-pole $\widehat{\widehat{Y}}$ produces the 2$^{nd}$ contributing term $\mathbf{T}^{(2e)}$ to the full electric dipole moment, the toroidal magnetic quadrupole $\widehat{\widehat{\overline{\overline{T}}}}^{(Qm)}$ and the toroidal electric octupole $\widehat{\widehat{\overline{\overline{T}}}}^{(Oe)}$. The detailed derivation is given in the Supplementary.

**Table 2.** The set of the toroidal moments

| Order | Toroidal electric multipoles | Toroidal magnetic multipoles |
|---|---|---|
| 1 | $T_j^{(e)} = \dfrac{1}{10} \int (\mathbf{J} \cdot \mathbf{r}) r_j - 2r^2 J_j \, dv$ <br> $T_j^{(2e)} = \dfrac{1}{10} \int 3r^4 J_j - 2r^2 (\mathbf{r} \cdot \mathbf{J}) r_j \, dv$ | - |
| 2 | $\overline{\overline{T}}_{jk}^{(Qe)} = \dfrac{1}{42} \int 4(\mathbf{r} \cdot \mathbf{J}) r_j r_k + 2(\mathbf{J} \cdot \mathbf{r}) r^2 \delta_{jk} -$ <br> $-5r^2 \left( r_j J_k + r_k J_j \right) dv$ | $T_j^{(m)} = -\dfrac{i\omega}{20} \int r^2 (\mathbf{r} \times \mathbf{J})_j \, dv$ |
| 3 | $\overline{\overline{T}}_{jkl}^{(Oe)} = \dfrac{1}{300} \int 35(\mathbf{r} \cdot \mathbf{J}) r_j r_k r_l -$ <br> $+ 20 r^2 (J_l r_k r_j + J_k r_l r_j + J_j r_l r_k) +$ <br> $+ (\delta_{lj} \delta_{kp} + \delta_{lk} \delta_{jp} + \delta_{lp} \delta_{jk}) \cdot$ <br> $\cdot [(\mathbf{r} \cdot \mathbf{J}) r^2 r_p + 4 J_p r^4] dv$ | $\overline{\overline{T}}_{jp}^{(Qm)} = \dfrac{i\omega}{42} \int r^2 [r_j (\mathbf{r} \times \mathbf{J})_p +$ <br> $+ (\mathbf{r} \times \mathbf{J})_j r_p] dv$ |

It is worth noting that not all the residual parts of the primitive multipoles affect scattering. The most known example is the residual part of the primitive electric quadrupole $\widehat{\widehat{\overline{\overline{Q}}}}^{(e)}$ [42,50]. The residual part is scalar not providing any contribution to the scattered **E**-field (see Supplementary). Moreover, most of the terms obtained via detracing do not influence the **E**-field. There are several works devoted to the investigation of this problem [40,53]. In the supplementary materials (section 3.8) we derive and show the nonradiative parts of the scalar and vector potentials, but their detailed analysis is a subject for further investigations.

As it seen from Equation (6)-(12) and Table 2, the toroidal moments contribute to a scattered electric field with the same propagator as their basic counterparts and the difference is in the space-time symmetry (see **Figure 2** and 1). All the electric (magnetic) toroidal moments have the same space-time symmetry as the toroidal electric (magnetic) dipole (Figure 1).

In the Section 3 we will analyze the particular poloidal currents configurations corresponding to the toroidal multipole moments (the schematics is given in Figure 2). The corresponding scattering calculated via Equation 6 and Table 2 are given in Figure 2(c,f) to prove that the toroidal multipoles give the same far-fields maps as the basic ones [16].

2.2.3. Scattering efficiency

For calculating the scattering power for a harmonic signal ($e^{-i\omega t}$), we consider an arbitrary multipole tensor $\hat{A}$ of rank ($\alpha + 1$), without any specification of its symmetry or trace value. This multipole contributes to the **E**-field as vectors after the contraction with vectors **n** as following:

$$A_j^{(\alpha+1)} = (-i\omega)^{\alpha+1} A_{jk..\alpha} n_k \dots n_\alpha, \tag{17}$$

where the superscript in $A_j^{(\alpha+1)}$ denotes a time derivative and a contraction result of the tensor $\hat{A}$ with vectors **n**. Thus, an arbitrary rank multipole $\hat{A}$ contributes to the electric field as:

$$E_i^{(A)} = \frac{1}{4\pi\varepsilon_0 c^2 R}\left[(n_i n_j - n^2 \delta_{ij})A_j^{(\alpha+1)}\right] =$$
$$= B(n_i n_j - n^2 \delta_{ij})A_j = B\varepsilon_{i\beta\gamma}\varepsilon_{\gamma\alpha j}n_\alpha n_\beta A_j, \tag{18}$$

the coefficient $B$ is denoted as

$$B = \frac{(-i\omega)^{\alpha+1}}{4\pi\varepsilon_0 c^2 R}. \tag{19}$$

A scattering efficiency is proportional to $|\mathbf{E}_{scat}|^2$. Taking into account that a dot product of any two basic multipoles (besides the multiplication on itself) is equal to zero [38,42] we obtain an independent contribution of each multipole to the scattering power. Toroidal moments and basic multipole moments interfere as $(\hat{A} + \frac{ik}{c}\hat{T}^{(A)})$, where $\hat{T}^{(A)}$ is a toroidal moment corresponding to a basic moment $\hat{A}$. Therefore, the influence of an interfering toroidal moment could be estimated as the difference of the scattered energies $\sim (\hat{A} + \frac{ik}{c}\hat{T}^{(A)})^2 - (\hat{A})^2$.

Similarly, the impact of $\widehat{T}^{(2A)} \sim \left(\widehat{A} + \frac{ik}{c}\widehat{T}^{(A)} + \frac{ik^3}{c}\widehat{T}^{(2A)}\right)^2 - \left(\widehat{A} + \frac{ik}{c}\widehat{T}^{(A)}\right)^2$, etc. Thus, the individual contribution of $A_j^{\alpha+1}$ is

$$|\mathbf{E}_{scat}|^2 = B^2 \cdot \varepsilon_{i\beta\gamma}\varepsilon_{\alpha j\gamma}n_\alpha n_\beta A_j \cdot \varepsilon_{i\varphi\theta}\varepsilon_{\xi l\theta}n_\xi n_\varphi A_l^*. \tag{20}$$

Taking into account the contraction from Equation (17) we introduce an individual contribution of the multipole $\widehat{A}$ to the $|\mathbf{E}_{scat}|^2$ and following [38,42] obtain the very useful relation:

$$|\mathbf{E}_{scat}|^2 = B^2 \left[\frac{(\alpha-1)!\,(\alpha+1)}{(2\alpha+1)!!}\right] |A_{jk..\alpha}|^2. \tag{21}$$

The $|A_{jk..\alpha}|^2$ in (21) defines the sum of squares of all components $|A_{jk..\alpha}|^2 = A_{jk..\alpha} \cdot A_{jk..\alpha}^*$, where the Einstein's notation for repeating indices has been applied. Therefore, the far-field full scattered power is [37]:

$$P_{scat} = \frac{1}{2}\sqrt{\frac{\varepsilon_0\varepsilon_d}{\mu_0}} \oint |\mathbf{E}_{scat}|^2 d\Omega. \tag{22}$$

Using the integration by solid angle and Equation (6) we obtain the multipole decomposition of the scattered power

$$\begin{aligned}
P_{scat} = & \frac{k_0^4\sqrt{\varepsilon_d}}{12\pi\varepsilon_0^2 c\mu_0}|p_i + \frac{ik}{c}T_i^{(e)} + \frac{ik^3}{c}T_i^{(2e)}|^2 + \frac{k_0^4\varepsilon_d\sqrt{\varepsilon_d}}{12\pi\varepsilon_0 c}|m_i + \frac{ik}{c}T_i^{(m)}|^2 + \\
& + \frac{k_0^6\varepsilon_d\sqrt{\varepsilon_d}}{160\pi\varepsilon_0^2 c\mu_0}|\bar{\bar{Q}}_{ij}^{(e)} + \frac{ik}{c}\bar{\bar{T}}_{ij}^{(Qe)}|^2 + \frac{k_0^6\varepsilon_d^2\sqrt{\varepsilon_d}}{160\pi\varepsilon_0 c}|\bar{\bar{Q}}_{ij}^{(m)} + \frac{ik}{c}\bar{\bar{T}}_{ij}^{(Qm)}|^2 + \\
& + \frac{k_0^8\varepsilon_d^2\sqrt{\varepsilon_d}}{3780\pi\varepsilon_0^2 c\mu_0}|\bar{\bar{O}}_{ijk}^{(e)} + \frac{ik}{c}\bar{\bar{T}}_{ijk}^{(Oe)}|^2 + \frac{k_0^8\varepsilon_d^3\sqrt{\varepsilon_d}}{3780\pi\varepsilon_0 c}|\bar{\bar{O}}_{ijk}^{(m)}|^2 + \\
& + \frac{k_0^{10}\varepsilon_d^3\sqrt{\varepsilon_d}}{145152\pi\varepsilon_0^2 c\mu_0}|\bar{\bar{S}}_{ijkl}|^2 + \frac{k_0^{10}\varepsilon_d^4\sqrt{\varepsilon_d}}{145152\pi\varepsilon_0 c}|\bar{\bar{Y}}_{ijkl}|^2 + \\
& + \frac{k_0^{12}\varepsilon_d^4\sqrt{\varepsilon_d}}{831600\pi\varepsilon_0^2 c\mu_0}|\bar{\bar{X}}_{ijkla}|^2.
\end{aligned} \tag{23}$$

The scattering cross-section is obtained via the normalization of the scattered power to the incident energy flux:

$$\sigma_{scat} = 2\sqrt{\frac{\varepsilon_0 \varepsilon_d}{\mu_0}} \frac{P_{scat}}{|\mathbf{E}_{inc}|^2}, \tag{24}$$

and the scattering efficiency is given by the normalization of the scattering cross-section on the geometric cross-section of a scatterer

$$Q_{scat} = \frac{\sigma_{scat}}{\sigma_{geom}}. \tag{25}$$

Equation (23-25) enables the analysis of the individual contributions of the full multipoles (basic + toroidal) to the scattering efficiency. Note, that the original formula (21) allows to calculate the contribution of any order multipole to a scattered field in a simple manner in contrast to the cumbersome routine calculations employed, e.g. in [42].

Summarizing, we have performed the Cartesian multipole decomposition of a scattered field up to the magnetic 16-pole and the electric 32-pole utilizing retarded potentials. To avoid the mutual interference of the primitive multipoles, we have found the irreducible basic high-order Cartesian multipoles and the family of toroidal moments being independent on the basic counterparts on the SO(3) algebraic group. Finally, we have derived the contributions of the multipoles to the scattered power, scattering cross-section, and scattering efficiency.

### 3. Numerical verification, and toroidal currents observation

Now, we compare the multipole decomposition of the scattering efficiency $Q_{scat}$ with numerical simulations in COMSOL Multiphysics software and Mie theory calculations in "Scatnlay" [54–56]. Plane wave incidents on a nondispersive nanosphere with refractive index n=4 and maximum size parameter $\chi = kr = \frac{2\pi r}{\lambda} \lesssim 2$. The particle's material imitate weakly absorbing Si or Ge nanoparticles in the optical or IR spectrum regions.

The simulations of the scattering efficiency are presented in **Figure 3**. Figure 3(a) shows a typical assembly of peaks associated with either electric or magnetic Mie resonances of a dielectric spherical subwavelength particle [57]. The full width at half maximum (FWHM) of multipole resonances sufficiently decreases with increasing the multipole order. In the spectral region associated with the high-order resonances (octupole and 16-pole resonances), wide and substantially asymmetric second resonances of the low-order multipoles (dipoles and quadrupoles) are excited (see e.g., peaks of solid blue and red lines at $kr$=1.853 and $kr$=1.568 corresponding to the electric and magnetic dipoles, peaks of solid purple and green lines at $kr$=2.162 and $kr$=1.898 related to the electric and magnetic quadrupoles).

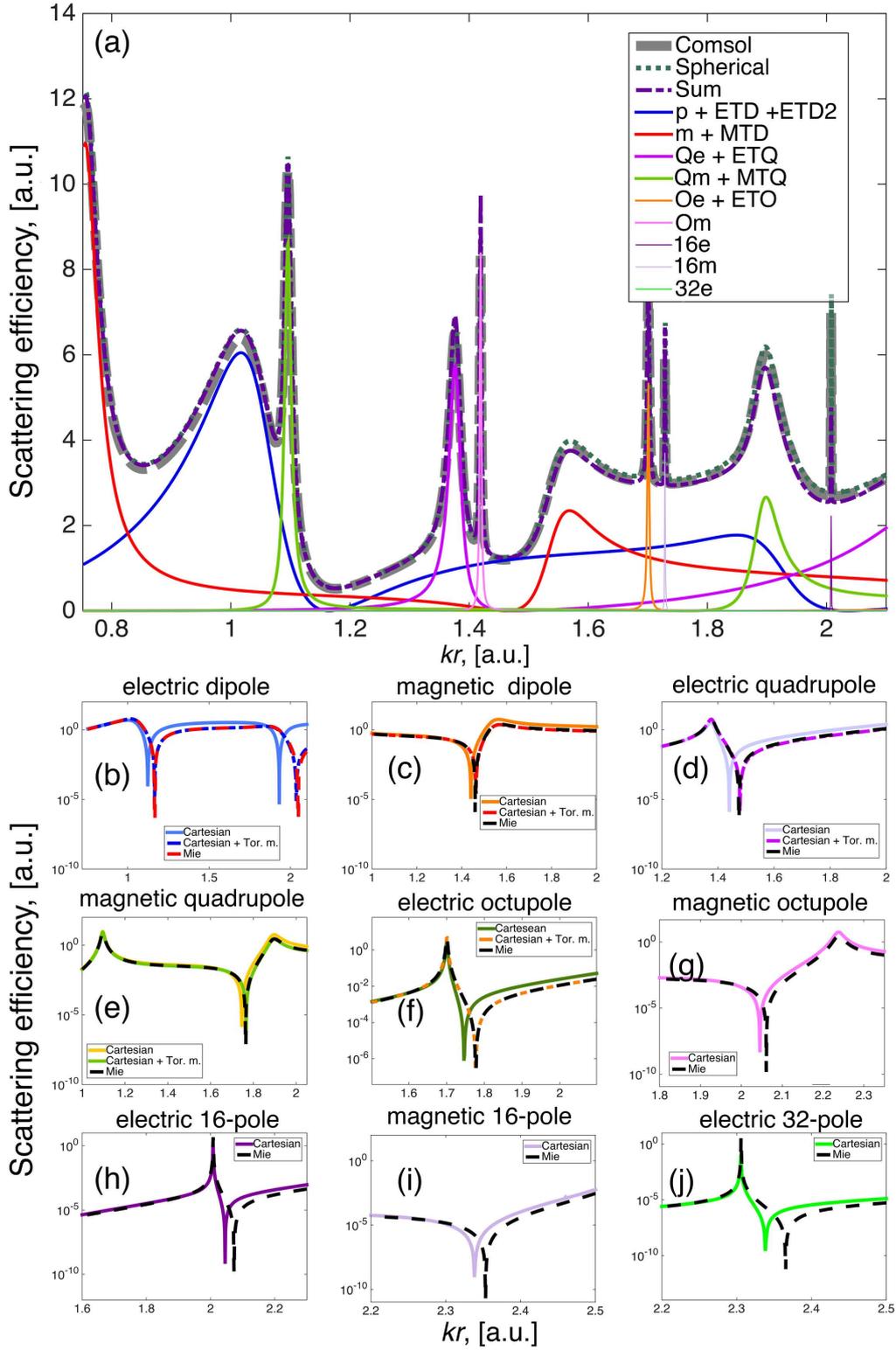

**Figure 3.** (a) the Cartesian multipole decomposition of the scattering efficiency for a plane wave scattered on a spherical particle with the refractive index *n* = 4 in comparison with the Mie theory (dashed green line) and numerical calculations in COMSOL Multiphysics (gray dashed line). The multipoles contributions to the scattering efficiency of the irreducible Cartesian multipoles with and without toroidal terms compared with Mie multipoles (b-j).

We find an excellent agreement of the scattering efficiency from the COMSOL full-field simulations and the Mie theory with the sum of all the Cartesian contributions with taking into account the toroidal terms from Equation (23-25). The lack of the toroidal multipoles for the 4$^{th}$ and 5$^{th}$ orders of the Cartesian multipoles leads to a worse coincidence of the plots in the region of high size parameters. To fix it we should derive 6$^{th}$ and 7$^{th}$ order multipoles, which is beyond the scope of this paper. However, even this approximation allows us to obtain pretty well coincidence as it is seen from Figures 3(a,g-j).

The semi-logarithmic scale in Figure 3(b-j) provides an additional information about the behavior of the irreducible Cartesian multipoles, e.g., that the minima of the basic Cartesian multipoles are always shifted from the spherical multipoles minima Figure 3(b-j).

At the points of the minimal contributions of the basic multipoles, toroidal moments reach their maxima and could exceed the basic one more than in 4-5 orders. Other interesting points are the minima of the full Cartesian multipoles (coinciding with the spherical ones) which will be discussed in details in Section 4.

The mismatch of the high-order Cartesian multipoles and Mie resonances (Figure 3(g-j) ) should be addressed. Before we mentioned that this behavior takes place because of the lack of the toroidal moments, however it is to be proved hereinafter. Inspired by the schematic Figure 2 originally presented in [58] we study the spatial distribution of the fields for the multipoles for different size parameters *kr*, when the toroidal moments either insufficient or demonstrate high enough values in comparison with the basic multipoles. The whole set of the field maps for the multipoles from the electric dipole up to the electric octupole with the corresponding clearly observable toroidal currents configurations are presented in the Supplementary. Here we consider only the very similar pictures for the electric octupole with the strong (**Figure 4**(a)) or weak (Figure 4(b)) contributions of the toroidal term, where the points are picked up from Figure 3(f), and the magnetic octupole for the two points from Figure 3(g), where either good (Figure 4(d)) or bad (Figure 4(c)) agreement of the Mie and Cartesian multipoles takes place. Note, that Figure 4(a,b) demonstrates the normalized electric field map, but Figure 4(c,d) – the magnetic one in the perpendicular plane. The size parameters are given in the figure caption.

Comparing Figure 3(f,g) with Figure 4 we should note that the differences between the basic Cartesian multipoles and the spherical multipoles are because of the non-negligible values of the toroidal terms provided by the toroidal behavior of the fields inside the particles (curl-like streamlines in the cross-sections). Further increasing the size parameter *kr* obviously lead to the excitation of the next generation of toroidal terms with the corresponding curls on a field map (see an example for a dipole with two sufficient toroidal moments in Figure 2 in the Supplementary) and so on. This is the very important feature of the Cartesian multipoles allowing to estimate the behavior of the displacement currents or fields inside a particle by the values of basic and toroidal terms.

Therefore, considering the similar situations in Figure 4(a-d), we can conclude that the observed mismatch takes place because of the absent toroidal term for the magnetic octupole, but not owing to its incorrectness. Following the same logic with the field maps for the rest of the high-order multipoles (see Supplementary) we should underline the importance of the toroidal moments for an accurate analysis.

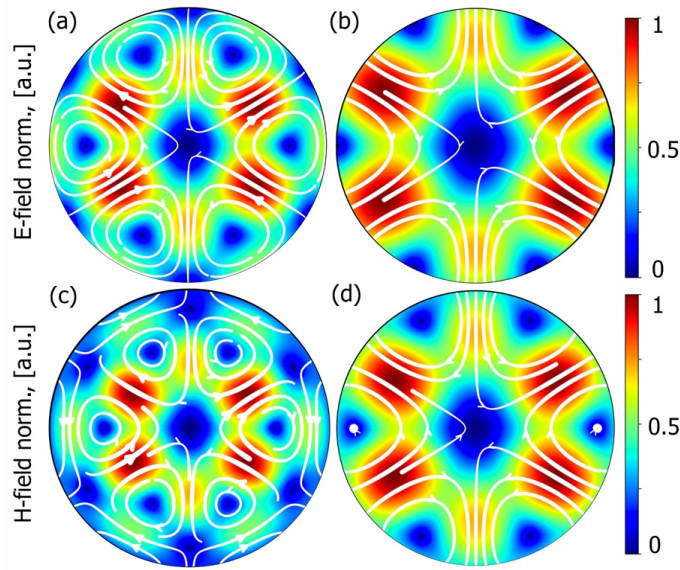

**Figure 4.** The normalized electric and magnetic fields corresponding to the spherical Mie electric and magnetic octupoles inside a particle with the refraction index *n* = 4 and (a) *kr* = 1.714 and (c) for *kr* =2.011 corresponding to the points, where the far-field is governed predominantly by the toroidal responses. The field maps depicted in Figure (b,d) and plotted for *kr* = 1.256 correspond to the near-zero toroidal contributions and do not show curl-like configurations of fields.

### 4. High order anapole states

Recently, numerous studies have been devoted to the far-field destructive interference of an electric and toroidal dipoles and forming an "anapole" state [33,52,59] being an auxiliary tool for the selective control of the scattering. Moreover, a strong electric field [33,52] or, vice versa – vanishing electric field accompanied with an intense magnetic field [35] were recently observed at "anapole" states allowing to realize either lasing or an electro-magnetic fields separation.

Moreover, the efforts for achieving higher order anapole states have been undertaken [25], but only the first zeroes of the electric and magnetic dipoles were discussed in details and explained in terms of the interfering basic and toroidal moments [25,40]. Our approach allows to identify the crucial role of the higher toroidal multipoles in the formation of nonradiating anapole states of an arbitrary order. It includes zeroes of radiation from electrical and magnetic dipoles, as well as from higher order multipoles, such as quadrupoles and octupole. Although, these anapole states could be identified by using the spherical multipole decomposition, their origin cannot be explained utilizing conventional approaches based on either spherical or Cartesian multipoles. And only the consideration of the higher order toroidal multipoles provides the full and consistent description presented hereinafter.

Our approach developed here based on the irreducible Cartesian multipoles considerably extends the border of a multipole decomposition and provides a useful tool for quantitative and qualitative analysis of the near and far-fields of arbitrary shape scatterers.

**Figure 5**(a) shows the irreducible Cartesian multipoles together with the toroidal terms contributing to the scattering efficiency spectrum. Note, the anapole states appear pairwise depending on the order: i.e. magnetic dipole and electric quadrupole, magnetic quadrupole and electric octupole. Mie coefficients allow to identify anapole state only with help of the following relation [33]:

$$K = \begin{cases} \dfrac{|d| - |a|}{|d| + |a|} & \text{for electric type multipoles,} \\ \dfrac{|c| - |b|}{|c| + |b|} & \text{for magnetic type multipoles.} \end{cases} \qquad (26)$$

$K$ is unity, when a multipole is excited (near-fields are not zero), but the radiation in the far-zone is vanishing. Here $(a, b)$ correspond to the Mie scattering coefficients, and $(d, c)$ - to the coefficients for the field inside a particle. The spectral positions of the anapole states and the peaks of $K$ coincide by definition of an anapole state (see Figure 5(b)). From Figure 5(b) it can be seen that $K$ maxima are shifted from far-field multipole resonances in Figure 5(a), that shows a non-resonant origin of the anapole phenomenon. Figure 5(c) shows individual normalized **E**- (for the multipoles of electric type) and **H**-field (for the mutipoles of magnetic type) maps for every certain multipole up to electric octupole with *kr* corresponding to an anapole state. The fields on every map are normalized on their maxima.

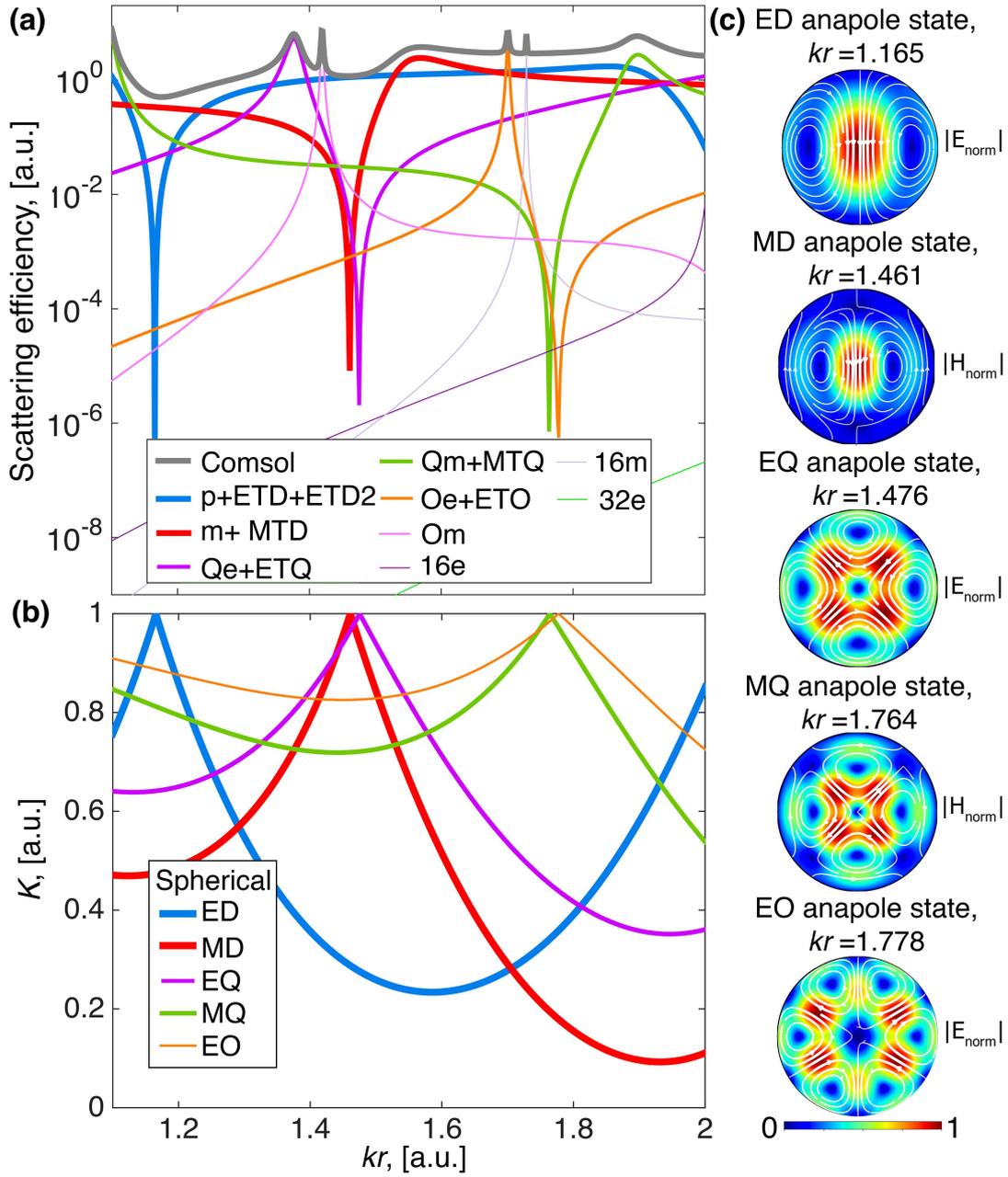

**Figure 5.** The scattering efficiency from Figure 3 of a spherical particle with $n = 4$ in semi-logarithmic scale (a) shows the first five anapole states in terms of Cartesian multipoles (a); the relation (26) for spherical multipoles (b); the visualization of the normalized internal fields corresponding to the certain spherical multipoles being on anapole states (c). Note, that for the electric multipoles we show the normalized (to a maximal value) electric field, and for the magnetic type – normalized magnetic field. The electric dipole contribution to the scattered efficiency approaches minimum at $kr = 1.165$, the magnetic dipole contribution - at $kr = 1.461$, the electric and magnetic quadrupoles contributions - at $kr = 1.476$ and $kr = 1.764$ correspondingly, the electric octupole contribution - at $kr = 1.778$.

Each field distribution corresponding to an anapole state shows the high contribution of a toroidal moment manifested by singular points, where the **E** or **H** field is vanishing (dark blue regions on Figure 5(c)), inside vortex-like streamlines. The number of singular points increases with growing of a multipole order. Anapoles of different orders give rise to the field concentration inside a particle just as it was shown in [33] for an electric dipole anapole. Moreover, the ability to realize anapoles of magnetic type opens a room of opportunities for magnetic field concentration, which is of high demand for a variety of applications [60–64].

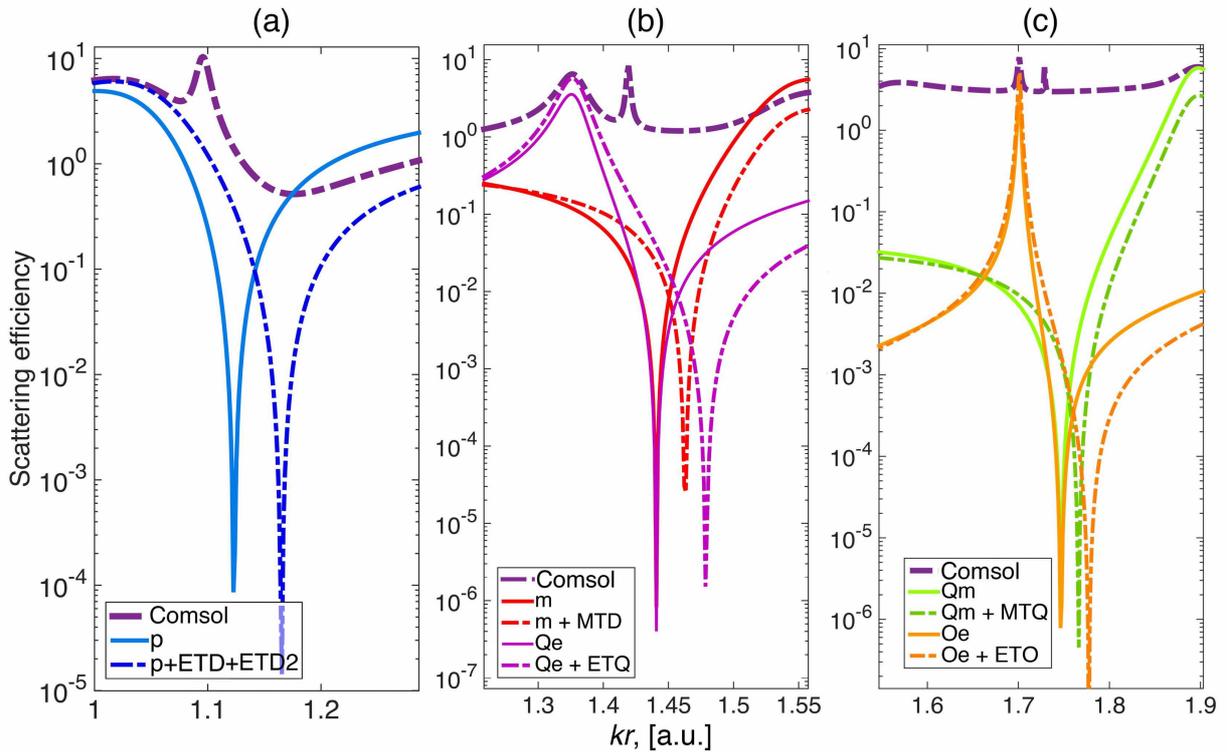

**Figure 6.** The spectra of the scattering efficiency of a spherical particle with $n = 4$ and the corresponding multipolar contributions. Solid lines denote the basic Cartesian multipoles without toroidal contribution, and dashed lines correspond to the full multipoles (basic + toroidal). (a) Electric dipole anapole state; (b) magnetic dipole and electric quadrupole anapole states; (c) magnetic quadrupole and electric octupole anapole states.

For analyzing the role of toroidal terms in the observed anapole states, let us consider **Figure 6** showing the basic multipoles together with the full ones in the close vicinity of anapole states.

The plots in Figure 6 split the spectrum in 3 parts corresponding to the anapoles of different orders: Figure 6(a) shows the basic electric dipole (**p**) (blue solid line), and the full electric dipole ($\mathbf{p} + \frac{ik}{c}\mathbf{T}^{(e)} + \frac{ik^3}{c}\mathbf{T}^{(2e)}$) (dashed dark blue line); Figure 6(b)) demonstrates the magnetic dipole ($\mathbf{m}$) (solid pink line), the full magnetic dipole ($\mathbf{m} + \frac{ik}{c}\mathbf{T}^{(m)}$) (dashed red line), the basic electric quadrupole ($\hat{\hat{Q}}^{(e)}$) (solid light purple line), and the full electric Cartesian quadrupole ($\hat{\hat{Q}}^{(e)} + \frac{ik}{c}\hat{\hat{T}}^{(Qe)}$) (purple dashed line); Figure 6(c) shows the 3$^{rd}$ order multipoles consisting of the basic magnetic quadrupole ($\hat{\hat{Q}}^{(m)}$) (light green solid line), basic electric octupole ($\hat{\hat{O}}^{(e)}$) (solid yellow line), and their full counterparts ($\hat{\hat{Q}}^{(m)} + \frac{ik}{c}\hat{\hat{T}}^{(Qm)}$) and ($\hat{\hat{O}}^{(e)} + \frac{ik}{c}\hat{\hat{T}}^{(Oe)}$) (dashed green and dashed orange lines correspondingly).

The minima of all the dashed lines in Figure 6 correspond to the anapole states. In these points, the basic, and toroidal multipoles have the same amplitudes (with taking into account constants like $\frac{ik^x}{c}$ from (6)) and $\pi$ - shifted phases giving rise to their destructive interference. Therefore, the contribution of the toroidal terms appears to be crucial for the elimination of the far-field from a given multipole without nullifying the corresponding near-field. It is also necessary to note that in our calculations anapoles of different orders usually lead to decreasing the total scattering efficiency (see Figure 5 and 6), however, this is not a common rule for other particles.

Moreover, the crossing points of the full multipoles near the anapole states are of particular interest. For example, at $kr = 1.468$ we have simultaneous vanishing full magnetic dipole and full electric quadrupole. The similar situation takes place for the magnetic quadrupole and electric octupole at $kr = 1.769$. According to our calculations the pairs of anapoles become closer to each other with increasing the refractive index of a sphere providing so called "hybrid anapole states" [25] corresponding to the simultaneous suppression of scattering from two multipoles.

Therefore, these results provided by the implementation of the high-order toroidal terms allows for a physical insight into the anapole formation and yields fundamental information on the light scattering and near-field distribution in subwavelength high-index particles.

## 5. Conclusion

The recent progress in all-dielectric nanophotonics leads to an intense development of nanoscale light governing meta-structures providing a flexible control over all the degrees of freedom including phase, amplitude, polarization, scattering direction etc. The widely known auxiliary tool for analyzing and designing scattering properties of a particle or nanoantenna is the multipole decomposition being a subject of an intense research nowadays. It provides very useful link between displacement currents or fields configurations involved in a system and far-field scattering patterns, and is highly demanded for tailoring the optical signature of an object.

In this work, we have significantly extended the Cartesian multipole analysis utilizing the irreducible representation with respect to mirror reflections and rotations of the SO(3) group. By using this generalized approach, we have derived the Cartesian multipoles up to the magnetic 16-pole and electric 32-pole and have explicitly revealed the high order toroidal moments up to the magnetic quadrupole and electric octupole toroidal terms.

Excluding the nonradiative terms we have obtained the irreducible representation of the electric field in tensor and vector forms. To obtain contributions of the derived multipoles to the far-field, the explicit expressions for the scattering cross-section and scattered efficiency have been derived for arbitrary rank multipoles.

The obtained results have been verified via the comparison with numerical calculations, and nearly perfect agreement with both the Mie theory and COMSOL Multiphysics has been shown. The toroidal terms have been shown to be of ultimate importance for an adequate interpretation of scattering efficiency spectra. Peculiar field distributions inside particles having poloidal shape and corresponding to the presence of the toroidal terms have been revealed and discussed. We have shown and discussed the evolution of the internal field for different multipoles including the state of a basic multipole with insufficient toroidal response, and the state with the predominating toroidal moments contribution.

Due to the far-field interference of a basic multipole and its toroidal counterpart it is possible to construct the high-order anapole states up to electrical octupole anapole. We have shown that the two types of anapoles being of both electric and magnetic types could localize either electric or magnetic energy inside a particle. The conditions for the simultaneous realization of two anapoles of different types have been obtained providing an opportunity to excite a "hybrid" anapole state.

Our new findings contribute to the understanding of the fundamental aspects of the light scattering on high-index nanoparticles and pave ways to novel approaches to nanoatennas designs and metasurfaces architectures. In terms of practical and highly demanded applications, the developed approach can be utilized for elaborating new all-dielectric nanoelements and nanoresonators with particular optical signatures owing to a controllable formation of scattering patterns. Furthermore, high local field enhancement inside particles delivered by single and hybrid anapole states is valuable for tailoring a variety of nonlinear interactions.

**Acknowledgement**

A.S.S acknowledges the support of the Russian Fund for Basic Research within the projects 18-02-00414, 18-52-00005 and the support of the Ministry of Education and Science of the Russian Federation (GOSZADANIE Grant No. 3.4982.2017/6.7). The development of analytical approach and the calculations of multipole moments have been supported by the Russian Science Foundation Grant No. 16-12-10287. Support has been provided by the Government of the Russian Federation (Grant No. 08-08). The work of AEM was supported by the Australian Research Council and UNSW Scientia Fellowship.

**Keywords:** All-dielectric nanophotonics, anapole, toroidal moment, Cartesian multipole, multipole decomposition, Mie resonance

# Supplementary materials

## 1 Derivation of multipole contributions to the scattered electric field in far-field zone

### 1.1 Definitions. Scalar and vector potentials

We use decomposition of retarded scalar and vector potentials to perform the multipole decomposition of the E-field in Raab's notation [1]

$$\Phi(\mathbf{R},t) = \frac{1}{4\pi\varepsilon_0} \int_V \frac{\rho(\mathbf{r}, t' + \Delta t)}{|\mathbf{R} - \mathbf{r}|} dv, \qquad (1.1)$$

$$A(\mathbf{R},t) = \frac{\mu_0}{4\pi} \int_V \frac{\mathbf{J}(\mathbf{r}, t' + \Delta t)}{|\mathbf{R} - \mathbf{r}|} dv, \qquad (1.2)$$

for time-dependent charge and current densities $\rho(\mathbf{r}, t)$ and $\mathbf{J}(\mathbf{r}, t)$ distributed in a finite arbitrary domain (Fig.1). The spatial retardation is described by difference in distances between an arbitrary point of the domain and point P as $|\mathbf{R} - \mathbf{r}|$. A radius vector from the origin $O$ to a point outside $P$ is defined as $\mathbf{R}$ and a radius vector from the origin to an arbitrary point of the domain is denoted as $\mathbf{r}$.

Thus, $\rho(\mathbf{r}, t)$ and $\mathbf{J}(\mathbf{r}, t)$ are time dependent, the temporal retardation of electromagnetic responses in the $P$ should be expressed as a function of $|\mathbf{R} - \mathbf{r}|$ and speed of light $c$: $t' + \Delta t = t - |\mathbf{R} - \mathbf{r}|/c$. For an multipole analysis of charge/current we consider series of $|\mathbf{R} - \mathbf{r}|$.

### 1.2 Scalar potential expansion

Expressing the distance $|\mathbf{R} - \mathbf{r}|$ in the form of

$$|\mathbf{R} - \mathbf{r}| = \sqrt{R^2 + (\mathbf{r} \cdot \mathbf{R}) + r^2} = R \cdot \left[1 + \frac{1}{R^2}((\mathbf{r} \cdot \mathbf{R}) + r^2)\right]^{0.5} \qquad (1.3)$$

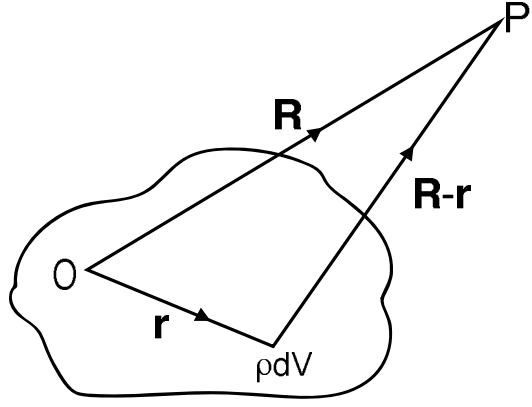

Figure 1: The scheme of charges and currents in the arbitrary finite domain referring to a point P outside the domain.

we consider the identity (1.3) and make a binomial expansion

$$|\mathbf{R} - \mathbf{r}| = R + \frac{1}{2}\frac{1}{R}(r^2 - 2(\mathbf{R}\cdot\mathbf{r})) - \frac{1}{8}\frac{1}{R^3}(r^2 - 2(\mathbf{R}\cdot\mathbf{r}))^2 + \\ + \frac{1}{16}\frac{1}{R^5}(r^2 - 2(\mathbf{R}\cdot\mathbf{r}))^3 - \frac{5}{128}\frac{1}{R^7}(r^2 - 2(\mathbf{R}\cdot\mathbf{r}))^4 - \frac{7}{256}\frac{1}{R^9}(r^2 - 2(\mathbf{R}\cdot\mathbf{r}))^5. \quad (1.4)$$

After some algebra and neglecting the terms with $r^x$, $x > 5$:

$$|\mathbf{R} - \mathbf{r}| = R - \frac{(\mathbf{R}\cdot\mathbf{r})}{R} + \frac{1}{2}\frac{r^2R^2 - (\mathbf{R}\cdot\mathbf{r})^2}{R^3} + \frac{1}{2}\frac{(\mathbf{R}\cdot\mathbf{r})r^2R^2 - (\mathbf{R}\cdot\mathbf{r})^3}{R^5} + \\ + \frac{1}{8}\frac{6(\mathbf{R}\cdot\mathbf{r})^2 r^2 R^2 - r^4 R^4 - 5(\mathbf{R}\cdot\mathbf{r})^4}{R^7} + \frac{1}{8}\frac{10(\mathbf{R}\cdot\mathbf{r})^3 r^2 R^2 - 3(\mathbf{R}\cdot\mathbf{r}) r^4 R^4 - 7(\mathbf{R}\cdot\mathbf{r})^5}{R^9} \quad (1.5)$$

$$\Delta t = \frac{R}{c} - \frac{|\mathbf{R} - \mathbf{r}|}{c} = \frac{1}{c}\Big[\frac{(\mathbf{R}\cdot\mathbf{r})}{R} + \frac{1}{2}\frac{(\mathbf{R}\cdot\mathbf{r})^2 - R^2 r^2}{R^3} + \frac{1}{2}\frac{(\mathbf{R}\cdot\mathbf{r})^3 - R^2(\mathbf{R}\cdot\mathbf{r})r^2}{R^5} + \\ \frac{1}{8}\frac{5(\mathbf{R}\cdot\mathbf{r})^4 - 6R^2(\mathbf{R}\cdot\mathbf{r})^2 r^2 - R^4 r^4}{R^7} + \frac{1}{8}\frac{+7(\mathbf{R}\cdot\mathbf{r})^5 - 10R^2(\mathbf{R}\cdot\mathbf{r})^3 r^2 + 3R^4(\mathbf{R}\cdot\mathbf{r})r^4}{R^9}\Big]. \quad (1.6)$$

Using the below table, the equation (1.6) could be written in the tensor form, taking into account that repeating indices assuming Einstein summation and the definition of a delta Kronecker symbol $\delta_{ij}$.

Table.1 Correspondence of vector/dyadic notation to tensor notation

$$(\mathbf{R} \cdot \mathbf{r}) = R_i r_i;$$

$$(\mathbf{R} \cdot \mathbf{r})^2 = R_i R_j r_i r_j;$$

$$(\mathbf{R} \cdot \mathbf{r})^3 = R_i R_j R_k r_i r_j r_k;$$

$$(\mathbf{R} \cdot \mathbf{r})^4 = R_i R_j R_k R_l r_i r_j r_k r_l;$$

$$(\mathbf{R} \cdot \mathbf{r})^5 = R_i R_j R_k R_l R_p r_i r_j r_k r_l r_p;$$

$$R^2 r^2 = R^2 \delta_{ij} r_i r_j;$$

$$(\mathbf{R} \cdot \mathbf{r}) R^2 r^2 = \tfrac{1}{3} R^2 (R_i \delta_{jk} + R_j \delta_{ik} + R_k \delta_{ij}) r_i r_j r_k;$$

$$R^4 r^4 = \tfrac{1}{3} R^4 (\delta_{ij} \delta_{kl} + \delta_{ik} \delta_{jl} + \delta_{il} \delta_{jk}) r_i r_j r_k r_l;$$

$$(\mathbf{R} \cdot \mathbf{r})^2 R^2 r^2 = \tfrac{1}{6} R^2 (\delta_{ij} R_k R_l + \delta_{ik} R_j R_l + \delta_{il} R_j R_k + \delta_{jk} R_i R_l + \delta_{jl} R_i R_k + \delta_{kl} R_i R_j) r_i r_j r_k r_l;$$

$$(\mathbf{R} \cdot \mathbf{r})^3 R^2 r^2 = \frac{1}{10} R_\alpha R_\beta R_\gamma R^2 \Big[ \delta_{\alpha l} \delta_{\beta m} \delta_{\gamma k} \delta_{ij} + \delta_{\alpha l} \delta_{\beta j} \delta_{\gamma m} \delta_{ik} + \delta_{\alpha m} \delta_{\beta j} \delta_{\gamma k} \delta_{il} + \delta_{\alpha l} \delta_{\beta j} \delta_{\gamma k} \delta_{im} +$$
$$\delta_{\alpha i} \delta_{\beta l} \delta_{\gamma m} \delta_{jk} + \delta_{\alpha i} \delta_{\beta l} \delta_{\gamma k} \delta_{jm} + \delta_{\alpha i} \delta_{\beta m} \delta_{\gamma k} \delta_{jl} + \delta_{\alpha i} \delta_{\beta j} \delta_{\gamma m} \delta_{kl} +$$
$$+ \delta_{\alpha i} \delta_{\beta j} \delta_{\gamma l} \delta_{km} + \delta_{\alpha i} \delta_{\beta j} \delta_{\gamma k} \delta_{lm} \Big] r_i r_j r_k r_l r_m;$$

$$(\mathbf{R} \cdot \mathbf{r}) R^4 r^4 = \frac{1}{15} R_\alpha R^4 \Big[ \delta_{ij} \delta_{kl} \delta_{\alpha p} + \delta_{ij} \delta_{pl} \delta_{\alpha k} + \delta_{ij} \delta_{kp} \delta_{\alpha l} + \delta_{ik} \delta_{jp} \delta_{\alpha l} + \delta_{ik} \delta_{lp} \delta_{\alpha j} +$$
$$\delta_{ik} \delta_{lj} \delta_{\alpha p} + \delta_{il} \delta_{jp} \delta_{\alpha k} + \delta_{il} \delta_{kp} \delta_{\alpha j} + \delta_{il} \delta_{jk} \delta_{\alpha p} + \delta_{ip} \delta_{jl} \delta_{\alpha k} + \delta_{ip} \delta_{jk} \delta_{\alpha l} + \delta_{ip} \delta_{kl} \delta_{\alpha j} + \delta_{jk} \delta_{lp} \delta_{\alpha i} +$$
$$\delta_{jl} \delta_{kp} \delta_{\alpha i} + \delta_{jp} \delta_{kl} \delta_{\alpha i} \Big] r_i r_j r_k r_l r_p.$$

Performing Taylor series expansion of $\rho(r, t' + \Delta t)$ over $t'$ that is

$$\rho(\vec{r}, t' + \Delta t) = \rho(\mathbf{r}, t') + \frac{d\rho(\mathbf{r}, t')}{dt} \Delta t + \frac{1}{2!} \frac{d(\rho(\mathbf{r}, t'))^2}{d^2 t} \Delta t^2 + \frac{1}{3!} \frac{d(\rho(\mathbf{r}, t'))^3}{d^3 t} \Delta t^3 +$$
$$+ \frac{1}{4!} \frac{d(\rho(\mathbf{r}, t'))^4}{d^4 t} \Delta t^4 + \frac{1}{5!} \frac{d(\rho(\mathbf{r}, t'))^5}{d^5 t} \Delta t^5 ..$$
(1.7)

and performing series of $\frac{1}{|\mathbf{R}-\mathbf{r}|}$ similarly with (1.4) we found:

$$\frac{1}{|\mathbf{R} - \mathbf{r}|} = R^{-1} + \frac{(\mathbf{R} \cdot \mathbf{r})}{R^3} + \frac{1}{2} \frac{-r^2 R^2 + 3 (\mathbf{R} \cdot \mathbf{r})^2}{R^5} + \frac{1}{2} \frac{-3 (\mathbf{R} \cdot \mathbf{r}) r^2 R^2 + 5 (\mathbf{R} \cdot \mathbf{r})^3}{R^7} +$$
$$+ \frac{1}{16} \frac{6 r^4 R^4 - 60 (\mathbf{R} \cdot \mathbf{r})^2 r^2 R^2 + 70 (\mathbf{R} \cdot \mathbf{r})^4}{R^9} + \frac{1}{8} \frac{15 (\mathbf{R} \cdot \mathbf{r}) r^4 R^4 - 70 (\mathbf{R} \cdot \mathbf{r})^3 r^2 R^2 + 63 (\mathbf{R} \cdot \mathbf{r})^5}{R^7}$$
(1.8)

After multiplication of (1.7) and (1.8) we obtained the expansion of scalar potential $\Phi$:

$$\Phi = \frac{1}{4\pi\varepsilon_0} \int \frac{1}{R} \Big[ \frac{\dot{\overset{\cdots}{\rho}}(\mathbf{R}\cdot\mathbf{r})^5}{120R^5c^5} + \frac{\dot{\rho}(\mathbf{R}\cdot\mathbf{r})}{Rc} + 3/2\frac{\dot{\rho}(\mathbf{R}\cdot\mathbf{r})^2}{R^3c} - 1/2\frac{\dot{\rho}r^2}{Rc} +$$
$$+5/2\frac{\dot{\rho}(\mathbf{R}\cdot\mathbf{r})^3}{R^5c} + 3/8\frac{\dot{\rho}r^4}{R^3c} + \frac{35\dot{\rho}(\mathbf{R}\cdot\mathbf{r})^4}{8R^7c} + \frac{497\rho(1)(\mathbf{R}\cdot\mathbf{r})^5}{64R^9c}$$
$$+1/2\frac{\rho(2)(\mathbf{R}\cdot\mathbf{r})^2}{R^2c^2} + \frac{\ddot{\rho}(\mathbf{R}\cdot\mathbf{r})^3}{R^4c^2} + 1/8\frac{\ddot{\rho}r^4}{R^2c^2} + \frac{15\ddot{\rho}(\mathbf{R}\cdot\mathbf{r})^4}{8R^6c^2} +$$
$$+7/2\frac{\ddot{\rho}(\mathbf{R}\cdot\mathbf{r})^5}{R^8c^2} + 1/6\frac{\overset{\cdots}{\rho}(\mathbf{R}\cdot\mathbf{r})^3}{R^3c^3} + \frac{5\overset{\cdots}{\rho}(\mathbf{R}\cdot\mathbf{r})^4}{12R^5c^3} + \frac{7\rho(3)(\mathbf{R}\cdot\mathbf{r})^5}{8R^7c^3}$$
$$+1/24\frac{\rho(4)(\mathbf{R}\cdot\mathbf{r})^4}{R^4c^4} + 1/8\frac{\rho(4)(\mathbf{R}\cdot\mathbf{r})^5}{R^6c^4} + \rho - 3/2\frac{\dot{\rho}(\mathbf{R}\cdot\mathbf{r})r^2}{R^3c} - \frac{15\dot{\rho}(\mathbf{R}\cdot\mathbf{r})^2r^2}{4R^5c} +$$
$$+\frac{105\dot{\rho}(\mathbf{R}\cdot\mathbf{r})r^4}{64R^5c} - \frac{35\dot{\rho}(\mathbf{R}\cdot\mathbf{r})^3r^2}{4R^7c} - 1/2\frac{\ddot{\rho}(\mathbf{R}\cdot\mathbf{r})r^2}{R^2c^2} - 3/2\frac{\ddot{\rho}(\mathbf{R}\cdot\mathbf{r})^2r^2}{R^4c^2} +$$
$$+3/4\frac{\ddot{\rho}(\mathbf{R}\cdot\mathbf{r})r^4}{R^4c^2} - \frac{15\ddot{\rho}(\mathbf{R}\cdot\mathbf{r})^3r^2}{4R^6c^2} - 1/4\frac{\overset{\cdots}{\rho}(\mathbf{R}\cdot\mathbf{r})^2r^2}{R^3c^3} - 5/6\frac{\overset{\cdots}{\rho}(\mathbf{R}\cdot\mathbf{r})^3r^2}{R^5c^3} + 1/8\frac{\overset{\cdots}{\rho}(\mathbf{R}\cdot\mathbf{r})r^4}{R^3c^3} -$$
$$-1/12\frac{\overset{\cdots\cdot}{\rho}(\mathbf{R}\cdot\mathbf{r})^3r^2}{R^4c^4} + 3/8\frac{\rho r^4}{R^4} + \frac{35\rho(\mathbf{R}\cdot\mathbf{r})^4}{8R^8} - \frac{35\rho(\mathbf{R}\cdot\mathbf{r})^3r^2}{4R^8} + \frac{15\rho(\mathbf{R}\cdot\mathbf{r})r^4}{8R^6} - \frac{15\rho(\mathbf{R}\cdot\mathbf{r})^2r^2}{4R^6} -$$
$$-3/2\frac{\rho(\mathbf{R}\cdot\mathbf{r})r^2}{R^4} + \frac{\rho(\mathbf{R}\cdot\mathbf{r})}{R^2} + 3/2\frac{\rho(\mathbf{R}\cdot\mathbf{r})^2}{R^4} - 1/2\frac{\rho r^2}{R^2} + 5/2\frac{\rho(\mathbf{R}\cdot\mathbf{r})^3}{R^6} + \frac{63\rho(\mathbf{R}\cdot\mathbf{r})^5}{8R^{10}} \Big] dv \qquad (1.9)$$

taking into account that dots above the $\rho$ denote time derivatives and $\dot{\overset{\cdots}{\rho}}$ - is five order derivative. Considering only $1/R$ components contributed to the far-field we obtained scalar potential $\Phi$ for the far-field:

$$\Phi = \frac{1}{4\pi\varepsilon_0} \int \Big[ \frac{\rho}{R} + \frac{\dot{\rho}(\mathbf{R}\cdot\mathbf{r})}{R^2c} + \frac{1}{2}\frac{\ddot{\rho}(\mathbf{R}\cdot\mathbf{r})^2}{R^3c^2} + \frac{1}{6}\frac{\overset{\cdots}{\rho}(\mathbf{R}\cdot\mathbf{r})^3}{R^4c^3} +$$
$$+ \frac{1}{24}\frac{\overset{\cdots\cdot}{\rho}(\mathbf{R}\cdot\mathbf{r})^4}{R^5c^4} + \frac{1}{120}\frac{\dot{\overset{\cdots}{\rho}}(\mathbf{R}\cdot\mathbf{r})^5}{R^6c^5} \Big] dv \qquad (1.10)$$

with the definition of electric multipoles as

$$\begin{aligned} p_i &= \int \rho r_i dv, \\ Q^{(e)}_{ij} &= \int \rho r_i r_j dv, \\ O^{(e)}_{ijk} &= \int \rho r_i r_j r_k dv, \\ S_{ijkl} &= \int \rho r_i r_j r_k r_p dv, \\ X_{ijklp} &= \int \rho r_i r_j r_k r_l r_p dv, \end{aligned} \qquad (1.11)$$

or alternatively, through the expansion of continuity equation (5.3-5.9) and expression of volume charge density using current density **J** at $dv$:

$$
\begin{aligned}
p_i &= \frac{i}{\omega}\int J_i dv, \\
Q^{(e)}_{ij} &= \frac{i}{\omega}\int r_j J_i + r_i J_j dv, \\
O^{(e)}_{ijk} &= \frac{i}{\omega}\int r_k r_j J_i + r_k r_i J_j + r_i r_j J_k dv, \\
S_{ijkl} &= \frac{i}{\omega}\int r_l r_k r_j J_i + r_l r_k r_i J_j + r_l r_i r_j J_k + r_i r_k r_j J_k dv, \\
X_{ijklp} &= \frac{i}{\omega}\int r_p r_l r_k r_j J_i + r_p r_l r_k r_i J_j + r_p r_l r_i r_j J_k + r_p r_i r_k r_j J_l + r_i r_l r_k r_j J_p dv.
\end{aligned}
\tag{1.12}
$$

We are able to represent the scalar potential 1.19 in the tensor form:

$$
\begin{aligned}
\Phi = \frac{1}{4\pi\varepsilon_0}\Big[&\frac{q}{R} + \frac{\dot{p}_i R_i}{R^2 c} + \frac{1}{2}\frac{\ddot{Q}^{(e)}_{ij} R_i R_j}{R^3 c^2} + \frac{1}{6}\frac{\dddot{O}^{(e)}_{ijk} R_i R_j R_k}{R^4 c^3} + \\
& + \frac{1}{24}\frac{\ddddot{S}_{ijkl} R_i R_j R_k R_l}{R^5 c^4} + \frac{1}{120}\frac{\dot{\hat{X}}_{ijklp} R_i R_j R_k R_l R_p}{R^6 c^5}\Big].
\end{aligned}
\tag{1.13}
$$

## 1.3 Vector potential expansion

Similarly to the derivation of the scalar potential, we use the same steps for the vector potential:

$$
\mathbf{A} = \frac{\mu_0}{4\pi}\int\Big[\frac{\mathbf{J}}{R} + \frac{\dot{\mathbf{J}}(\mathbf{R}\cdot\mathbf{r})}{R^2 c} + \frac{1}{2}\frac{\ddot{\mathbf{J}}(\mathbf{R}\cdot\mathbf{r})^2}{R^3 c^2} + \frac{1}{6}\frac{\dddot{\mathbf{J}}(\mathbf{R}\cdot\mathbf{r})^3}{R^4 c^3} + \\
+ \frac{1}{24}\frac{\ddddot{\mathbf{J}}(\mathbf{R}\cdot\mathbf{r})^4}{R^5 c^4} + \ldots\Big]dv.
\tag{1.14}
$$

in the limitation of the $5^{th}$ order expansion. Taking into account series of $(\mathbf{R}\cdot\mathbf{r})^n$ terms in table (1) and substituting equations (5.31-5.35) into (1.14) we got:

$$
\begin{aligned}
A_i = \frac{\mu_0}{4\pi}\Big[&\frac{\dot{p}_i}{R} + \frac{1}{2}\frac{(\ddot{Q}^{(e)}_{ik} - \varepsilon_{ikj}\dot{m}_j)}{R^2 c}R_k + \Big(\frac{1}{6}\frac{\dddot{O}^{(e)}_{ikp}}{R^3 c^2} - \frac{1}{2}\frac{\varepsilon_{ikj}\ddot{Q}^{(m)}_{jp}}{R^3 c^2}\Big)R_k R_p + \\
& + \Big(\frac{1}{24}\frac{\ddddot{S}_{ikpl}}{R^4 c^3} - \frac{1}{6}\frac{\varepsilon_{ikj}\dddot{O}^{(m)}_{jpl}}{R^4 c^3}\Big)R_k R_p R_l + \\
& + \Big(\frac{1}{120}\frac{\dot{\hat{X}}_{ikplt}}{R^5 c^4} - \frac{1}{24}\frac{\varepsilon_{ikj}\ddddot{Y}_{jplt}}{R^5 c^4}\Big)R_k R_p R_l R_t\Big],
\end{aligned}
\tag{1.15}
$$

where magnetic multipoles are denoted as:

$$m_j = \frac{1}{2}\int (\mathbf{r}\times\mathbf{J})_j dv,$$
$$Q^{(m)}_{jp} = \frac{2}{3}\int r_p(\mathbf{r}\times\mathbf{J})_j dv,$$
$$O^{(m)}_{jpl} = \frac{3}{4}\int r_l r_p(\mathbf{r}\times\mathbf{J})_j dv, \tag{1.16}$$
$$Y_{jplt} = \frac{4}{5}\int r_t r_l r_p(\mathbf{r}\times\mathbf{J})_j dv,$$

and $\varepsilon_{ijk}$ is a Levi-Chivita symbol.

## 1.4 The multipole expansion of the E-field in far-zone

The scalar potential contributes to the E-field as $\mathbf{E} = -\nabla(\Phi)$, using identity

$$\begin{aligned}\nabla_i\Big[\frac{1}{R}f(t')\Big] &= -\frac{1}{R^2}\frac{R_i}{R}f(t') + \frac{1}{R}\frac{d(t')}{dR_i}\dot{f}(t') = \\ &= -\frac{R_i}{R^3}f(t') + \frac{1}{R}\frac{d(t-R/c)}{dR_i}\dot{f}(t') = -\frac{R_i}{R^3}f(t') - \frac{R_i}{cR^2}\dot{f}(t'),\end{aligned} \tag{1.17}$$

and the far-field approximation:

$$\nabla_i\Big[\frac{1}{R}f(t')\Big] \approx -\frac{R_i}{cR^2}\dot{f}(t'), \tag{1.18}$$

we find the contribution of scalar potential:

$$\begin{aligned}E^\Phi_i = \frac{1}{4\pi\varepsilon_0 c^2}\frac{1}{R}\Big[&\frac{R_i R_j}{R^2}\ddot{p}_j + \frac{1}{2}\frac{R_i R_j R_k}{R^3 c}\dddot{Q}^{(e)}_{jk} + \frac{1}{6}\frac{R_i R_j R_k R_p}{R^4 c^2}\ddddot{O}^{(e)}_{jkp}+ \\ &+\frac{1}{24}\frac{R_i R_j R_k R_p R_l}{R^5 c^3}\dddddot{S}_{jkpl} + \frac{1}{120}\frac{R_i R_j R_k R_l R_p R_t}{R^6 c^4}\ddddddot{X}_{jklpt}\Big].\end{aligned} \tag{1.19}$$

For the vector potential contribution to the E-field in the far zone we have

$$\begin{aligned}E^A_i = -\dot{A}_i = -\frac{1}{4\pi\varepsilon_0 c^2}\frac{1}{R}\Big[&\ddot{p}_i + \frac{1}{2}\frac{(\dddot{Q}^{(e)}_{ik}-\varepsilon_{ikj}\ddot{m}_j)}{Rc}R_k + \Big(\frac{1}{6}\frac{\ddddot{O}^{(e)}_{ikp}}{R^2 c^2} - \frac{1}{2}\frac{\varepsilon_{ikj}\dddot{Q}^{(m)}_{jp}}{R^2 c^2}\Big)R_k R_p+ \\ &+\Big(\frac{1}{24}\frac{\dddddot{S}_{ikpl}}{R^3 c^3} - \frac{1}{6}\frac{\varepsilon_{ikj}\ddddot{O}^{(m)}_{jpl}}{R^3 c^3}\Big)R_k R_p R_l+ \\ &+\Big(\frac{1}{120}\frac{\ddddddot{X}_{ikplt}}{R^4 c^4} - \frac{1}{24}\frac{\varepsilon_{ikj}\dddddot{Y}_{jplt}}{R^4 c^4}\Big)R_k R_p R_l R_t\Big].\end{aligned} \tag{1.20}$$

and finally we obtain

$$E_i = E_i^{\Phi} + E_i^{A} = \frac{1}{4\pi\varepsilon_0 c^2 R}\left[\frac{(R_i R_j - R^2 \delta_{ij})}{R^2}\left(\ddot{p}_j + \frac{1}{2}\frac{R_k}{Rc}\dddot{Q}_{jk}^{(e)} + \frac{1}{6}\frac{R_k R_p}{R^2 c^2}\ddddot{O}_{jkp}^{(e)} + \right.\right.$$
$$\left.+ \frac{1}{24}\frac{R_k R_l R_p}{R^3 c^3}\dot{\hat{S}}_{jkpl} + \frac{1}{120}\frac{R_k R_l R_p R_t}{R^4 c^4}\ddot{\hat{X}}_{jkplt}\right) +$$
$$+\left(\frac{\varepsilon_{ikj}\ddot{m}_j}{Rc}R_k + \frac{1}{2}\frac{\varepsilon_{ikj}\dddot{Q}_{jp}^{(m)}}{R^2 c^2}R_k R_p + \frac{1}{6}\frac{\varepsilon_{ikj}\ddddot{O}_{jpl}^{(m)}}{R^3 c^3}R_k R_p R_l + \frac{1}{24}\frac{\varepsilon_{ikj}\dot{Y}_{jplt}}{R^4 c^4}R_k R_p R_l R_t\right)\Bigg],$$
(1.21)

Taking into account the definition $n_i = R_i/|\mathbf{R}|$ and taking derivatives we simplify equation (1.21) to

$$E_i = \frac{k^2}{4\pi\varepsilon_0 R}\left[\left(n_i n_j - n^2 \delta_{ij}\right)\left(-p_j + \frac{ik}{2}(Q_{jk}^{(e)} n_k) + \frac{k^2}{6}(O_{jkp}^{(e)} n_k n_p) + \right.\right.$$
$$\left.+ \frac{-ik^3}{24}(S_{jkpl} n_k n_p n_l) + \frac{-k^4}{120}(X_{jkplt} n_k n_p n_l n_t)\right) +$$
$$+\left(\frac{-1}{c}\varepsilon_{ikj} n_k m_j + \frac{ik}{2c}\varepsilon_{ikj} n_k (Q_{jp}^{(m)} n_p) + \frac{k^2}{6c}\varepsilon_{ikj} n_k (O_{jpl}^{(m)} n_p n_l) + \frac{-ik^3}{24c}\varepsilon_{ikj} n_k (Y_{jplt} n_p n_l n_t)\right)\Bigg],$$
(1.22)

and transforming tensors to vector/dyadic notation

$$\mathbf{E} = \frac{k^2}{4\pi\varepsilon_0}\frac{1}{R}\Bigg(\left[\mathbf{n}\times[\mathbf{p}\times\mathbf{n}]\right] + \frac{ik}{2}[\mathbf{n}\times[\mathbf{n}\times(\overline{\mathbf{Q}}^{(e)}\cdot\mathbf{n})]] +$$
$$+\frac{k^2}{6}[\mathbf{n}\times[\mathbf{n}\times(\overline{\mathbf{O}}^{(e)}\cdot\mathbf{n}\cdot\mathbf{n})]] + \frac{ik^3}{24}[\mathbf{n}\times[\overline{\mathbf{S}}\cdot\mathbf{n}\cdot\mathbf{n}\cdot\mathbf{n})\times\mathbf{n}]] + \frac{k^4}{120}[\mathbf{n}\times[(\overline{\mathbf{X}}\cdot\mathbf{n}\cdot\mathbf{n}\cdot\mathbf{n}\cdot\mathbf{n})\times\mathbf{n}]]$$
$$+\frac{1}{c}[\mathbf{m}\times\mathbf{n}] + \frac{ik}{2c}[\mathbf{n}\times(\mathbf{Q}^{(m)}\cdot\mathbf{n})] + \frac{k^2}{6c}[\mathbf{n}\times(\mathbf{O}^{(m)}\cdot\mathbf{n}\cdot\mathbf{n})]$$
$$+\frac{k^3}{24c}[(\mathbf{Y}\cdot\mathbf{n}\cdot\mathbf{n}\cdot\mathbf{n})\times\mathbf{n}]\Bigg).$$
(1.23)

## 2 Symmetrization and detracing

### 2.1 Used notations

In our notation a tensor denoted without a line over a letter is not symmetrical and traceless, for example $A_{ijk}$. One line above a tensor $\overline{A}_{ijk}$ denotes symmetrical and not traceless tensor. The two times overlined tensor $\overline{\overline{A}}_{ijk}$ defines a traceless and symmetrical tensor. Traceless and symmetrical tensors fulfill the requirement of irreducibility.

All primitive electrical type multipoles are symmetrical, and not traceless. Starting with the magnetic octupole all primitive magnetic type multipoles are not traceless and symmetrical, with the exception of the magnetic quadrupole which is traceless and non-symmetrical. For the definition of tensors with contracted indices and decreased rank by 2 we use undeline, for example $\overline{\underline{X}}_{vvjkl}$. In similar way, two index contractions of a $5^{th}$-rank results in vector $\overline{\underline{\underline{X}}}_{vvuuj}$.

## 2.2 Detracing

We consider detracing process using Applequist's theorem [2] for a symmetric tensor $\overline{A}$:

$$\overline{\overline{A}}^{(n)}_{\alpha_1...\alpha_n} = \sum_{m=0}^{\lfloor n/2 \rfloor} (-1)^m (2n - 2m - 1)!! \sum_{T_{\{\alpha\}}} \delta_{\alpha_1 \alpha_2}...\delta_{\alpha_{(2m-1)}\alpha_{(2m)}} \overline{A}^{n:m}_{\alpha_{(2m-1)}\alpha_{(n)}}. \tag{2.1}$$

Using (2.1) we show procedure for detracing of a tensor $\overline{\overline{A}}$ till the $5^{th}$ order:
for the $2^{nd}$ rank tensor $\overline{A}$:

$$\overline{\overline{A}}_{ij} = \frac{1}{3}(3 \cdot \overline{A}_{ij} - \delta_{ij}\overline{A}_{vv}), \tag{2.2}$$

for the $3^{rd}$ rank tensor $\overline{A}$:

$$\overline{\overline{A}}_{ijk} = \frac{1}{3 \cdot 5}\left[3 \cdot 5 \cdot \overline{A}_{ijk} - 3 \cdot (\delta_{ij}\overline{A}_{vvk} + \delta_{ik}\overline{A}_{vvj} + \delta_{jk}\overline{A}_{vvi})\right], \tag{2.3}$$

for the $4^{th}$ rank tensor $\overline{A}$:

$$\overline{\overline{A}}_{ijkl} = \frac{1}{3 \cdot 5 \cdot 7}\Big[3 \cdot 5 \cdot 7 \cdot \overline{A}_{ijkl} - 3 \cdot 5 \cdot (\delta_{ij}\overline{A}_{vvkl} + \delta_{ik}\overline{A}_{vvjl} + \delta_{jk}\overline{A}_{vvil} + \\ + \delta_{jl}\overline{A}_{vvik} + \delta_{kl}\overline{A}_{vvij} + \delta_{il}\overline{A}_{vvjk}) + 3 \cdot (\delta_{ij}\delta_{kl}\overline{A}_{uuvv} + \delta_{ik}\delta_{jl}\overline{A}_{uuvv} + \delta_{il}\delta_{jk}\overline{A}_{uuvv})\Big], \tag{2.4}$$

and the for $5^{th}$ rank tensor $\overline{A}$:

$$\overline{\overline{A}}_{ijkla} = \Big[\overline{A}_{ijkla} - \frac{1}{9}(\delta_{ij}\overline{A}_{vvkla} + \delta_{ik}\overline{A}_{vvjla} + \\ \delta_{il}\overline{A}_{vvjak} + \delta_{ia}\overline{A}_{vvjkl} + \delta_{jk}\overline{A}_{vvial} + \delta_{jl}\overline{A}_{vviak} + \delta_{ja}\overline{A}_{vvikl} + \delta_{kl}\overline{A}_{vvija} + \delta_{ka}\overline{A}_{vvijl} + \delta_{la}\overline{A}_{vvijk}) + \\ + \frac{1}{7 \cdot 9}(\delta_{ij}\delta_{kl}\overline{A}_{uuvva} + \delta_{ik}\delta_{jl}\overline{A}_{uuvva} + \delta_{il}\delta_{jk}\overline{A}_{uuvva} + \\ + \delta_{aj}\delta_{kl}\overline{A}_{uuvvi} + \delta_{ak}\delta_{jl}\overline{A}_{uuvvi} + \delta_{al}\delta_{jk}\overline{A}_{uuvvi} + \\ + \delta_{ia}\delta_{kl}\overline{A}_{uuvvj} + \delta_{ik}\delta_{al}\overline{A}_{uuvvj} + \delta_{il}\delta_{ak}\overline{A}_{uuvvj} + \\ + \delta_{ij}\delta_{al}\overline{A}_{uuvvk} + \delta_{ia}\delta_{jl}\overline{A}_{uuvvk} + \delta_{il}\delta_{ja}\overline{A}_{uuvvk} + \\ + \delta_{ij}\delta_{ka}\overline{A}_{uuvvl} + \delta_{ik}\delta_{ja}\overline{A}_{uuvvl} + \delta_{ia}\delta_{jk}\overline{A}_{uuvvl})\Big]. \tag{2.5}$$

We express primitive electrical multipoles in their irreducible form using obtained equations (2.2 - 2.5), with substitution index set from {i,j,k,l,a} to {j,k,p,l,t}, for direct implementation of the obtained tensors in **E**-field equation (1.22). Applying equation (2.2) for the electric

quadrupole we have:

$$\overline{Q}^{(e)}_{jk} = \overline{\overline{Q}}^{(e)}_{jk} + \frac{1}{3}\delta_{jk}\overline{Q}^{(e)}_{vv}, \qquad (2.6)$$

where scalar $Q^{(e)}_{vv}$ is

$$\overline{Q}^{(e)}_{vv} = \frac{i}{\omega}\int 2(\mathbf{r}\cdot\mathbf{J})dv. \qquad (2.7)$$

Using equation (2.3) for the electric octupole we find:

$$\overline{O}^{(e)}_{jkl} = \overline{\overline{O}}^{(e)}_{jkl} + \frac{1}{5}(\delta_{jk}\overline{O}^{(e)}_{vvl} + \delta_{jl}\overline{O}^{(e)}_{vvk} + \delta_{lk}\overline{O}^{(e)}_{vvj}), \qquad (2.8)$$

where the tensor $\overline{O}^{(e)}_{vvj}$ is defined as

$$\overline{O}^{(e)}_{vvj} = \frac{i}{\omega}\int 2(\mathbf{r}\cdot\mathbf{J})r_j + r^2 J_j dv. \qquad (2.9)$$

To obtain traceless electric 16-pole firstly we applied equation (2.4) and afterwards (2.2) detracing 2-rank tensors obtained from the first detracing:

$$\overline{S}_{jkpl} = \overline{\overline{S}}_{jkpl} + \frac{1}{7}(\delta_{pj}\overline{\overline{S}}_{vvkl} + \delta_{pk}\overline{\overline{S}}_{vvjl} + \delta_{jk}\overline{\overline{S}}_{vvpl} + \delta_{jl}\overline{\overline{S}}_{vvpk} + \delta_{kl}\overline{\overline{S}}_{vvpj} + \delta_{pl}\overline{\overline{S}}_{vvjk}) + $$
$$+ (2\frac{1}{3}\frac{1}{7} - \frac{1}{5\cdot 7})(\delta_{pj}\delta_{kl}\overline{S}_{uuvv} + \delta_{pk}\delta_{jl}\overline{S}_{uuvv} + \delta_{pl}\delta_{jk}\overline{S}_{uuvv}), \qquad (2.10)$$

where in in the first term of brackets $(2\frac{1}{3}\frac{1}{7} - \frac{1}{5\cdot 7})$ the number 2 is obtained from the number of 2-rank tensors in equation (2.4), fraction $\frac{1}{7}$ comes from factor of the brackets with 2-rank tensors in the equation (2.4), and fraction $\frac{1}{3}$ is the number obtained during detracing in equation (2.2) The obtained tensors such as $\overline{\overline{S}}_{vvjp}$ have integral form:

$$\overline{\overline{S}}_{vvjp} = \frac{i}{\omega}\int 2(\mathbf{r}\cdot\mathbf{J})r_j r_p + r^2 J_j r_p + r^2 J_p r_j - \frac{4}{3}(\mathbf{r}\cdot\mathbf{J})r^2 dv \qquad (2.11)$$

and obtained scalars such $\overline{S}_{vvuu}$ produce integral:

$$\overline{S}_{vvvv} = \frac{i}{\omega}\int 4(\mathbf{r}\cdot\mathbf{J})r^2 dv. \qquad (2.12)$$

Applying the same steps for the electric 32-pole we obtain:

$$\overline{X}_{jkplt} = \Big[\overline{\overline{X}}_{jkplt} + \frac{1}{9}(\delta_{jk}\overline{\overline{X}}_{vvplt} + \delta_{jp}\overline{\overline{X}}_{vvklt}+$$

$$\delta_{jl}\overline{\overline{X}}_{vvktp} + \delta_{jt}\overline{\overline{X}}_{vvkpl} + \delta_{kp}\overline{\overline{X}}_{vvjtl} + \delta_{kl}\overline{\overline{X}}_{vvjtp} + \delta_{kt}\overline{\overline{X}}_{vvjpl} + \delta_{pl}\overline{\overline{X}}_{vvjkt} + \delta_{pt}\overline{\overline{X}}_{vvjkl} + \delta_{lt}\overline{\overline{X}}_{vvjkp}) +$$

$$(2\frac{1}{5}\frac{1}{9} - \frac{1}{7\cdot 9})(\delta_{jk}\delta_{pl}\overline{X}_{uuvvt} + \delta_{jp}\delta_{kl}\overline{X}_{uuvvt} + \delta_{jl}\delta_{kp}\overline{X}_{uuvvt}+$$

$$+\delta_{tk}\delta_{pl}\overline{X}_{uuvvj} + \delta_{tp}\delta_{kl}\overline{X}_{uuvvj} + \delta_{tl}\delta_{kp}\overline{X}_{uuvvj}+$$

$$+\delta_{jt}\delta_{pl}\overline{X}_{uuvvk} + \delta_{jp}\delta_{tl}\overline{X}_{uuvvk} + \delta_{jl}\delta_{tp}\overline{X}_{uuvvk}+$$

$$+\delta_{jk}\delta_{tl}\overline{X}_{uuvvp} + \delta_{jt}\delta_{kl}\overline{X}_{uuvvp} + \delta_{jl}\delta_{kt}\overline{X}_{uuvvp}+$$

$$+\delta_{jk}\delta_{pt}\overline{X}_{uuvvl} + \delta_{jp}\delta_{kt}\overline{X}_{uuvvl} + \delta_{jt}\delta_{kp}\overline{X}_{uuvvl})\Big],$$

(2.13)

where $3^{rd}$ rank tensors such as $\overline{\overline{X}}_{vvjkl}$ have the following form:

$$\overline{\overline{X}}_{vvjkl} = \frac{i}{\omega}\int 2(\mathbf{r}\cdot\mathbf{J})r_jr_kr_l + r^2(r_jr_kJ_l + r_lr_kJ_j + r_jr_lJ_k)$$
$$-\frac{1}{5}\Big[(4r^2(\mathbf{r}\cdot\mathbf{J})r_p + r^4J_p)(\delta_{jk}\delta_{lp} + \delta_{jl}\delta_{kp} + \delta_{lk}\delta_{jp})\Big]dv$$

(2.14)

and $1^{st}$ rank tensors are denoted as:

$$\overline{X}_{vvuuj} = \frac{i}{\omega}\int 4(\mathbf{r}\cdot\mathbf{J})r^2r_j + r^4J_jdv.$$

(2.15)

In equations (2.6, 2.8, 2.10, 2.13) we introduce once overlined tensors through double overlined tensors (irreducible tensor) further substituting them into the **E** far-field equation (1.22).

## 2.3 Symmetrization

To simplify the process of derivation of equations for symmetrization of high order tensors preventing lose of generality we preserve names and appearance of indices as in the equation for the **E**-field for magnetic multipoles.

2.3 Symmetrization of 2nd rank tensor. Magnetic dipole

To consider a symmetrization of 2-rank tensor, we write the identity $A_{jp} = A_{jp}$ and represent it as following:

$$A_{jp} = \frac{1}{2}\big[A_{jp} + A_{pj}\big] + \frac{1}{2}\big[A_{jp} - A_{pj}\big].$$

(2.16)

The first bracket is the symmetrized term and the second one could be represented using the Levi-Chivita antisymmetric tensor:

$$A_{jp} - A_{pj} = \left[\delta_{uj}\delta_{vp} - \delta_{vj}\delta_{up}\right]A_{uv} = \varepsilon_{\beta jp}(\varepsilon_{uv\beta}A_{uv}) = \varepsilon_{\beta jp}N_{\beta}. \tag{2.17}$$

In result we obtained identity

$$A_{jp} = \overline{A}_{jp} + \frac{1}{2}\varepsilon_{\beta jp}N_{\beta}, \tag{2.18}$$

with two terms in the right side - the symmetrical $2^{nd}$ rank tensor and the antisymmetric one composing from the Levi-Chivita 3-rank tensor and a vector. Also we introduce **N**-vector:

$$\mathbf{N} = \frac{2}{3}\int r^2\mathbf{J} - (\mathbf{r}\cdot\mathbf{J})\mathbf{r}dv, \tag{2.19}$$

which will be used for the derivation of the electric toroidal moment.

2.3 Symmetrization of a 3th rank tensor. Magnetic octupole

For a nonsymmetric tensor $A_{jpl}$ its symmetrical form is obtained using summation of all possible permutations:

$$\overline{A}_{jpl} = \frac{1}{6}(A_{jpl} + A_{jlp} + A_{pjl} + A_{plj} + A_{ljp} + A_{lpj}). \tag{2.20}$$

Taking into account the partial symmetry of the magnetic octupole moment

$$O^{(m)}_{jpl} = \frac{3}{4}\int r_l r_p (\vec{r}\times\vec{J})_j dv, \tag{2.21}$$

which is expressed in following relations:

$$O^{(m)}_{jpl} = O^{(m)}_{jlp} = \frac{3}{4}\int(\mathbf{r}\times\mathbf{J})_j r_p r_l dv = \frac{3}{4}\int(\mathbf{r}\times\mathbf{J})_j r_l r_p dv, \tag{2.22}$$

$$O^{(m)}_{pjl} = O^{(m)}_{plj} = \frac{3}{4}\int(\mathbf{r}\times\mathbf{J})_p r_j r_l dv = \frac{3}{4}\int(\mathbf{r}\times\mathbf{J})_j r_l r_j dv, \tag{2.23}$$

$$O^{(m)}_{ljp} = O^{(m)}_{lpj} = \frac{3}{4}\int(\mathbf{r}\times\mathbf{J})_l r_j r_p dv = \frac{3}{4}\int(\mathbf{r}\times\mathbf{J})_l r_p r_j dv. \tag{2.24}$$

We assume that the tensor $A$ has integral form of the magnetic octupole for simplification only, and derive all the following expressions in the tensor form without loss of generality. Due to the fact that the magnetic octupole has partial symmetry (2.22-2.24), the equation (2.20) could

be reduced to:
$$\overline{A}_{jpl} = \frac{1}{3}\Big(A_{jpl} + A_{pjl} + A_{lpj}\Big) \tag{2.25}$$

To find the irreducible form for the magnetic octupole, we write the identity $A_{jpl} = A_{jpl}$ and represent it as sums and differences as in the equation (2.16):

$$A_{jpl} = \Big[\frac{1}{3}(A_{jpl} + A_{pjl} + A_{ljp}) + \frac{1}{3}(A_{jpl} - A_{pjl}) + \frac{1}{3}(A_{jpl} - A_{lpj})\Big]. \tag{2.26}$$

We denote the gathered terms of (2.26) and express using Levi-Civita symbols:

$$A_{jpl} = \overline{A}_{jpl} + \frac{1}{3}\varepsilon_{\alpha jp}R_{\alpha l} + \frac{1}{3}\varepsilon_{\alpha jl}R_{\alpha p}, \tag{2.27}$$

where $R$ is a 2-rank tensor defined as:

$$R_{\alpha l} = \varepsilon_{\alpha\beta\gamma}A_{\beta\gamma l} = \frac{3}{4}\int\Big[J_\alpha r_l r^2 - r_\alpha r_l(\mathbf{r}\cdot\mathbf{J})\Big]dv, \tag{2.28}$$

and $\overline{A}$ is symmetrical, and not traceless 3-rank tensor

$$\overline{A}_{jpl} = \frac{1}{3}(A_{jpl} + A_{pjl} + A_{ljp}). \tag{2.29}$$

Using equation (2.3) we find symmetrized and traceless $3^{rd}$ order tensor $\overline{\overline{A}}_{jpl}$ as following:

$$\overline{\overline{A}}_{jpl} = \overline{A}_{jpl} - \frac{1}{5}\Big(\delta_{jp}\overline{A}_{vvl} + \delta_{jl}\overline{A}_{vvp} + \delta_{pl}\overline{A}_{vvj}\Big), \tag{2.30}$$

where vectors in integrals such as $\overline{A}_{vvj}$ are

$$\overline{A}_{vvj} = \frac{1}{3}\frac{3}{4}\int 2[(\mathbf{r}\times\mathbf{J})\cdot\mathbf{r}]r_j + r^2(\mathbf{r}\times\mathbf{J})_j dv = \frac{1}{4}\int r^2(\mathbf{r}\times\mathbf{J})_j dv. \tag{2.31}$$

Due to the partial symmetry of the magnetic octupole moment (see formulas (2.22)-(2.24) ) $R$ -tensor has zero trace:

$$R_{\alpha\alpha} = \underline{A_{231}} - \underline{\underline{A_{321}}} + \underline{\underline{A_{312}}} - A_{132} + A_{123} - \underline{A_{213}} = 0, \tag{2.32}$$

but it remains nonsymmetrical. Symmetrizing 2-rank tensor $\overline{R}$

$$R_{\alpha l} = \frac{1}{2}(R_{\alpha l} + R_{l\alpha}) + \frac{1}{2}(R_{\alpha l} - R_{l\alpha}), \tag{2.33}$$

we obtain its integral form:

$$\overline{\overline{R}}_{\alpha l} = \frac{1}{2}(R_{\alpha l} + R_{l\alpha}) = \frac{3}{8} \int r^2 \Big((J_l r_\alpha + J_\alpha r_l)\Big) - 2(\mathbf{J} \cdot \mathbf{r}) r_\alpha r_l dv. \qquad (2.34)$$

Denoting the difference using Levi-Chivita symbol we derive:

$$R_{\alpha l} - R_{l\alpha} = \varepsilon_{\beta \alpha l} G_\beta, \qquad (2.35)$$

where

$$G_\beta = \varepsilon_{\beta xs}\varepsilon_{xvw} A_{vws} = -(\delta_{\beta v}\delta_{sw} - \delta_{\beta w}\delta_{sv})A_{vws} = \\ = A_{s\beta s} - A_{\beta ss} = -\frac{3}{4}\int (\mathbf{r} \times \mathbf{J}) r^2 dv. \qquad (2.36)$$

Applying to equation (2.26) equations (2.30, 2.34, 2.35) we finally have:

$$A_{jpl} = \overline{\overline{A}}_{jpl} + \frac{1}{5}\Big(\delta_{jp}\overline{A}_{vvl} + \delta_{jl}\overline{A}_{vvp} + \delta_{pl}\overline{A}_{vvj}\Big) + \\ + \frac{1}{3}\varepsilon_{\alpha jp}(\overline{\overline{R}}_{\alpha l} + \frac{1}{2}\varepsilon_{\beta \alpha l}G_\beta) + \frac{1}{3}\varepsilon_{\alpha jl}(\overline{\overline{R}}_{\alpha p} + \frac{1}{2}\varepsilon_{\beta \alpha p}G_\beta). \qquad (2.37)$$

2.3 Symmetrization of 4th rank tensor. Magnetic 16-pole

Using the same steps as in equations (2.16) and (2.26) we obtain for the $4^{th}$ rank tensor:

$$A_{jplt} = \Big[\frac{1}{4}(A_{jplt} + A_{pjlt} + A_{ljpt} + A_{tjpl}) + \frac{1}{4}(A_{jplt} - A_{pjlt}) \\ + \frac{1}{4}(A_{jplt} - A_{lpjt}) + \frac{1}{4}(A_{jplt} - A_{tlpj})\Big] = \\ = \frac{1}{4}\overline{A_{jplt}} + \frac{1}{4}\varepsilon_{\gamma jp}\varepsilon_{\alpha\beta\gamma}A_{\alpha\beta lt} + \frac{1}{4}\varepsilon_{\gamma jl}\varepsilon_{\alpha\beta\gamma}A_{\alpha\beta pt} + \frac{1}{4}\varepsilon_{\gamma jt}\varepsilon_{\alpha\beta\gamma}A_{\alpha\beta pl}. \qquad (2.38)$$

Substituting the definition of the magnetic 16-pole into term $\varepsilon_{\alpha\beta\gamma}A_{\alpha\beta lt}$ and expressing Levi-Chivita symbols as differences of delta Kroneker's symbols

$$\varepsilon_{\alpha\beta\gamma}A_{\alpha\beta lt} = \frac{4}{5}\int \varepsilon_{\alpha\beta\gamma}(\mathbf{r} \times \mathbf{J})_\alpha r_\beta r_l r_t dv = \\ = \frac{4}{5}\int \varepsilon_{\alpha\beta\gamma}\varepsilon_{\alpha xy}r_x J_y r_\beta r_l r_t dv = \frac{4}{5}\int (\delta_{\beta x}\delta_{\gamma y} - \delta_{\beta y}\delta_{\gamma x})r_x J_y r_\beta r_l r_t dv = \\ = \frac{4}{5}\int r_\beta J_\gamma r_\beta r_l r_t - r_\gamma J_\beta r_\beta r_l r_t dv = \frac{4}{5}\int r^2 J_\gamma r_l r_t - r_\gamma(\mathbf{r} \cdot \mathbf{J})r_l r_t dv = C_{\gamma lt} + D_{\gamma lt} \qquad (2.39)$$

we decomposed the term $\varepsilon_{\alpha\beta\gamma}A_{\alpha\beta lt}$ into two $3^{rd}$ rank tensors and denoted them as $C$ and $D$:

$$C_{\gamma lt} = \frac{4}{5}\int r^2 J_\gamma r_l r_t dv,$$
$$D_{\gamma lt} = -\frac{4}{5}\int r_\gamma (\mathbf{r}\cdot\mathbf{J})r_l r_t dv. \tag{2.40}$$

The tensor $C$ similarly to $A_{jplt}$ is not symmetrical and traceless as a magnetic octupole and the tensor $D$ is symmetrical and not traceless as an electric octupole. This separation will help us simplify equations. With substitution equation (2.39) in (2.38) we derive:

$$A_{jplt} = \overline{A}_{jplt} + \frac{1}{4}\varepsilon_{\gamma jp}(C_{\gamma lt} + D_{\gamma lt}) + \frac{1}{4}\varepsilon_{\gamma jl}(C_{\gamma pt} + D_{\gamma lt}) + \frac{1}{4}\varepsilon_{\gamma jt}(C_{\gamma pl} + D_{\gamma lt}). \tag{2.41}$$

As it was said above, the tensor $D_{\gamma lt}$ is symmetrical and not traceless, i.e. $D_{\gamma lt} = \overline{D}_{\gamma lt}$. Then using the equation (2.3) we have:

$$\overline{D}_{\gamma lt} = \overline{\overline{D}}_{\gamma lt} + \frac{1}{5}(\delta_{\gamma l}\overline{D}_{vvt} + \delta_{tl}\overline{D}_{vv\gamma} + \delta_{\gamma t}\overline{D}_{vvl}), \tag{2.42}$$

where the tensor $\overline{\overline{D}}_{\gamma lt}$ has the integral form:

$$\overline{\overline{D}}_{\gamma lt} = -\frac{4}{5}\int (\mathbf{r}\cdot\mathbf{J})\Big[r_\gamma r_t r_l - \frac{1}{5}r^2(\delta_{\gamma t}\delta_{lp} + \delta_{\gamma l}\delta_{tp} + \delta_{\gamma p}\delta_{tp})r_p\Big], dv \tag{2.43}$$

and the vectors such as $\overline{D}_{vvt}$ are defined as:

$$\overline{D}_{vvt} = -\frac{4}{5}\int (\mathbf{r}\cdot\mathbf{J})r^2 r_t dv. \tag{2.44}$$

Using equation (2.37) to decompose $C$ we obtain:

$$C_{\gamma lt} = \overline{\overline{C}}_{\gamma lt} + \frac{1}{5}\Big(\delta_{\gamma l}\overline{C}_{vvt} + \delta_{\gamma t}\overline{C}_{vvl} + \delta_{lt}\overline{C}_{vv\gamma}\Big) +$$
$$+ \frac{1}{3}\varepsilon_{\alpha\gamma l}(\overline{F}_{\alpha t} + \frac{1}{2}\varepsilon_{\beta\alpha t}H_\beta) + \frac{1}{3}\varepsilon_{\alpha\gamma t}(\overline{F}_{\alpha l} + \frac{1}{2}\varepsilon_{\beta\alpha l}H_\beta), \tag{2.45}$$

where the symmetrical and traceless tensor $\overline{\overline{C}}$ is defined as:

$$\overline{\overline{C}}_{\gamma lt} = \frac{1}{3}\frac{4}{5}\int r^2\Big[(J_\gamma r_l r_t + J_l r_\gamma r_t + J_t r_\gamma r_l) -$$
$$- \frac{1}{5}(\delta_{\gamma t}\delta_{lp} + \delta_{\gamma l}\delta_{tp} + \delta_{\gamma p}\delta_{tp})(2(\mathbf{r}\cdot\mathbf{J})r_p + J_p r^2)\Big]dv, \tag{2.46}$$

and vectors such as $\overline{\overline{C}}_{vvt}$ have integral form:

$$\overline{C}_{vvt} = \frac{1}{3}\frac{4}{5} \int \left[ r^4 J_t + 2(\mathbf{r} \cdot \mathbf{J})r_t \right] dv. \tag{2.47}$$

Similarly to the tensor $R$ we define $F$:

$$F_{\alpha t} = \varepsilon_{\alpha\beta x} C_{\beta x t} = \frac{4}{5} \int r^2 \varepsilon_{\alpha\beta x} J_\beta r_x r_t dv = -\frac{4}{5} \int r^2 (\mathbf{r} \times \mathbf{J})_\alpha r_t dv, \tag{2.48}$$

the tensor $\overline{\overline{F}}$:

$$\overline{\overline{F}}_{\alpha t} = -\frac{1}{2}\frac{4}{5} \int r^2 \left[ (\mathbf{r} \times \mathbf{J})_t r_\alpha + (\mathbf{r} \times \mathbf{J})_\alpha r_t \right] dv, \tag{2.49}$$

and similarly to the $G_\beta$ we define the tensor $H_\beta$:

$$H_\beta = \varepsilon_{\beta pt}\varepsilon_{pvw} C_{vwt} = -(\delta_{\beta v}\delta_{tw} - \delta_{\beta w}\delta_{tv})C_{vwt} = C_{t\beta t} - C_{\beta tt} =$$
$$= \frac{4}{5} \int r^2 (\mathbf{J} \cdot \mathbf{r}) r_\beta dv - \frac{4}{5} \int r^4 J_\beta dv. \tag{2.50}$$

Expressing from (2.4) only once overlined the $4^{th}$ rank tensor

$$\overline{A}_{jplt} = \overline{\overline{A}}_{jplt} + \frac{1}{7}(\delta_{jp}\overline{A}_{vvlt} + \delta_{jl}\overline{A}_{vvpt} + \delta_{pl}\overline{A}_{vvjt}+$$
$$+ \delta_{pt}\overline{A}_{vvjl} + \delta_{lt}\overline{A}_{vvjp} + \delta_{jt}\overline{A}_{vvpl})- \tag{2.51}$$
$$- \frac{1}{5 \cdot 7}(\delta_{jp}\delta_{lt}\overline{A}_{uuvv} + \delta_{jl}\delta_{pt}\overline{A}_{uuvv} + \delta_{jt}\delta_{pl}\overline{A}_{uuvv}),$$

and taking into account, that all terms in (2.51) such as of $\overline{A}_{vvlt}$ are symmetrical and traceless so we are able to write (the contribution of term in square brackets in (2.52) is zero):

$$\overline{\overline{A}}_{vvlt} = \overline{A}_{vvlt} = \frac{1}{4}\frac{4}{5} \int \left[ 2((\mathbf{r} \times \mathbf{J}) \cdot \mathbf{r})r_l r_t \right] + r^2((\mathbf{r} \times \mathbf{J})_l r_t + (\mathbf{r} \times \mathbf{J})_t r_l) dv =$$
$$= \frac{1}{4}\frac{4}{5} \int r^2 \left( (\mathbf{r} \times \mathbf{J})_l r_t + (\mathbf{r} \times \mathbf{J})_t r_l \right) dv, \tag{2.52}$$

we substitute equations (2.51) and (2.45) (2.42) into equation (2.41) and find the final equation for identity (2.38)

$$A_{jplt} = \left[ \overline{\overline{A}}_{jplt} + \frac{1}{7}(\delta_{jp}\overline{\overline{A}}_{vvlt} + \delta_{jl}\overline{\overline{A}}_{vvpt} + \delta_{pl}\overline{\overline{A}}_{vvjt} + \delta_{pt}\overline{\overline{A}}_{vvjl} + \delta_{lt}\overline{\overline{A}}_{vvjp} + \delta_{jt}\overline{\overline{A}}_{vvpl}) - \right.$$
$$\left. - \frac{1}{5 \cdot 7}(\delta_{jp}\delta_{lt}\overline{A}_{uuvv} + \delta_{jl}\delta_{pt}\overline{A}_{uuvv} + \delta_{jt}\delta_{pl}\overline{A}_{uuvv}) \right] +$$
$$+ \frac{1}{4}\varepsilon_{\gamma jp} \left[ \overline{\overline{C}}_{\gamma lt} + \frac{1}{5}\left(\delta_{\gamma l}\overline{C}_{vvt} + \delta_{\gamma t}\overline{C}_{vvl} + \delta_{lt}\overline{C}_{vv\gamma}\right) + \frac{1}{3}\varepsilon_{\alpha\gamma l}(\overline{F}_{\alpha t} + \frac{1}{2}\varepsilon_{\beta\alpha t}H_\beta) + \frac{1}{3}\varepsilon_{\alpha\gamma t}(\overline{F}_{\alpha l} + \frac{1}{2}\varepsilon_{\beta\alpha l}H_\beta) + \right.$$
$$\left. + \overline{\overline{D}}_{\gamma lt} + \frac{1}{5}(\delta_{\gamma l}\overline{D}_{vvt} + \delta_{tl}\overline{D}_{vv\gamma} + \delta_{\gamma t}\overline{D}_{vvl}) \right] +$$
$$+ \frac{1}{4}\varepsilon_{\gamma jl} \left[ \overline{\overline{C}}_{\gamma pt} + \frac{1}{5}\left(\delta_{\gamma p}\overline{C}_{vvt} + \delta_{\gamma t}\overline{C}_{vvp} + \delta_{pt}\overline{C}_{vv\gamma}\right) + \frac{1}{3}\varepsilon_{\alpha\gamma p}(\overline{F}_{\alpha t} + \frac{1}{2}\varepsilon_{\beta\alpha t}H_\beta) + \frac{1}{3}\varepsilon_{\alpha\gamma t}(\overline{F}_{\alpha p} + \frac{1}{2}\varepsilon_{\beta\alpha p}H_\beta) + \right.$$
$$\left. + \overline{\overline{D}}_{\gamma pt} + \frac{1}{5}(\delta_{\gamma p}\overline{D}_{vvt} + \delta_{tp}\overline{D}_{vv\gamma} + \delta_{\gamma t}\overline{D}_{vvp}) \right] +$$
$$+ \frac{1}{4}\varepsilon_{\gamma jt} \left[ \overline{\overline{C}}_{\gamma pl} + \frac{1}{5}\left(\delta_{\gamma p}\overline{C}_{vvl} + \delta_{\gamma l}\overline{C}_{vvp} + \delta_{pl}\overline{C}_{vv\gamma}\right) + \frac{1}{3}\varepsilon_{\alpha\gamma p}(\overline{F}_{\alpha l} + \frac{1}{2}\varepsilon_{\beta\alpha l}H_\beta) + \frac{1}{3}\varepsilon_{\alpha\gamma l}(\overline{F}_{\alpha p} + \frac{1}{2}\varepsilon_{\beta\alpha p}H_\beta) + \right.$$
$$\left. + \overline{\overline{D}}_{\gamma pl} + \frac{1}{5}(\delta_{\gamma l}\overline{D}_{vvp} + \delta_{pl}\overline{D}_{vv\gamma} + \delta_{\gamma p}\overline{D}_{vvl}) \right].$$

(2.53)

## 3 Definition of radiative and non-radiative terms in E-field

Substitution of equations for symmetric and traceless multipoles (2.8-2.13), (2.18), (2.37), and (2.53) into the **E** far-field equation (1.22) leads to extraction of lower rank tensors. Some of tensors provide a contribution to the electric field and some of them remain non-zero only for scalar and vector potentials. To avoid confusion coming from the number of indices and their permutations we simplify all tensors equation to the dyadic/vector form. That operation helps to simplify and gather tensors, which have same radiation pattern.

Before we proceed to consideration of radiative and non-radiative tensors, it should be noted that propagator of electrical multipoles could be written using Levi-Civita symbols and **n**-vectors as:

$$\frac{(R_i R_j - R^2 \delta_{ij})}{R^2} \frac{R_k}{R} = \varepsilon_{w\alpha j}\varepsilon_{wi\beta}n_\alpha n_\beta n_k. \tag{3.1}$$

Further it sufficiently reduces the equations.

## 3.1 Electric Quadrupole

Firstly we consider an extra term for the electric quadrupole (2.6)

$$\varepsilon_{waj}\varepsilon_{wi\beta}n_\alpha n_\beta n_k \delta_{jk} Q^{(e)}_{vv} = \varepsilon_{wak}\varepsilon_{wi\beta}n_\alpha n_\beta n_k Q^{(e)}_{vv}. \tag{3.2}$$

The Kronoker's delta symbol allows to change the indices in Levi-Chivita symbol. Thus, Levi-Chivita symbol contracts two times with **n** vectors that is equal in vector form to vector product $\varepsilon_{waj}n_i n_j = (\mathbf{n} \times \mathbf{n})_w = 0$. Therefore

$$\varepsilon_{wak}\varepsilon_{wi\beta}n_\alpha n_\beta n_k Q^{(e)}_{vv} = 0 \tag{3.3}$$

the scalar $Q^{(e)}_{vv}$ does not provide any contribution to the **E**-field in the far-field zone.

## 3.2 Electric Octupole

The electric octupole produces only one vector contributing to the far-field:

$$\begin{aligned}
\varepsilon_{waj}\varepsilon_{wi\beta}n_\alpha n_\beta n_k n_l \delta_{kl} O^{(e)}_{vvj} = \\
= \varepsilon_{waj}\varepsilon_{wi\beta}n_\alpha n_\beta (n_k n_l \delta_{kl}) O^{(e)}_{vvj} = \varepsilon_{waj}\varepsilon_{wi\beta}n_\alpha n_\beta O^{(e)}_{vvj} = \\
\varepsilon_{waj}\varepsilon_{wi\beta}n_\alpha n_\beta O^{(e)}_{vvj} = \mathbf{n} \times (\mathbf{n} \times \overline{\mathbf{O}}^{(e)}) = \mathbf{n} \times (-\overline{\mathbf{O}}^{(e)} \times \mathbf{n}).
\end{aligned} \tag{3.4}$$

It contribute to $1^{st}$ order electric dipole toroidal moment. Other two terms are non-radiative:

$$\begin{aligned}
\varepsilon_{waj}\varepsilon_{wi\beta}n_\alpha n_\beta n_k n_l \delta_{jk} O^{(e)}_{vvl} = \\
= (\varepsilon_{waj}n_\alpha n_k \delta_{jk})\varepsilon_{wi\beta}n_\beta n_l O^{(e)}_{vvl} = 0 \cdot \varepsilon_{wi\beta}n_\beta n_l O^{(e)}_{vvl} = 0,
\end{aligned} \tag{3.5}$$

$$\begin{aligned}
\varepsilon_{waj}\varepsilon_{wi\beta}n_\alpha n_\beta n_k n_l \delta_{jl} O^{(e)}_{vvk} = \\
= (\varepsilon_{waj}n_\alpha n_l \delta_{jl})\varepsilon_{wi\beta}n_\beta n_k O^{(e)}_{vvl} = 0 \cdot \varepsilon_{wi\beta}n_\beta n_k O^{(e)}_{vvk} = 0.
\end{aligned} \tag{3.6}$$

## 3.3 Electric 16-pole

Detracing the electric 16-pole we obtain three nonradiative terms for $\overline{\overline{\overline{S}}}_{vvlp}$:

$$\begin{aligned}
\varepsilon_{waj}\varepsilon_{wi\beta}n_\alpha n_\beta n_k n_l n_p \delta_{jk}\overline{\overline{\overline{S}}}_{vvlp} = 0, \\
\varepsilon_{waj}\varepsilon_{wi\beta}n_\alpha n_\beta n_k n_l n_p \delta_{jl}\overline{\overline{\overline{S}}}_{vvkp} = 0, \\
\varepsilon_{waj}\varepsilon_{wi\beta}n_\alpha n_\beta n_k n_l n_p \delta_{jp}\overline{\overline{\overline{S}}}_{vvlp} = 0,
\end{aligned} \tag{3.7}$$

and three radiative terms:

$$\varepsilon_{w\alpha j}\varepsilon_{wi\beta}n_\alpha n_\beta n_k n_l n_p \delta_{lp}\overline{\overline{S}}_{vvjk} = \varepsilon_{w\alpha j}\varepsilon_{wi\beta}n_\alpha n_\beta n_k(n_l n_p \delta_{lp})\overline{\overline{S}}_{vvjk} = \mathbf{n}\times(\mathbf{n}\times(\overline{\overline{\mathbf{S}}}\cdot\mathbf{n})),$$
$$\varepsilon_{w\alpha j}\varepsilon_{wi\beta}n_\alpha n_\beta n_k n_l \delta_{lk}\overline{\overline{S}}_{vvjp}n_p = \varepsilon_{w\alpha j}\varepsilon_{wi\beta}n_\alpha n_\beta(n_l n_k \delta_{lk})\overline{\overline{S}}_{vvjp}n_p = \mathbf{n}\times(\mathbf{n}\times(\overline{\overline{\mathbf{S}}}\cdot\mathbf{n})), \qquad (3.8)$$
$$\varepsilon_{w\alpha j}\varepsilon_{wi\beta}n_\alpha n_\beta n_k n_p \delta_{pk}\overline{\overline{S}}_{vvjl}n_l = \varepsilon_{w\alpha j}\varepsilon_{wi\beta}n_\alpha n_\beta(n_p n_k \delta_{pk})\overline{\overline{S}}_{vvjl}n_l = \mathbf{n}\times(\mathbf{n}\times(\overline{\overline{\mathbf{S}}}\cdot\mathbf{n})).$$

Other residual terms of the electric 16-pole are zero, as same as in the case of electric quadrupole. Scalars do not provide a contribution to the dynamical fields.

## 3.4 Electric 32-pole

Residual terms in electric 32 pole are separated on the contribution to the electric octupole and electric dipole. Terms related to the electric octupole have 4 nonradiative terms:

$$\begin{aligned}
\varepsilon_{w\alpha j}\varepsilon_{wi\beta}n_\alpha n_\beta n_k n_p n_l n_t \delta_{jk}\overline{\overline{X}}_{vvplt} &= 0, \\
\varepsilon_{w\alpha j}\varepsilon_{wi\beta}n_\alpha n_\beta n_k n_p n_l n_t \delta_{jp}\overline{\overline{X}}_{vvklt} &= 0, \\
\varepsilon_{w\alpha j}\varepsilon_{wi\beta}n_\alpha n_\beta n_k n_p n_l n_t \delta_{jl}\overline{\overline{X}}_{vvktk} &= 0, \\
\varepsilon_{w\alpha j}\varepsilon_{wi\beta}n_\alpha n_\beta n_k n_p n_l n_t \delta_{jt}\overline{\overline{X}}_{vvkpl} &= 0,
\end{aligned} \qquad (3.9)$$

and 6 radiative terms contributing to the electric octupole toroidal moment:

$$\begin{aligned}
\varepsilon_{w\alpha j}\varepsilon_{wi\beta}n_\alpha n_\beta n_k n_p n_l n_t \delta_{kp}\overline{\overline{X}}_{vvjtl} &= \mathbf{n}\times(\mathbf{n}\times(\underline{\overline{\overline{X}}}\cdot\mathbf{n}\cdot\mathbf{n})); \\
\varepsilon_{w\alpha j}\varepsilon_{wi\beta}n_\alpha n_\beta n_k n_p n_l n_t \delta_{kl}\overline{\overline{X}}_{vvjtp} &= \mathbf{n}\times(\mathbf{n}\times(\underline{\overline{\overline{X}}}\cdot\mathbf{n}\cdot\mathbf{n})); \\
\varepsilon_{w\alpha j}\varepsilon_{wi\beta}n_\alpha n_\beta n_k n_p n_l n_t \delta_{kt}\overline{\overline{X}}_{vvjpl} &= \mathbf{n}\times(\mathbf{n}\times(\underline{\overline{\overline{X}}}\cdot\mathbf{n}\cdot\mathbf{n})); \\
\varepsilon_{w\alpha j}\varepsilon_{wi\beta}n_\alpha n_\beta n_k n_p n_l n_t \delta_{pl}\overline{\overline{X}}_{vvjkt} &= \mathbf{n}\times(\mathbf{n}\times(\underline{\overline{\overline{X}}}\cdot\mathbf{n}\cdot\mathbf{n})); \\
\varepsilon_{w\alpha j}\varepsilon_{wi\beta}n_\alpha n_\beta n_k n_p n_l n_t \delta_{pt}\overline{\overline{X}}_{vvjkl} &= \mathbf{n}\times(\mathbf{n}\times(\underline{\overline{\overline{X}}}\cdot\mathbf{n}\cdot\mathbf{n})); \\
\varepsilon_{w\alpha j}\varepsilon_{wi\beta}n_\alpha n_\beta n_k n_p n_l n_t \delta_{lt}\overline{\overline{X}}_{vvjkp}) &= \mathbf{n}\times(\mathbf{n}\times(\underline{\overline{\overline{X}}}\cdot\mathbf{n}\cdot\mathbf{n})),
\end{aligned} \qquad (3.10)$$

where the underlined tensor $\underline{\overline{\overline{X}}}$ denotes contracted by two indices $\overline{\overline{X}}$. Terms related to the electric dipole are part of the $2^{nd}$ order electric dipole toroidal moment. It produces 12 nonradiative

terms:

$$\varepsilon_{w\alpha j}\varepsilon_{wi\beta}n_\alpha n_\beta n_k n_p n_l n_t \delta_{jk}\delta_{pl}\overline{\overline{X}}_{uuvvt} = 0;$$

$$\varepsilon_{w\alpha j}\varepsilon_{wi\beta}n_\alpha n_\beta n_k n_p n_l n_t \delta_{jp}\delta_{kl}\overline{\overline{X}}_{uuvvt} = 0;$$

$$\varepsilon_{w\alpha j}\varepsilon_{wi\beta}n_\alpha n_\beta n_k n_p n_l n_t \delta_{jl}\delta_{kp}\overline{\overline{X}}_{uuvvt} = 0;$$

$$\varepsilon_{w\alpha j}\varepsilon_{wi\beta}n_\alpha n_\beta n_k n_p n_l n_t \delta_{jt}\delta_{pl}\overline{\overline{X}}_{uuvvk} = 0;$$

$$\varepsilon_{w\alpha j}\varepsilon_{wi\beta}n_\alpha n_\beta n_k n_p n_l n_t \delta_{jp}\delta_{tl}\overline{\overline{X}}_{uuvvk} = 0;$$

$$\varepsilon_{w\alpha j}\varepsilon_{wi\beta}n_\alpha n_\beta n_k n_p n_l n_t \delta_{jl}\delta_{tp}\overline{\overline{X}}_{uuvvk} = 0;$$

$$\varepsilon_{w\alpha j}\varepsilon_{wi\beta}n_\alpha n_\beta n_k n_p n_l n_t \delta_{jk}\delta_{tl}\overline{\overline{X}}_{uuvvp} = 0; \quad (3.11)$$

$$\varepsilon_{w\alpha j}\varepsilon_{wi\beta}n_\alpha n_\beta n_k n_p n_l n_t \delta_{jt}\delta_{kl}\overline{\overline{X}}_{uuvvp} = 0;$$

$$\varepsilon_{w\alpha j}\varepsilon_{wi\beta}n_\alpha n_\beta n_k n_p n_l n_t \delta_{jl}\delta_{kt}\overline{\overline{X}}_{uuvvp} = 0;$$

$$\varepsilon_{w\alpha j}\varepsilon_{wi\beta}n_\alpha n_\beta n_k n_p n_l n_t \delta_{jk}\delta_{pt}\overline{\overline{X}}_{uuvvl} = 0;$$

$$\varepsilon_{w\alpha j}\varepsilon_{wi\beta}n_\alpha n_\beta n_k n_p n_l n_t \delta_{jp}\delta_{kt}\overline{\overline{X}}_{uuvvl} = 0;$$

$$\varepsilon_{w\alpha j}\varepsilon_{wi\beta}n_\alpha n_\beta n_k n_p n_l n_t \delta_{jt}\delta_{kp}\overline{\overline{X}}_{uuvvl} = 0,$$

and 3 radiative ones:

$$\varepsilon_{w\alpha j}\varepsilon_{wi\beta}n_\alpha n_\beta n_k n_p n_l n_t \delta_{tk}\delta_{pl}\overline{X}_{uuvvj} = \mathbf{n}\times(\mathbf{n}\times(\underline{\underline{\overline{X}}})) = \mathbf{n}\times(-\underline{\underline{\overline{X}}}\times\mathbf{n});$$

$$\varepsilon_{w\alpha j}\varepsilon_{wi\beta}n_\alpha n_\beta n_k n_p n_l n_t \delta_{tp}\delta_{kl}\overline{X}_{uuvvj} = \mathbf{n}\times(\mathbf{n}\times(\underline{\underline{\overline{X}}})) = \mathbf{n}\times(-\underline{\underline{\overline{X}}}\times\mathbf{n}); \quad (3.12)$$

$$\varepsilon_{w\alpha j}\varepsilon_{wi\beta}n_\alpha n_\beta n_k n_p n_l n_t \delta_{tl}\delta_{kp}\overline{X}_{uuvvj} = \mathbf{n}\times(\mathbf{n}\times(\underline{\underline{\overline{X}}})) = \mathbf{n}\times(-\underline{\underline{\overline{X}}}\times\mathbf{n}),$$

where the 2 times underlined tensor $\underline{\underline{\overline{X}}}$ denotes two times contracted by two indices $\overline{X}$ tensor.

## 3.5 Magnetic Quadrupole

After the magnetic quadrupole symmetrization, the residual terms produce a contribution to the electric dipole toroidal moment:

$$\frac{\varepsilon_{ikj}\varepsilon_{\beta jl}N_\beta}{R^2}R_l R_k = \varepsilon_{ikj}\varepsilon_{\beta jl}N_\beta n_l n_k = \quad (3.13)$$
$$= (\mathbf{N}\times\mathbf{n})\times\mathbf{n} = \mathbf{n}\times(-\mathbf{N}\times\mathbf{n})$$

## 3.6 Magnetic Octupole

We have the same situation for the residual terms obtained after symmetrization process of the magnetic octupole. Only one of three terms gives a contribution to the **E**-field

$$\frac{\varepsilon_{ikj}\delta_{lp}\overline{O}^{(m)}_{vvj}}{R^3}R_l R_k R_p = \varepsilon_{ikj}\delta_{lp}\overline{O}^{(m)}_{vvj}n_l n_k n_p = \\ = \varepsilon_{ikj}(\delta_{lp}n_l n_p)\overline{O}^{(m)}_{vvj}n_k = \varepsilon_{ikj}\overline{O}^{(m)}_{vvj}n_k = (\mathbf{n} \times \underline{\mathbf{\overline{O}}}^{(m)}) = (-\underline{\mathbf{\overline{O}}}^{(m)} \times \mathbf{n}), \tag{3.14}$$

other ones

$$\frac{\varepsilon_{ikj}\delta_{jl}\overline{O}_{vvp}}{R^3}R_l R_k R_p = (\varepsilon_{ikj}\delta_{jl}n_l n_k)\overline{O}_{vvp}n_p = 0 \tag{3.15}$$

$$\frac{\varepsilon_{ikj}\delta_{jp}\overline{O}_{vvl}}{R^3}R_l R_k R_p = (\varepsilon_{ikj}\delta_{jp}n_j n_p)\overline{O}_{vvk}n_k = 0 \tag{3.16}$$

represent nonradiative terms. Similarly, we obtain the one nonradiative term

$$\varepsilon_{ikj}\varepsilon_{\alpha jp}\overline{\overline{R}}_{\alpha l}n_l n_k n_p = \mathbf{n} \times (\mathbf{n} \times (\overline{\overline{\mathbf{R}}} \cdot \mathbf{n})) \tag{3.17}$$

and radiative term extracted from the magnetic octupole

$$\varepsilon_{ikj}\varepsilon_{\alpha jl}\overline{\overline{R}}_{\alpha p}n_l n_k n_p = \mathbf{n} \times (\mathbf{n} \times (\overline{\overline{\mathbf{R}}} \cdot \mathbf{n})) \tag{3.18}$$

which contributes to the electric quadrupole moment

$$\varepsilon_{ikj}\varepsilon_{\alpha jp}\varepsilon_{\beta\alpha l}G_\beta n_l n_k n_p = -(\delta_{i\alpha}\delta_{kp} - \delta_{ip}\delta_{k\alpha})\varepsilon_{\beta\alpha l}G_\beta n_l n_k n_p = \\ = -\delta_{i\alpha}(\delta_{kp}n_k n_p)\varepsilon_{\beta\alpha l}G_\beta n_l + \delta_{ip}\delta_{k\alpha}\varepsilon_{\beta\alpha l}G_\beta n_l n_k n_p = (\mathbf{G} \times \mathbf{n}) \tag{3.19}$$

and similarly

$$\varepsilon_{ikj}\varepsilon_{\alpha jl}\varepsilon_{\beta\alpha p}G_\beta n_l n_k n_p = -(\delta_{i\alpha}\delta_{kl} - \delta_{ip}\delta_{l\alpha})\varepsilon_{\beta\alpha l}G_\beta n_l n_k n_p = \\ = -\delta_{i\alpha}(\delta_{lp}n_l n_p)\varepsilon_{\beta\alpha k}G_\beta n_n + \delta_{ip}\delta_{l\alpha}\varepsilon_{\beta\alpha k}G_\beta n_l n_k n_p = (\mathbf{G} \times \mathbf{n}) \tag{3.20}$$

## 3.7 Magnetic 16-pole

The reduction of the magnetic 16-pole is the most difficult part of the mathematics presented here. The magnetic 16-pole produces the contribution to the magnetic quadrupole, the electric octupole, and the electric dipole. After detracing of the $4^{th}$ rank tensor $\overline{Y}$ we have the non-zero contribution to the magnetic quadrupole:

$$\varepsilon_{ikj}n_k n_p n_l n_t \delta_{pl}\overline{\overline{Y}}_{vvjt} = \mathbf{n} \times (\overline{\overline{Y}} \cdot \mathbf{n});$$
$$\varepsilon_{ikj}n_k n_p n_l n_t \delta_{pt}\overline{\overline{Y}}_{vvjl} = \mathbf{n} \times (\overline{\overline{Y}} \cdot \mathbf{n}); \quad (3.21)$$
$$\varepsilon_{ikj}n_k n_p n_l n_t \delta_{lt}\overline{\overline{Y}}_{vvjp} = \mathbf{n} \times (\overline{\overline{Y}} \cdot \mathbf{n}),$$

and as in the all previous cases the contribution of scalars to the far-field is zero:

$$\varepsilon_{ikj}n_k n_p n_l n_t \delta_{jt}\overline{\overline{Y}}_{vvpl} = 0;$$
$$\varepsilon_{ikj}n_k n_p n_l n_t \delta_{jp}\overline{\overline{Y}}_{vvlt} = 0; \quad (3.22)$$
$$\varepsilon_{ikj}n_k n_p n_l n_t \delta_{jl}\overline{\overline{Y}}_{vvpt} = 0.$$

Consideration of tensors $\overline{\overline{C}}$ and $\overline{\overline{D}}$ produces the same set of radiative and non-radiative terms. Therefore, further we derive the equations only for $C$. The tensor $C$ produces radiation terms only for electric octupole:

$$\varepsilon_{ikj}n_k n_p n_l n_t \varepsilon_{\gamma jp}\overline{\overline{C}}_{\gamma lt} = \mathbf{n} \times [\mathbf{n} \times (\overline{\overline{C}} \cdot \mathbf{n} \cdot \mathbf{n})];$$
$$\varepsilon_{ikj}n_k n_p n_l n_t \varepsilon_{\gamma jl}\overline{\overline{C}}_{\gamma pt} = \mathbf{n} \times [\mathbf{n} \times (\overline{\overline{C}} \cdot \mathbf{n} \cdot \mathbf{n})]; \quad (3.23)$$
$$\varepsilon_{ikj}n_k n_p n_l n_t \varepsilon_{\gamma jt}\overline{\overline{C}}_{\gamma pl} = \mathbf{n} \times [\mathbf{n} \times (\overline{\overline{C}} \cdot \mathbf{n} \cdot \mathbf{n})].$$

As it was shown for electric octupole, only one of three terms after detracing produces contribution to the electric field. Taking into account the number of terms in the equation (2.53) we have three radiative terms:

$$\varepsilon_{ikj}n_k n_p n_l n_t \varepsilon_{\gamma jp}\delta_{lt}\overline{C}_{vv\gamma} = \mathbf{n} \times [-\overline{\mathbf{C}} \times \mathbf{n}];$$
$$\varepsilon_{ikj}n_k n_l n_p n_t \varepsilon_{\gamma jl}\delta_{pt}\overline{C}_{vv\gamma} = \mathbf{n} \times [-\overline{\mathbf{C}} \times \mathbf{n}]; \quad (3.24)$$
$$\varepsilon_{ikj}n_k n_t n_p n_l \varepsilon_{\gamma jt}\delta_{pl}\overline{C}_{vv\gamma} = \mathbf{n} \times [-\overline{\mathbf{C}} \times \mathbf{n}]$$

and six non-radiative:

$$\varepsilon_{ikj}n_k n_p n_l n_t \varepsilon_{\gamma jp}\delta_{\gamma l}\overline{C}_{vvt} = 0;$$
$$\varepsilon_{ikj}n_k n_p n_l n_t \varepsilon_{\gamma jp}\delta_{\gamma t}\overline{C}_{vvl} = 0;$$
$$\varepsilon_{ikj}n_k n_l n_p n_t \varepsilon_{\gamma jl}\delta_{\gamma p}\overline{C}_{vvt} = 0;$$
$$\varepsilon_{ikj}n_k n_l n_p n_t \varepsilon_{\gamma jl}\delta_{\gamma t}\overline{C}_{vvp} = 0; \quad (3.25)$$
$$\varepsilon_{ikj}n_k n_t n_p n_l \varepsilon_{\gamma jt}\delta_{\gamma p}\overline{C}_{vvl} = 0;$$
$$\varepsilon_{ikj}n_k n_t n_p n_l \varepsilon_{\gamma jt}\delta_{\gamma l}\overline{C}_{vvp} = 0.$$

The $\overline{\overline{F}}$ tensor contributes radiative terms only to the magnetic quadrupole, same as $\overline{\overline{C}}$ and $\overline{\overline{D}}$ tensors:

$$\varepsilon_{ikj} n_k n_p n_l n_t \varepsilon_{\gamma jp} \varepsilon_{\alpha \gamma l} \overline{\overline{F}}_{\alpha t} = \mathbf{n} \times (-\overline{\overline{\mathbf{F}}} \cdot \mathbf{n});$$

$$\varepsilon_{ikj} n_k n_p n_l n_t \varepsilon_{\gamma jp} \varepsilon_{\alpha \gamma t} \overline{\overline{F}}_{\alpha l} = \mathbf{n} \times (-\overline{\overline{\mathbf{F}}} \cdot \mathbf{n});$$

$$\varepsilon_{ikj} n_k n_l n_p n_t \varepsilon_{\gamma jl} \varepsilon_{\alpha \gamma p} \overline{\overline{F}}_{\alpha t} = \mathbf{n} \times (-\overline{\overline{\mathbf{F}}} \cdot \mathbf{n}); \qquad (3.26)$$

$$\varepsilon_{ikj} n_k n_l n_p n_t \varepsilon_{\gamma jl} \varepsilon_{\alpha \gamma t} \overline{\overline{F}}_{\alpha p} = \mathbf{n} \times (-\overline{\overline{\mathbf{F}}} \cdot \mathbf{n});$$

$$\varepsilon_{ikj} n_k n_t n_p n_l \varepsilon_{\gamma jt} \varepsilon_{\alpha \gamma p} \overline{\overline{F}}_{\alpha l} = \mathbf{n} \times (-\overline{\overline{\mathbf{F}}} \cdot \mathbf{n});$$

$$\varepsilon_{ikj} n_k n_t n_p n_l \varepsilon_{\gamma jt} \varepsilon_{\alpha \gamma l} \overline{\overline{F}}_{\alpha p} = \mathbf{n} \times (-\overline{\overline{\mathbf{F}}} \cdot \mathbf{n}).$$

The **H**-vector obtained during symmetrization of magnetic 16 pole (see Equation (2.50)) provide only radiative contributions to the electric dipole :

$$\varepsilon_{ikj} n_k n_t n_p n_l \varepsilon_{\gamma jp} \varepsilon_{\alpha \gamma l} \varepsilon_{\beta \alpha t} H_\beta = \mathbf{n} \times [\mathbf{H} \times \mathbf{n}];$$

$$\varepsilon_{ikj} n_k n_t n_p n_l \varepsilon_{\gamma jp} \varepsilon_{\alpha \gamma t} \varepsilon_{\beta \alpha l} H_\beta = \mathbf{n} \times [\mathbf{H} \times \mathbf{n}];$$

$$\varepsilon_{ikj} n_k n_t n_l n_p \varepsilon_{\gamma jl} \varepsilon_{\alpha \gamma p} \varepsilon_{\beta \alpha t} H_\beta = \mathbf{n} \times [\mathbf{H} \times \mathbf{n}]; \qquad (3.27)$$

$$\varepsilon_{ikj} n_k n_t n_l n_p \varepsilon_{\gamma jl} \varepsilon_{\alpha \gamma t} \varepsilon_{\beta \alpha p} H_\beta = \mathbf{n} \times [\mathbf{H} \times \mathbf{n}];$$

$$\varepsilon_{ikj} n_k n_l n_t n_p \varepsilon_{\gamma jt} \varepsilon_{\alpha \gamma p} \varepsilon_{\beta \alpha l} H_\beta = \mathbf{n} \times [\mathbf{H} \times \mathbf{n}];$$

$$\varepsilon_{ikj} n_k n_l n_t n_p \varepsilon_{\gamma jt} \varepsilon_{\alpha \gamma l} \varepsilon_{\beta \alpha p} H_\beta = \mathbf{n} \times [\mathbf{H} \times \mathbf{n}].$$

## 3.8 Equations of the E-field with radiative multipoles and nonradiative scalar and vector potentials

Substituting final expressions obtained in the previous section into equation (1.22), reducing number of terms using calculations (3.3 -3.27), and gathering terms having a same propagation pattern, we obtained explicit, irreducible multipole decomposition of the **E**-field in the far-zone:

$$E_i = \frac{k^2}{4\pi\varepsilon_0^2 R}\left[\left(n_i n_j - n^2 \delta_{ij}\right)\left(-p_j + \frac{ik}{2}(\overline{\overline{Q}}^{(e)}_{jk} n_k) + \frac{k^2}{6}([\overline{\overline{O}}^{(e)}_{jkp} + \frac{1}{5}\overline{\overline{O}}^{(e)}_{vvj}\delta_{kp}]n_k n_p) + \right.\right.$$
$$+\frac{-ik^3}{24}([\overline{\overline{S}}_{jkpl} + \frac{3}{7}\delta_{kl}\overline{\overline{S}}_{vvpj}]n_k n_p n_l) + \frac{-k^4}{120}([\overline{\overline{X}}_{jkplt} + \frac{6}{9}\delta_{pt}\overline{\overline{X}}_{vvjkl} + \frac{3}{35}\delta_{kp}\delta_{lt}\overline{X}_{jvvuu}]n_k n_p n_l n_t)\Big) +$$
$$+\left(\frac{-1}{c}\varepsilon_{ikj}n_k m_j + \frac{ik}{2c}\varepsilon_{ikj}n_k([\overline{\overline{Q}}^{(m)}_{jp} + \frac{1}{2}\varepsilon_{\beta jp}N_\beta]n_p) + \right.$$
$$+\frac{k^2}{6c}\varepsilon_{ikj}n_k\Big([\overline{\overline{O}}^{(m)}_{jpl} + \frac{1}{5}\delta_{pl}\overline{O}^{(m)}_{vvj} + \frac{2}{3}\varepsilon_{\alpha jp}(\overline{\overline{R}}_{\alpha l} + \frac{1}{2}\varepsilon_{\beta\alpha l}G_\beta)\Big]n_p n_l\Big) +$$
$$+\frac{-ik^3}{24c}\varepsilon_{ikj}n_k\Big(\Big(\overline{\overline{Y}}_{jplt} + \frac{3}{7}(\delta_{lt}\overline{Y}_{vvjp}) +$$
$$+\frac{3}{4}\varepsilon_{\gamma jp}\Big[\overline{\overline{C}}_{\gamma lt} + \overline{\overline{D}}_{\gamma lt} + \frac{1}{5}\delta_{lt}\Big(\overline{C}_{vv\gamma} + \overline{D}_{vv\gamma}\Big) + \frac{2}{3}\varepsilon_{\alpha\gamma l}(\overline{\overline{F}}_{\alpha t} + \frac{1}{2}\varepsilon_{\beta\alpha t}H_\beta)\Big]\Big)n_p n_l n_t)\Big)\right]$$
(3.28)

Nonradiative part of the scalar potential:

$$\Phi^{(nonrad)} = \frac{1}{4\pi\varepsilon_0}\left[\frac{(-ik)^2}{6}\frac{\delta_{jk}Q^{(e)}_{vv}R_j R_k}{R^3} + \frac{(-ik)^3}{30}\frac{(\delta_{jl}O^{(e)}_{vvk} + \delta_{lk}O^{(e)}_{vvj})R_j R_k R_l}{R^4} + \right.$$
$$+\frac{(-ik)^4}{24}\Big[\frac{1}{7}(\delta_{pj}\overline{\overline{S}}_{vvkl} + \delta_{jk}\overline{\overline{S}}_{vvpl} + \delta_{jl}\overline{\overline{S}}_{vvpk}) +$$
$$+\frac{1}{15}(\delta_{pj}\delta_{kl}\overline{S}_{uuvv} + \delta_{pk}\delta_{jl}\overline{S}_{uuvv} + \delta_{pl}\delta_{jk}\overline{S}_{uuvv})\Big]\frac{R_j R_k R_l R_p}{R^5} +$$
$$+\frac{(-ik)^5}{120}\Big[\frac{1}{9}(\delta_{jk}\overline{\overline{X}}_{vvplt} + \delta_{jp}\overline{\overline{X}}_{vvklt}\delta_{jl}\overline{\overline{X}}_{vvktp} + \delta_{jt}\overline{\overline{X}}_{vvkpl}) +$$
$$\frac{1}{35}(\delta_{jk}\delta_{pl}\overline{X}_{uuvvt} + \delta_{jp}\delta_{kl}\overline{X}_{uuvvt} + \delta_{jl}\delta_{kp}\overline{X}_{uuvvt} +$$
$$+\delta_{jt}\delta_{pl}\overline{X}_{uuvvk} + \delta_{jp}\delta_{tl}\overline{X}_{uuvvk} + \delta_{jl}\delta_{tp}\overline{X}_{uuvvk} +$$
$$+\delta_{jk}\delta_{tl}\overline{X}_{uuvvp} + \delta_{jt}\delta_{kl}\overline{X}_{uuvvp} + \delta_{jl}\delta_{kt}\overline{X}_{uuvvp} +$$
$$+\delta_{jk}\delta_{pt}\overline{X}_{uuvvl} + \delta_{jp}\delta_{kt}\overline{X}_{uuvvl} + \delta_{jt}\delta_{kp}\overline{X}_{uuvvl})\Big)\frac{R_j R_k R_l R_p R_t}{R^6}\right].$$
(3.29)

and nonradiative vector potential

$$A_j^{(nonrad)} = \frac{1}{4\pi\varepsilon_0}\left(\frac{(-ik)^2}{6c}\frac{\delta_{jk}\overline{Q}_{vv}^{(e)}R_k}{R^3} + \frac{(-ik)^3}{30c}\frac{(\delta_{jl}\overline{O}_{vvk}^{(e)} + \delta_{lk}\overline{O}_{vvj}^{(e)})R_kR_l}{R^4}+\right.$$

$$\left[-\frac{1}{30}\frac{(-ik)^3}{c^2}\varepsilon_{jk\alpha}\left(\delta_{\alpha l}\overline{O}_{vvp}^{(m)} + \delta_{\alpha p}\overline{O}_{vvl}^{(m)}\right) + \frac{(-ik)^4}{24c}\left(\frac{1}{7}(\delta_{pj}\overline{\overline{S}}_{vvkl} + \delta_{jk}\overline{\overline{S}}_{vvpl} + \delta_{jl}\overline{\overline{S}}_{vvpk})+\right.\right.$$

$$\left.\left.+\frac{1}{15}(\delta_{pj}\delta_{kl}\overline{S}_{uuvv} + \delta_{pk}\delta_{jl}\overline{S}_{uuvv} + \delta_{pl}\delta_{jk}\overline{S}_{uuvv})\right)\right]\frac{R_kR_lR_p}{R^5}+$$

$$\left[\frac{(-ik)^4}{24c^2}\varepsilon_{jk\alpha}\left(\frac{1}{7}(\delta_{\alpha p}\overline{\overline{Y}}_{vvlt} + \delta_{\alpha l}\overline{\overline{Y}}_{vvpt} + \delta_{\alpha t}\overline{\overline{Y}}_{vvpl}) - \frac{1}{5\cdot 7}(\delta_{\alpha p}\delta_{lt}\overline{Y}_{uuvv} + \delta_{\alpha l}\delta_{pt}\overline{Y}_{uuvv} + \delta_{\alpha t}\delta_{pl}\overline{Y}_{uuvv})+\right.\right.$$

$$+\frac{1}{4}\varepsilon_{\gamma\alpha p}\left[\frac{1}{5}\left(\delta_{\gamma l}\overline{C}_{vvt} + \delta_{\gamma t}\overline{C}_{vvl} + \delta_{\gamma l}\overline{D}_{vvt} + \delta_{\gamma t}\overline{D}_{vvl}\right)\right]+$$

$$+\frac{1}{4}\varepsilon_{\gamma\alpha l}\left[\frac{1}{5}\left(\delta_{\gamma p}\overline{C}_{vvt} + \delta_{\gamma t}\overline{C}_{vvp} + \delta_{\gamma p}\overline{D}_{vvt} + \delta_{\gamma t}\overline{D}_{vvp}\right)\right]+$$

$$\left.+\frac{1}{4}\varepsilon_{\gamma\alpha t}\left[\frac{1}{5}\left(\delta_{\gamma p}\overline{C}_{vvl} + \delta_{\gamma l}\overline{C}_{vvp} + \delta_{\gamma l}\overline{D}_{vvp} + \delta_{\gamma p}\overline{D}_{vvl}\right)\right]\right)+$$

$$+\frac{(-ik)^5}{120c}\left((\frac{1}{9}(\delta_{jk}\overline{\overline{X}}_{vvplt} + \delta_{jp}\overline{\overline{X}}_{vvklt}\delta_{jl}\overline{\overline{X}}_{vvktp} + \delta_{jt}\overline{\overline{X}}_{vvkpl})+\right.$$

$$\frac{1}{35}(\delta_{jk}\delta_{pl}\overline{X}_{uuvvt} + \delta_{jp}\delta_{kl}\overline{X}_{uuvvt} + \delta_{jl}\delta_{kp}\overline{X}_{uuvvt}+$$

$$+\delta_{jt}\delta_{pl}\overline{X}_{uuvvk} + \delta_{jp}\delta_{tl}\overline{X}_{uuvvk} + \delta_{jl}\delta_{tp}\overline{X}_{uuvvk}+$$

$$+\delta_{jk}\delta_{tl}\overline{X}_{uuvvp} + \delta_{jt}\delta_{kl}\overline{X}_{uuvvp} + \delta_{jl}\delta_{kt}\overline{X}_{uuvvp}+$$

$$\left.\left.+\delta_{jk}\delta_{pt}\overline{X}_{uuvvl} + \delta_{jp}\delta_{kt}\overline{X}_{uuvvl} + \delta_{jt}\delta_{kp}\overline{X}_{uuvvl})\right)\right]\frac{R_kR_lR_pR_t}{R^6}\right)$$

(3.30)

## 4 Toroidal moments evaluation

In this section we derive toroidal moments. The main idea is bring into accordance parts of the extracted tensors in equation (3.28), combine them, and gather in the terms having similar symmetry properties.

### 4.1 Toroidal electric dipole

We combine vectors $\overline{\mathbf{O}}$ and $\mathbf{N}$ extracted from the electric octupole and the magnetic quadrupole and have

$$\frac{ik}{2c}\frac{1}{2}\varepsilon_{ikj}\varepsilon_{\beta jl}N_\beta n_l n_k + \frac{k^2}{6}\frac{1}{5}\varepsilon_{w\alpha j}\varepsilon_{wi\beta}n_\alpha n_\beta \overline{O}_{vvj}^{(e)} =$$
$$= -\frac{1}{2}\frac{ik}{2c}\left[\mathbf{n}\times(\mathbf{N}\times\mathbf{n})\right] - \frac{k^2}{6}\frac{1}{5}\left[\mathbf{n}\times(\overline{\mathbf{O}}^{(e)}\times\mathbf{n})\right] = \mathbf{n}\times(\frac{ik}{c}\mathbf{T}^{(e)}\times\mathbf{n}).$$

(4.1)

With substitution of equations (2.9) and (2.19) in the previous equation

$$\frac{ik}{c}\mathbf{T}^{(e)} = -\frac{1}{2}\frac{ik}{2c}\frac{2}{3}\int r^2\mathbf{J} - (\mathbf{r}\cdot\mathbf{J})\mathbf{r}\,dv - \frac{k^2}{6}\frac{1}{5}\frac{i}{\omega}\int 2(\mathbf{r}\cdot\mathbf{J})r_j + r^2 J_j\,dv =$$
$$= \frac{ik}{c}\frac{1}{10}\int (\mathbf{J}\cdot\mathbf{r})\mathbf{r} - 2r^2\mathbf{J}\,dv, \tag{4.2}$$

where electric toroidal moment is

$$\boxed{T_j^{(e)} = \frac{1}{10}\int (\mathbf{J}\cdot\mathbf{r})r_j - 2r^2 J_j\,dv.} \tag{4.3}$$

## 4.2 Toroidal magnetic dipole

The magnetic toroidal moment derived from the parts of the magnetic octupole obtained from detracing of the octupole (2.30) and symmetrization of $2^{nd}$ order $\overline{R}$-tensor (2.35):

$$\frac{k^2}{6c}\varepsilon_{ikj}(n_l n_k n_p)\left[\frac{1}{5}\delta_{pl}\overline{O}^{(m)}_{vvj} + \frac{2}{3}\varepsilon_{\alpha jp}\frac{1}{2}\varepsilon_{\beta\alpha l}G_\beta\right] =$$
$$= -\frac{1}{c}\frac{i}{\omega}\frac{ik}{6c}\left(\mathbf{n}\times(-\frac{1}{5}\overline{\overline{\mathbf{O}}}^{(m)} + \frac{2}{6}\mathbf{G})\right) = -\frac{1}{c}\left(\mathbf{n}\times\frac{ik}{c}\mathbf{T}^{(m)}\right), \tag{4.4}$$

with definition of magnetic toroidal moment:

$$\boxed{T_j^{(m)} = -\frac{1}{20}\frac{i}{\omega}\int (\mathbf{r}\times\mathbf{J})_j r^2\,dv.} \tag{4.5}$$

## 4.3 Toroidal electric quadrupole

Similarly to the calculation of the toroidal electric dipole we found:

$$\frac{-ik^3}{24}\frac{3}{7}\varepsilon_{w\alpha j}\varepsilon_{wi\beta}n_\alpha n_\beta n_p[n_k n_l\delta_{lk}]S_{vvjp} + \frac{k^2}{6c}\varepsilon_{ikj}n_l n_k n_p\frac{2}{3}\varepsilon_{\alpha jl}\overline{\overline{R}}_{\alpha p} =$$
$$= \frac{-ik^3}{24}\frac{3}{7}\left(\mathbf{n}\times\left(\mathbf{n}\times(\overline{\overline{\mathbf{S}}}\cdot\mathbf{n})\right)\right) + \frac{k^2}{6c}\frac{2}{3}\left(\mathbf{n}\times\left(\mathbf{n}\times(\overline{\overline{\mathbf{R}}}\cdot\mathbf{n})\right)\right) = \tag{4.6}$$
$$= \frac{ik}{2}\left[\mathbf{n}\times\left[\mathbf{n}\times(\frac{ik}{c}\overline{\overline{\mathbf{T}}}^{(Qe)}\cdot\mathbf{n})\right]\right].$$

Substituting integral form (2.34) and (2.11) for $\overline{\overline{R}}$ and $\overline{\overline{S}}$ we derive toroidal electric quadrupole moment as:

$$\boxed{\overline{\overline{T}}^{(Qe)}_{jk} = \frac{1}{42}\int 4(\mathbf{r}\cdot\mathbf{J})r_j r_k + 2(\mathbf{J}\cdot\mathbf{r})r^2\delta_{jk} - 5r^2\left(r_j J_k + J_j r_k\right)dv.} \tag{4.7}$$

## 4.4 Toroidal magnetic quadrupole

Using the same steps we obtain the toroidal magnetic quadrupole moment taking into account integral forms (2.52) and (2.49) of tensors $\overline{\overline{Y}}$ and $\overline{\overline{F}}$ :

$$\frac{-ik^3}{24c}\varepsilon_{ikj}(n_l n_k n_p n_t)\left[\frac{3}{7}\delta_{lt}\overline{Y}_{vvjp} + \frac{3}{4}\varepsilon_{\gamma jp}\frac{2}{3}\varepsilon_{\alpha\gamma l}\overline{F}_{\alpha t}\right] = \\ = \frac{ik}{2c}\left(\mathbf{n} \times \left[-\frac{1}{12}\cdot\left(\frac{3}{7}\overline{\overline{Y}} - \frac{1}{2}\overline{\overline{F}}\right)\right]\right) = \frac{ik}{2c}\left(\mathbf{n} \times \frac{ik}{c}\mathbf{T}^{(Qm)}\right), \quad (4.8)$$

with the following definition of the toroidal magnetic quadrupole:

$$\boxed{\overline{\overline{T}}^{(Qm)}_{jp} = \frac{i\omega}{42}\int r^2\Big(r_j(\mathbf{r}\times\mathbf{J})_p + (r\times\mathbf{J})_j r_p\Big)dv}. \quad (4.9)$$

## 4.5 Toroidal electric octupole

Toroidal electric octupole moment obtained during detracing of electric 32-pole (2.14) and symmetrization and detracing of magnetic 16-pole (2.44) and (2.46):

$$\frac{k^2}{6}\left[\frac{-k^2}{20}\frac{6}{9}\varepsilon_{w\alpha j}\varepsilon_{wi\beta}(n_\alpha n_\beta n_k n_p n_l n_t)\delta_{pt}\overline{X}_{vvjkl} - \frac{ik}{4c}\varepsilon_{ikj}n_k\frac{3}{4}\varepsilon_{\gamma jp}(\overline{\overline{C}}_{\gamma lt} + \overline{\overline{D}}_{\gamma lt})n_p n_l n_t\right] = \\ = \frac{-ik}{c}\frac{k^2}{6}\frac{1}{240}\Big[\mathbf{n}\times[\mathbf{n}\times\big((60\overline{\overline{C}} + 35\overline{\overline{D}})\cdot\mathbf{n}\cdot\mathbf{n}\big)]\Big] = \frac{k^2}{6}\Big[\mathbf{n}\times[\mathbf{n}\times\big((\frac{ik}{c}\overline{\overline{T}}^{(eO)}\cdot\mathbf{n}\cdot\mathbf{n}\big)]\Big], \quad (4.10)$$

where toroidal electric octupole is defined as:

$$\boxed{\overline{\overline{T}}^{(eO)}_{jkp} = \frac{1}{300}\int\Big[35(\mathbf{r}\cdot\mathbf{J})r_j r_k r_p - 20r^2(J_j r_k r_p + J_k r_j r_p + J_p r_j r_k) + \\ + (\delta_{\gamma j}\delta_{kp} + \delta_{\gamma k}\delta_{jp} + \delta_{\gamma p}\delta_{jk})\big[(\mathbf{r}\cdot\mathbf{J})r^2 r_\gamma + 4r^4 J_\gamma\big]\Big]dv}. \quad (4.11)$$

## 4.6 The 2nd order toroidal electric dipole

The $2^{nd}$ order toroidal electric dipole is obtained from detracing of electric 32-pole equation (2.15) and symmetrization of the magnetic 16-pole and detracing of resulted $\overline{C}$ and $\overline{D}$ tensors (equations (2.47) and (2.44)):

$$\frac{k^4}{120}\frac{3}{35}\overline{\overline{X}} + \frac{ik^3}{24c}\frac{3}{4}\left(\frac{1}{5}(\overline{\overline{C}} + \overline{\overline{D}}) - \frac{2}{3}\frac{1}{2}H\right) = \\ = \frac{ik^3}{c}\frac{1}{280}\int 3r^4 \mathbf{J} - 2r^2(\mathbf{r}\cdot\mathbf{J})\mathbf{r}\, dv \quad (4.12)$$

and finally we derive

$$T_j^{(2e)} = \frac{1}{280} \int 3r^4 J_j - 2r^2(\mathbf{r} \cdot \mathbf{J}) r_j dv. \tag{4.13}$$

## 4.7 Equation of the E-field in the vector and dyadic forms

Applying the definitions of the toroidal moments (4.3, 4.5, 4.7, 4.9, 4.11, 4.13) and gathering tensors in the **E**-field (3.28) using previously obtained expressions we rewrite the **E**-field in a vector form:

$$\mathbf{E} = \frac{k^2}{4\pi\varepsilon_0} \frac{1}{R} \Bigg( [\mathbf{n} \times [(\mathbf{p} + \frac{ik}{c}\mathbf{T}^{(e)} + \frac{ik^3}{c}\mathbf{T}^{(2e)}) \times \mathbf{n}]] + \frac{ik}{2}[\mathbf{n} \times [\mathbf{n} \times ((\overline{\overline{\mathbf{Q}}}^{(e)} + \frac{ik}{c}\overline{\overline{\mathbf{T}}}^{(Qe)}) \cdot \mathbf{n})]] +$$

$$+ \frac{k^2}{6}[\mathbf{n} \times [\mathbf{n} \times ((\overline{\overline{\mathbf{O}}}^{(e)} + \frac{ik}{c}\overline{\overline{\mathbf{T}}}^{(Oe)}) \cdot \mathbf{n} \cdot \mathbf{n})]] + \frac{ik^3}{24}[\mathbf{n} \times [(\overline{\overline{\mathbf{S}}} \cdot \mathbf{n} \cdot \mathbf{n} \cdot \mathbf{n}) \times \mathbf{n}]] + \frac{k^4}{120}[\mathbf{n} \times [(\overline{\overline{\mathbf{X}}} \cdot \mathbf{n} \cdot \mathbf{n} \cdot \mathbf{n}) \times \mathbf{n}]]$$

$$+ \frac{1}{c}[(\mathbf{m} + \frac{ik}{c}\mathbf{T}^{(m)}) \times \mathbf{n}] + \frac{ik}{2c}[\mathbf{n} \times ((\overline{\overline{\mathbf{Q}}}^{(m)} + \frac{ik}{c}\overline{\overline{\mathbf{T}}}^{(Qm)}) \cdot \mathbf{n})] + \frac{k^2}{6c}[\mathbf{n} \times (\overline{\overline{\mathbf{O}}}^{(m)} \cdot \mathbf{n} \cdot \mathbf{n})]$$

$$+ \frac{k^3}{24c}[(\overline{\overline{\mathbf{Y}}} \cdot \mathbf{n} \cdot \mathbf{n} \cdot \mathbf{n}) \times \mathbf{n}] \Bigg). \tag{4.14}$$

## 5 Multipole formulation

### 5.1 Expansion of a continuity equation

For simplification of expansion terms in the vector potential **A**, we considered the law of continuity:

$$\nabla \cdot \mathbf{J}(\mathbf{r}, t) = -\dot{\rho}(t), \tag{5.1}$$

and the tensor identity found from the property of Levi-Civita antisymmetric tensor is:

$$J_j r_i - J_i r_j = \varepsilon_{ijk} (\mathbf{r} \times \mathbf{J})_k. \tag{5.2}$$

Using definition of Libniz's product rule $d(u \cdot v)/dx = v \cdot du/dx + u \cdot dv/dx$:

$$\nabla_j(r_i J_j) = J_i - r_i \dot{\rho}, \tag{5.3}$$

we multiply (5.3) by **r** - vector. Performing index permutations we have:

$$r_j(\nabla_k(r_i J_k)) = r_j J_i - r_j r_i \dot{\rho}, \tag{5.4}$$

or using Libniz's rule again we obtained:

$$-r_j J_k(\nabla_k r_i) + \nabla_k(r_i r_j J_k) = r_j J_i - r_j r_i \dot{\rho}. \tag{5.5}$$

Taking into account $r_j J_k(\nabla_k r_i) = r_j J_k \delta_{ki} = r_j J_i$ and move first term to the left side of the equation (5.5) we obtain:

$$\nabla_k(r_i r_j J_k) = r_i J_j + r_j J_i - r_j r_i \dot{\rho}. \tag{5.6}$$

Repeating the same steps we found the same equations for $3^{rd}$ order:

$$\nabla_p(r_i r_j r_k J_p) = r_i r_j J_k + r_i r_k J_j + r_j r_k J_i - r_i r_j r_k \dot{\rho}, \tag{5.7}$$

for $4^{th}$ order:

$$\nabla_a(r_i r_j r_k r_p J_a) = r_i r_j r_k J_p + r_j r_i r_p J_k + r_i r_k r_p J_j + r_j r_k r_p J_i - r_i r_j r_k r_p \dot{\rho}, \tag{5.8}$$

and $5^{th}$ order:

$$\nabla_b(r_i r_j r_k r_p r_a J_b) = r_i r_j r_k r_p J_a + r_a r_j r_k r_p J_i + r_i r_a r_k r_p J_j + \\ + r_i r_j r_a r_p J_k + r_i r_j r_k r_a J_p - r_i r_j r_k r_p r_a \dot{\rho}. \tag{5.9}$$

Left sides of the underlined equations (5.6-5.8) have terms with symmetrical to permutations indices such as $r_i J_j$ and $r_j J_i$ in (5.6). Preserving identity we add and substitute to the right side of the equation $r_i J_j$:

$$\nabla_k(r_i r_j J_k) = r_i J_j + (r_j J_i - r_i J_j) + r_i J_j - r_j r_i \dot{\rho}. \tag{5.10}$$

Combining identical terms and applying identity (5.2) for the $2^{nd}$ rank tensor we got:

$$\nabla_k(r_i r_j J_k) = 2r_i J_j + \varepsilon_{ijk}(\mathbf{r} \times \mathbf{J})_k - r_j r_i \dot{\rho}, \tag{5.11}$$

similarly for the $3^{rd}$ rank:

$$\nabla_p(r_i r_j r_k J_p) = \varepsilon_{ikl} r_j (\mathbf{r} \times \mathbf{J})_l + \varepsilon_{ijl} r_k (\mathbf{r} \times \mathbf{J})_l + \\ 3 J_i r_j r_k - r_i r_j r_k \dot{\rho}, \tag{5.12}$$

for the $4^{th}$ rank:
$$\nabla_a(r_i r_j r_k r_p J_a) = \varepsilon_{ipl} r_j r_k (\mathbf{r} \times \mathbf{J})_l + \varepsilon_{ikl} r_j r_p (\mathbf{r} \times \mathbf{J})_l + \\ + \varepsilon_{ijl} r_p r_k (\mathbf{r} \times \mathbf{J})_l + 4 J_i r_j r_k r_p - r_i r_j r_k r_p \dot{\rho}, \quad (5.13)$$

and for the $5^{th}$ rank:
$$\nabla_a(r_i r_j r_k r_p J_a) = \varepsilon_{ipl} r_j r_k r_a (\mathbf{r} \times \mathbf{J})_l + \varepsilon_{ikl} r_j r_p r_a (\mathbf{r} \times \mathbf{J})_l + \\ + \varepsilon_{ijl} r_p r_k r_a (\mathbf{r} \times \mathbf{J})_l + \varepsilon_{ial} r_i r_j r_k (\mathbf{r} \times \mathbf{J})_l + 5 J_i r_j r_k r_p r_a - r_i r_j r_k r_p r_a \dot{\rho}. \quad (5.14)$$

## 5.2 Multipole formulation:

Producing volume integration over the (5.3 and 5.11-5.13) it's clear to see that the integral in left side is zero. Therefore, we can move charge densities to the left side of equations. For a $1^{st}$ rank tensor we got:
$$\int r_i \dot{\rho} dv = \int J_i dv, \quad (5.15)$$

for a $2^{nd}$ rank:
$$\int r_j r_i \dot{\rho} dv = 2 \int r_i J_j dv + \varepsilon_{ijk} \int (\mathbf{r} \times \mathbf{J})_k dv, \quad (5.16)$$

for $3^{rd}$ rank:
$$\int r_i r_j r_k \dot{\rho} dv = \varepsilon_{ikl} \int r_j (\mathbf{r} \times \mathbf{J})_l dv + \varepsilon_{ijl} \int r_k (\mathbf{r} \times \mathbf{J})_l dv + 3 \int J_i r_j r_k dv, \quad (5.17)$$

for $4^{th}$ rank:
$$\int r_i r_j r_k r_p \dot{\rho} dv = \varepsilon_{ipl} \int r_j r_k (\mathbf{r} \times \mathbf{J})_l dv + \varepsilon_{ikl} \int r_j r_p (\mathbf{r} \times \mathbf{J})_l dv + \\ + \varepsilon_{ijl} \int r_p r_k (\mathbf{r} \times \mathbf{J})_l dv + 4 \int J_i r_j r_k r_p dv, \quad (5.18)$$

and for $5^{th}$ rank:
$$r_i r_j r_k r_p r_a \dot{\rho} = \varepsilon_{ipl} \int r_j r_k r_a (\mathbf{r} \times \mathbf{J})_l dv + \varepsilon_{ikl} \int r_j r_p r_a (\mathbf{r} \times \mathbf{J})_l dv + \\ + \varepsilon_{ijl} \int r_p r_k r_a (\mathbf{r} \times \mathbf{J})_l dv + \varepsilon_{ial} \int r_i r_j r_k (\mathbf{r} \times \mathbf{J})_l dv + 5 \int J_i r_j r_k r_p r_a dv. \quad (5.19)$$

And expressing from (5.15 - 5.18) integrals with current density we found for for $1^{st}$ rank:
$$\int J_i dv = \int r_i \dot{\rho} dv, \quad (5.20)$$

for $2^{nd}$ rank:
$$\int J_i r_j dv = \frac{1}{2}\Big[\int r_j r_i \dot{\rho} dv - \varepsilon_{ijk}\int (\mathbf{r}\times\mathbf{J})_k dv\Big], \tag{5.21}$$

for $3^{rd}$ rank:
$$\int J_i r_j r_k dv = \frac{1}{3}\Big[\int r_i r_j r_k \dot{\rho} dv - \varepsilon_{ikl}\int r_j(\mathbf{r}\times\mathbf{J})_l dv - \varepsilon_{ijl}\int r_k(\mathbf{r}\times\mathbf{J})_l dv\Big], \tag{5.22}$$

for $4^{th}$ rank:
$$\int J_i r_j r_k r_p dv = \frac{1}{4}\Big[\int r_i r_j r_k r_p \dot{\rho} dv - \varepsilon_{ipl}\int r_j r_k(\mathbf{r}\times\mathbf{J})_l dv - \\ -\varepsilon_{ikl}\int r_j r_p(\mathbf{r}\times\mathbf{J})_l dv - \varepsilon_{ijl}\int r_p r_k(\mathbf{r}\times\mathbf{J})_l dv\Big], \tag{5.23}$$

and for $5^{th}$ rank:

$$\int J_i r_j r_k r_p dv = \frac{1}{4}\Big[\int r_i r_j r_k r_p r_a \dot{\rho} dv - \varepsilon_{ipl}\int r_j r_k r_a(\mathbf{r}\times\mathbf{J})_l dv - \\ -\varepsilon_{ikl}\int r_j r_p r_a(\mathbf{r}\times\mathbf{J})_l dv - \varepsilon_{ijl}\int r_p r_k r_a(\mathbf{r}\times\mathbf{J})_l dv - \varepsilon_{ial}\int r_i r_j r_k(\mathbf{r}\times\mathbf{J})_l dv\Big]. \tag{5.24}$$

Denoting the integral in left sides of (5.20-5.24) and using definition of magnetic multipole moments we derived following equations:

$$\int J_i dv = \dot{p}_i \tag{5.25}$$

$$\int J_i r_j dv = \frac{1}{2}\Big[\dot{Q}^{(e)}_{ij} - \varepsilon_{ijk} m_k\Big], \tag{5.26}$$

$$\int J_i r_j r_k dv = \Big[\frac{1}{3}\dot{O}^{(e)}_{ijk} - \frac{1}{2}\varepsilon_{ikl} Q^{(m)}_{jl} - \frac{1}{2}\varepsilon_{ijl} Q^{(m)}_{kl}\Big], \tag{5.27}$$

$$\int J_i r_j r_k r_p dv = \Big[\frac{1}{4}\dot{S}_{ijkp} - \frac{1}{3}\varepsilon_{ipl} O^{(m)}_{jkl} - \frac{1}{3}\varepsilon_{ikl} O^{(m)}_{jpl} - \frac{1}{3}\varepsilon_{ijl} O^{(m)}_{pkl}\Big], \tag{5.28}$$

$$\int J_i r_j r_k r_p r_a dv = \Big[\frac{1}{5}\dot{X}_{ijkpa} - \frac{1}{4}\varepsilon_{ipl} Y_{jkal} - \frac{1}{4}\varepsilon_{ikl} Y_{jpal} - \frac{1}{4}\varepsilon_{ijl} Y_{pkal} - \frac{1}{4}\varepsilon_{ial} Y_{pjkl}\Big]. \tag{5.29}$$

The repeated terms of magnetic multipole moments with various combinations of indices in equations (5.27-5.29) are simplified, due to their **E**-field contributions differs only in dummy indices.

Let us consider for example contribution of magnetic octupoles to the **E**-field equation (from the first part of supplementary) and equation (5.28). Taking into account that $n_i = R_i/R = \{n_x, n_y, n_z\}$, where $R_i$ defines the vector from an origin to an observation point and $|n_i| \leq 1$.

$$\frac{1}{3}\Big[\varepsilon_{ipl}M_{jkl} + \varepsilon_{ikl}M_{jpl} + \varepsilon_{ijl}M_{pkl}\Big]n_j n_k n_p = \frac{1}{4}\Big[\varepsilon_{ipl}\int r_j r_k (\mathbf{r}\times\mathbf{J})_l dv +$$
$$+\varepsilon_{ikl}\int r_j r_p (\mathbf{r}\times\mathbf{J})_l dv + \varepsilon_{ijl}\int r_p r_k (\mathbf{r}\times\mathbf{J})_l dv\Big]n_j n_k n_p = \varepsilon_{ipl}M_{jkl}n_j n_k n_p. \quad (5.30)$$

In a result for the equations (5.25-5.29) we have:

$$\int J_i dv = \dot{p}_i \quad (5.31)$$

$$\int J_i r_j dv = \frac{1}{2}\Big[\dot{Q}^{(e)}_{ij} - \varepsilon_{ijk}m_k\Big], \quad (5.32)$$

$$\int J_i r_j r_k dv = \frac{1}{3}\dot{O}^{(e)}_{ijk} - \varepsilon_{ikl}Q^{(m)}_{jl}, \quad (5.33)$$

$$\int J_i r_j r_k r_p dv = \frac{1}{4}\dot{S}_{ijkp} - \varepsilon_{ipl}O^{(m)}_{jkl}, \quad (5.34)$$

$$\int J_i r_j r_k r_p r_a dv = \frac{1}{5}\dot{X}_{ijkpa} - \varepsilon_{ipl}Y_{jkal}. \quad (5.35)$$

# 6 Basic and toroidal fields distributions

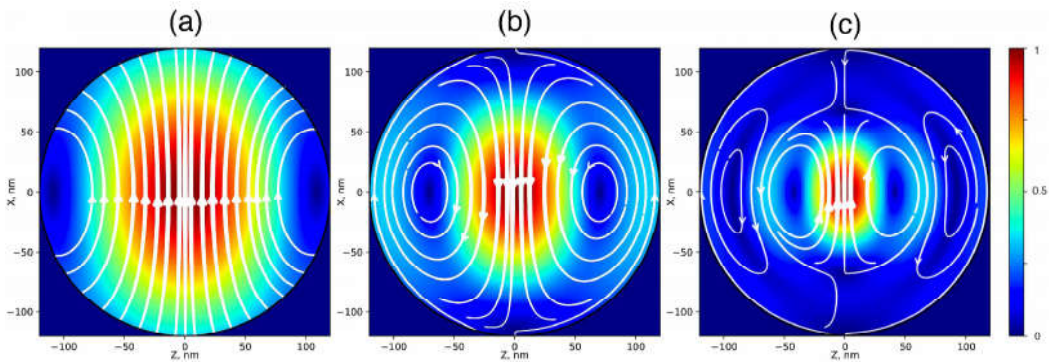

Figure 2: The normalized electric fields corresponding to the full electric dipole inside a particle with the refraction index n = 4 and (a) kr=0.754, (b) kr=1.091, (c) kr=1.909. Figure (a) does not demonstrate curl-like configurations of poloidal displacement currents and corresponds to the map of basic Cartesian dipole without toroidal contribution. Figure (b) and (c) shows field maps with a significant toroidal contribution of the first and second order toroidal moments.

For an analysis of field configurations inside a particle, we plot the fields distributions corresponding to individual Mie multipoles inside a particle with refractive index n=4 in the regions where toroidal moments sufficiently contribute to the far-field and in the region with the basic Cartesian multipoles only. In contrast to other toroidal moments, the $5^{th}$ order approximation of the Cartesian multipoles enables to obtain two toroidal terms for an electric dipole moment (Equation (4.3) and (4.13)). The Figure 2 (a) shows the electric field distribution (normalized to a maximal value) for a Cartesian dipole with the predominating role of the basic part.

The appearance of poloidal-like field configurations in Figure 2 (b) and (c) indicates the presence of the toroidal moments. The streamlines in Figure 2 (b) show the two singular points (deep blue) surrounded by vortices corresponding to the first order toroidal moment. In Figure 2 (c) one can see 4 singular points and the corresponding vortices of the $2^{nd}$ order toroidal dipole moment.

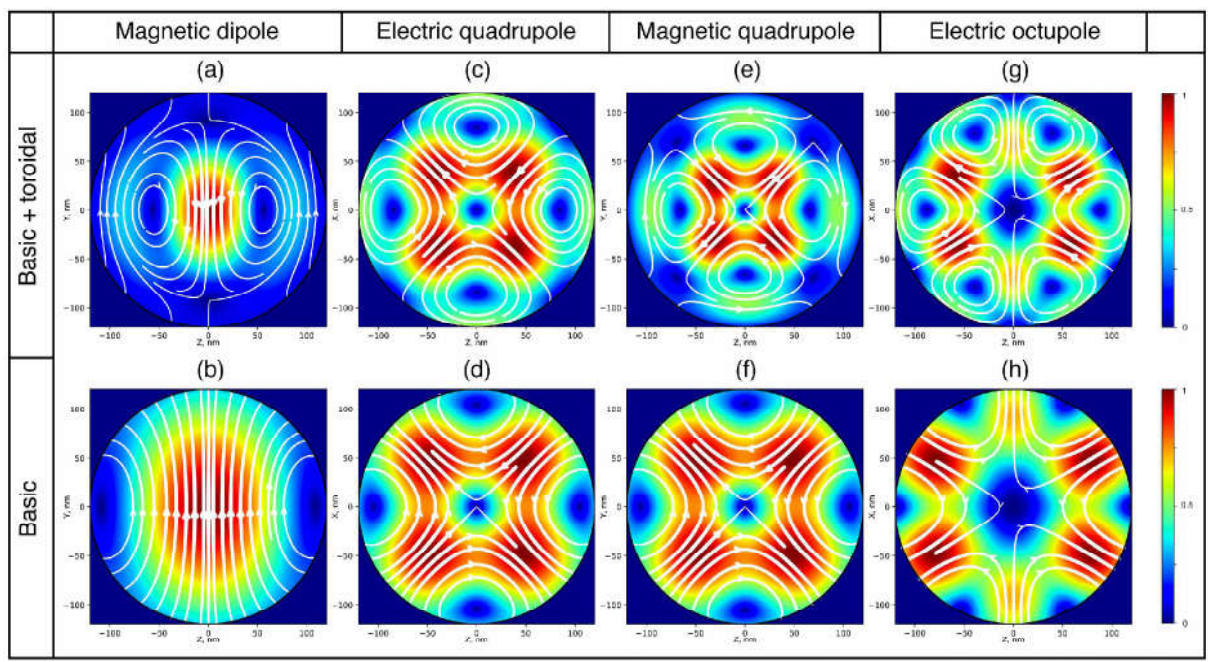

Figure 3: Normalized to a maximal value electric field (c,d,g,h), and normalized magnetic field (a,b,e,f) for the full multipoles calculated via Mie theory: magnetic dipole (a) kr=1.480 and (b) kr=0.754, electric quadrupole (c) kr=1.533 and (d) kr=1.109, magnetic quadrupole (e) kr=1.821 and (f) kr=1.109, and electric octupole (g) kr=1.934 and (h) kr=1.256. Field distributions (a), (c), (e), (g) are calculated for the size parameters allowing for sufficient toroidal terms, and demonstrate curl-like configurations. In contrast, Figure (b), (d), (f), (h) show the field configurations for the size parameters corresponding to the excitation of basic multipoles only.

Similarly with the electric dipole, higher order toroidal moments up to magnetic quadrupole and electric octupole are manifested by poloidal-like current configurations in the Figure 3 (a,c,e,g). The approximation of $5^{th}$ order Cartesian multipoles is not include the magnetic toroidal octupole moment and the higher toroidal multipoles (see Figure 4 (a,c,e,g)). The singular points and the toroidal-like field configurations indicate sufficient contribution of the toroidal moments in the regions close to anapole states. As it can be see, the number of singular points increases by 2 for each type of multipole (e.g.quadrupole, octupole, etc) and the scattered magnetic field concentrates closer to a center of the particle. Figure 3 (b,d,f,h) and 4 (b,d,f,h) corresponds to the maps of basic Cartesian multipoles without any toroidal contributions.

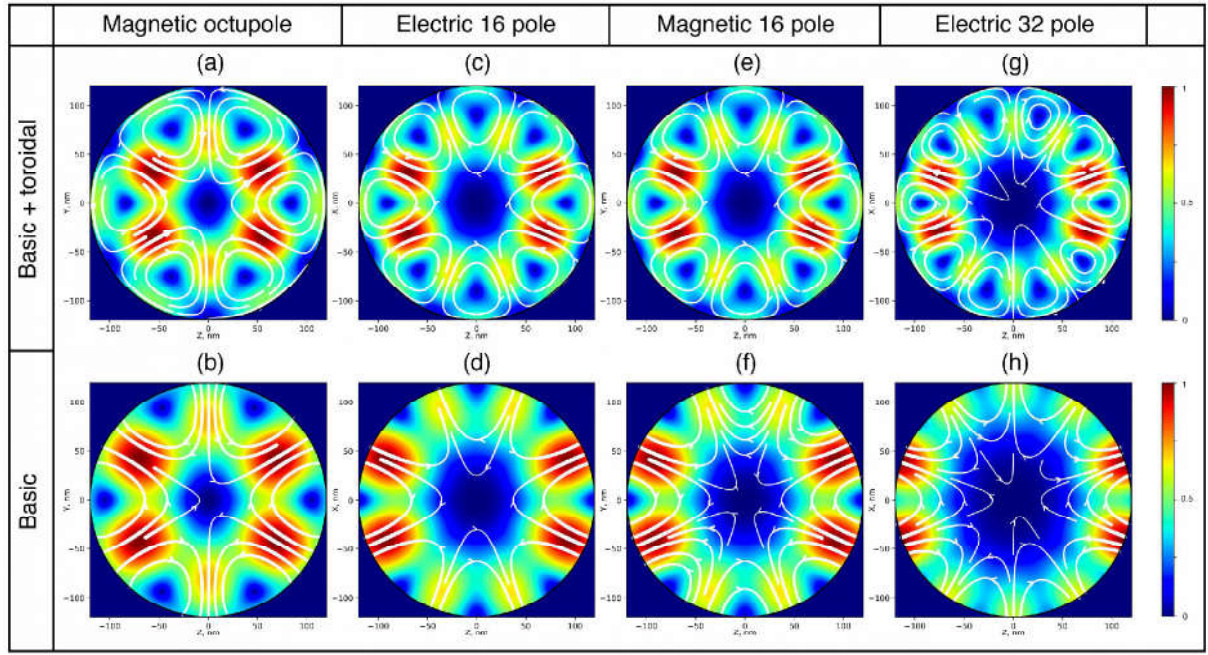

Figure 4: Normalized to a maximal value electric field (c,d,g,h), and normalized magnetic field (a,b,e,f) for the full multipoles calculated via Mie theory: magnetic octupole (a) kr=2.011 and (b) kr=1.256, electrtic 16 pole (c) kr=2.154 and (d) kr=1.508, magnetic 16 pole (e) kr=2.357 and (f) kr= 1.508, and electric 32 pole (g) kr=2.403 and (h) kr=1.508. Field distributions (a), (c), (e), (g) are calculated for the size parameters allowing for sufficient toroidal terms, and demonstrate curl-like configurations. In contrast, Figure (b), (d), (f), (h) show the field configurations for the size parameters corresponding to the excitation of basic multipoles only.